\documentclass[twocolumn]{aastex63}
\usepackage{amsmath,amssymb}
\usepackage{graphicx}

\shorttitle{Faint Asteroids with DECam}
\shortauthors{Heinze et al.}

\begin{document}

\title{The Flux Distribution and Sky Density of 25th Magnitude Main Belt Asteroids}

\author{A. N. Heinze}
\affiliation{Institute for Astronomy, University of Hawaii, 2680 Woodlawn, Honolulu, HI, 96822, USA; aheinze@hawaii.edu}

\author{Joseph Trollo}
\affiliation{Asana, 1550 Bryant St \#200, San Francisco, CA 94103, USA; joseph.trollo@gmail.com}

\author{Stanimir Metchev}
\affiliation{Department of Physics and Astronomy, University of Western Ontario, 1151 Richmond St, London, ON N6A 3K7, Canada; smetchev@uwo.ca}
\altaffiliation{Department of Physics and Astronomy, Stony Brook University, Stony Brook, NY 11794-3800, USA}

\correspondingauthor{A. N. Heinze} 

\begin{abstract}
Digital tracking enables telescopes to detect asteroids several times fainter than conventional techniques. We describe our optimized methodology to acquire, process, and interpret digital tracking observations, and we apply it to probe the apparent magnitude distribution of main belt asteroids fainter than any previously detected from the ground. All-night integrations with the Dark Energy Camera (DECam) yield 95\% completeness at $R$ magnitude 25.0, and useful sensitivity to $R=25.6$ mag when we use an analytical detection model to correct flux overestimation bias. In a single DECam field observed over two nights, we detect a total of 3234 distinct asteroids, of which 3123 are confirmed on both nights. At opposition from the Sun, we find a sky density of $697 \pm 15$ asteroids per square degree brighter than $R = 25.0$ mag, and $1031 \pm 23$ brighter than $R = 25.6$ mag. We agree with published results for the sky density and apparent magnitude distribution of asteroids brighter than $R=23$ mag. For a power law defined by $dN/dR \propto 10^{\alpha R}$, we find marginally acceptable fits with a constant slope $\alpha = 0.28 \pm 0.02$ from $R=20$ to 25.6 mag. Better fits are obtained for a broken power law with $\alpha=0.218 \pm 0.026$ for $R=20$ to 23.5 mag, steepening to $\alpha=0.340 \pm 0.025$ for $R = 23.5$ to 25.6 mag. The constant or steepening power law indicates asteroids fainter than $R = 23.5$ mag are abundant, contrary to some previous claims but consistent with theory. 
\end{abstract}


\section{Introduction}
The magnitude and size distribution of main-belt asteroids (MBAs) holds clues to their physical properties and collisional evolution \citep[e.g.][]{Cheng2004,Bottke2005b,Durda2007}. Digital tracking enables the detection of asteroids up to ten times fainter than conventional methods \citep[e.g.][]{Zhai2014,digitracks}, allowing us to probe the main belt population down to sizes as small as 100 meters with 4-meter class telescopes. 

A significant change in asteroid properties is believed to occur in this size range: the transition from large objects whose cohesive strength comes mainly from self-gravity to small ones held together by material tensile strength. Laboratory experiments and numerical hydrocode modeling indicate this transition occurs at a size somewhat larger than 100 meters \citep[][and references therein]{Bottke2005a}. These theoretical results are independently supported by measurements of asteroid rotation periods. Asteroids larger than about 300 meters observe a `spin barrier', almost never having periods shorter than two hours, while smaller objects can rotate much faster \citep[see, e.g.][]{Hergenrother2011}. The two-hour spin barrier corresponds to the rotation period at which centrifugal breakup occurs for a strengthless, gravitationally bound object with typical asteroidal density\footnote{Typical density is about 2 g cm$^{-3}$ \citep{Carry2012}. The rotation period for centrifugal breakup of a strengthless gravitationally bound object depends only on density, not size. This is because for a given density and rotation period, surface gravity and centrifugal force both increase linearly with the object's radius.}. Hence, asteroids larger than a few hundred meters probably are strengthless `rubble piles', while smaller asteroids have nonzero internal strength \citep{Hergenrother2011}. Based on the expected greater strength of the smallest asteroids, models predict that very small MBAs will be more abundant than naive extrapolation of power-law fits to the distribution of larger objects would indicate \citep{Bottke2005b,deElia2007}. While near-Earth objects (NEOs) are routinely observed in this size range (hence the discovery that they violate the spin barrier), the observations we report herein may be the first to probe extremely small MBAs with statistical power sufficient to test model predictions of an upturn in their size-frequency distribution.

Very small MBAs have short collisional lifetimes in spite of their greater tensile strength \citep{Bottke2005b,Henych2018}. Therefore they are younger than their larger counterparts, having been produced relatively recently by the breakup of larger bodies due to collision or (more rarely) to YORP-induced fission \citep[e.g.][]{Agarwal2013}. Their abundance and dynamical distribution can tell us about the recent collisional history of the main asteroid belt. Small MBAs also frequently impact larger objects. These impacts produce craters, change the spin axis of the target body, and in some cases even disrupt it. Hence, we must know the abundance of small MBAs to understand the collisional environment inhabited by larger asteroids.

Studying small MBAs can enhance our understanding of NEOs, since most \citep[about 94\%; see][]{Bottke2002} NEOs originate from the main belt \citep[e.g.][]{Morbidelli1999,OBrien2005,deElia2007,Minton2010}. Relative to large objects, small MBAs are more sensitive to the Yarkovsky effect \citep{Farinella1998,Nesvorny2004}, which is believed to be the main process that transports MBAs into the unstable orbital resonances from whence they evolve into Earth-crossing orbits \citep{Bottke2006}. Since near-Earth objects (NEOs) approach Earth more closely than MBAs, they can be detected down to small sizes: in fact, the vast majority of known NEOs are smaller than 1 km. Studying the same size cohort in the main belt will elucidate the selection effects involved in the main-belt to NEO transition and help us understand both populations in more detail.

Selection effects would be expected to include preferential transfer of the smallest objects, because they are most sensitive to the Yarkovsky effect \citep[e.g.][]{Nesvorny2004}. {This would suggest that the size distribution of small MBAs should be shallower than that of NEOs, since NEO dynamical lifetimes are far too short for them to reach a new collisional equilibrium after migrating from the main belt \citep{Bottke2002}. Another expected selection effect has to do with location in the main belt: NEOs should preferentially come from relatively small regions close to unstable orbital resonances. \citet{Bottke2002} predict that 61\% of NEOs are delivered from the main belt by just two strong resonances: the $\nu_6$ secular resonance\footnote{An asteroid in the $\nu_6$ secular resonance is dynamically unstable because its perihelion precesses at the same rate as that of Saturn, the 6th planet.} that sculpts the inner edge of the main belt, and the 3:2 mean-motion resonance with Jupiter that carves out the deepest Kirkwood Gap at $a=2.5$ AU. By measuring the abundance and size distribution of small MBAs dynamically near these resonances, we can assess the efficiency with which the Yarkovsky effect and resonant interactions move asteroids into near-Earth orbits. Intriguingly, \citet{Farnocchia2013} find a large majority of NEOs to have retrograde rotations. This implicates the $\nu_6$ secular resonance as a major source of NEOs, because it has the unique property that only MBAs with retrograde rotation can be transported into it by the Yarkovsky effect.

Herein, we describe the observational methodology for detecting extremely faint asteroids and probe the statistical distribution of their apparent magnitudes, thereby laying the foundation required in order to address the scientific questions raised above. Compared to our pilot project described in \citet{digitracks}, the current work reaches much fainter magnitudes, uses more far more mature methodology and statistical analysis, and most importantly includes an incompleteness correction to determine the apparent magnitude distribution of main belt asteroids fainter than have ever previously been systematically probed. Our current data are also sufficient to determine the absolute magnitude distribution of very small MBAs, which can be directly compared with that of NEOs to probe size-dependent selection effects in the MBA to NEO transition. We defer our analysis of these absolute magnitude distributions to our companion paper (Heinze et al., in prep.), which will rely heavily on the results presented herein. More detailed analysis of dynamical selection effects requires new data sets spanning more than two nights. The current work lays methodological foundations for such future research.

Previous studies of small MBAs include \citet{Gladman2009}, who used the 4m Mayall telescope at Kitt Peak to probe asteroids as faint as $R$ magnitude 23.5; \citet{Parker2008}, who measured the size distributions of asteroid collisional families using data from the Sloan Digital Sky Survey; \citet{Yoshida2003} and \citet{Yoshida2007}, who used the 8m Subaru telescope to probe to $r$ magnitude $\sim 24.4$; \citet{Terai2007} and \citet{Terai2013}, who also probed magnitude $24$ asteroids with Subaru, specifically targeting off-ecliptic fields to study high inclination objects; and \citet{Wiegert2007}, who probed the $g$ and $r$ colors of asteroids down to magnitude $23$ using CFHT.

The observations we present herein are 95\% complete at $R=25.0$ and reach their 50\% completeness limit at $R=25.3$. Hence, they constitute the most sensitive ground-based asteroid survey to date. Since we observed the same field on two consecutive nights, we are able to confirm the reality of most of our asteroids by detecting the same objects in two independent data sets. Two-night detection will also enable us to obtain distances with $\sim 1.5$\% accuracy (and hence absolute magnitudes) using the RRV method of \citet{curves} and \citet{LinRRV}. We reserve our analysis of the distance determinations and the absolute magnitude and size distributions for a companion paper (Heinze et al., in prep.), which will rely heavily on the analyses of completeness and flux overestimation bias we develop herein.

We describe our observations and image processing in Sections \ref{sec:obs} and \ref{sec:dataproc}. In Sections \ref{sec:digitracks} and \ref{sec:post} we present the digital tracking methodology that allows us to find and measure extremely faint asteroids. Section \ref{sec:fp} quantifies the expected number of false detections among our confirmed asteroids, demonstrating that it is negligible. In Section \ref{sec:fn} we analyze our detection completeness as a function of magnitude using a sophisticated fake-asteroid simulation. Section \ref{sec:2night} describes how we link detections of the same asteroid from one night to the next. Our main results are in Sections \ref{sec:counts} and \ref{sec:statbias}: the latter describes how we model and correct flux overestimation bias, and presents the first-ever determination of the sky density and apparent magnitude distribution of MBAs down to $R$ magnitude 25.6. We offer our conclusions in Section \ref{sec:conc}.

\section{Observations} \label{sec:obs}
We observed using the Dark Energy Camera (DECam) on the 4 meter Blanco Telescope at Cerro Tololo Inter-American Observatory (CTIO) on UT 2014 March 30, March 31, April 07, and April 08. Our primary objective was to detect and characterize extremely faint MBAs using digital tracking. 

The DECam field of view is 2 degrees in diameter and comprises 60 active science CCDs with dimensions of $4096\times 2048$ pixels, for a total area of 503 megapixels (see Figure \ref{fig:DECamdet}). The pixel scale varies from 0.2637 arcsec/pixel at field center to 0.2626 arcsec/pixel at the edge, and the horizontal and vertical gaps between CCDs are 201 pixels and 153 pixels, respectively (NOAO Data Handbook\footnote{The NOAO Data Handbook is available at \url{http://ast.noao.edu/sites/default/files/NOAO\_DHB\_v2.2.pdf}}). A single exposure captures 2.7 square degrees of sky onto active science CCDs, while a simple three-exposure dither pattern will fill in the gaps between detectors and deliver contiguous coverage over 3 square degrees.

\subsection{Choice of Target Fields}

We planned our observations to achieve two primary requirements: to detect the smallest asteroids possible in the main belt, and to recover as many as possible of them on multiple nights to constrain their distances and orbits. The first two nights (March 30-31) constituted a `discovery run' in which thousands of new asteroids would be found and measured over two nights. Measurements on two consecutive nights confirm the reality of faint objects and enable the calculation of accurate distances using the RRV method \citep{curves,LinRRV}. We intended the second pair of nights (April 07-08) as a `recovery run' in which most of the asteroids would be recovered and approximate orbits could be calculated from the resulting 8-9 day arcs. 

The requirement of detecting the smallest MBAs dictates targeting observations near the antisolar point, where asteroids are at their brightest because of their low phase angles. The requirement of recovering the same asteroids on additional nights dictates offsetting the target field from night to night in order to follow the mean sky motion of MBAs. Since the antisolar point moves eastward while MBAs near opposition are moving westward in their retrograde loop, our target field could be centered accurately on the anti-solar point for only one night. 

We chose this to be the second night of our discovery run because the antisolar field on that night had fewer bright ($\lesssim$8th magnitude) stars that could interfere with asteroid detection. Hence, our target field for the first night was centered not on the antisolar point, but rather on a set of asteroids whose average position on the following night would coincide with the antisolar point. As the antisolar point moves eastward by about one degree per day, while MBAs near opposition move about 0.2 degrees westward every day, our March 30 field was centered roughly 1.2 degrees east of the antisolar point on that date.

On our first night, UT March 30, we centered our observations at RA 12:38:12.51, Dec -04:07:40.4. This is 1.22 degrees away from the antisolar point at 04:00 UT March 30 (approximately local midnight at CTIO). The average sky motion of MBAs in this field on this date is $-0.215\deg$/day in RA and $0.092\deg$/day in Dec. Applying these offsets to our March 30 field gives the target for our March 31 observations: RA 12:37:20.84, Dec -04:02:08.7. This location is within one arcminute of the antisolar point at local midnight on March 31: hence, our targeting was almost perfectly optimized. The target fields for the two nights overlap heavily since the DECam field of view is much larger than the offset between them, but the offset was important because it minimized the loss of asteroids off the edges of the field from one night to the next. 

To detect the faintest possible asteroids we targeted just a single DECam field on each night. During brief intervals near the beginning and end of the night when our target field was at airmass greater than 2.0, we observed photometric calibration fields and targets of opportunity. 

\subsection{Observing Conditions and Acquired Data}

On March 30, we acquired a total of 147 images of the target field. All were 90-second exposures: 131 in the wide VR filter for maximum sensitivity, and 16 more in the r filter for calibration purposes. On March 31 we acquired 225 images of our primary field, each one a 90-second exposure. Of these, 210 were in the VR filter and 5 each were in the g, r, and i filters. The reason for the smaller number of images on March 30 was the failure of one of the DECam instrument computers, which took about three hours to recover. On both nights we dithered the pointing after every exposure, following a quasi-random pattern to ensure maximum cancellation of instrumental systematics and artifacts. We used a large number of DECam observing scripts that each took a few images dithered on a hexagon or linear sequence, and we built up the quasi-random pattern by continually changing the scale and orientation of these regular sequences. Our dither amplitudes were chosen to fill in the gaps between detectors seamlessly, while keeping to the same field as much as possible.


As described above, the nights of March 30-31 constituted our discovery run. Based on extrapolating the power laws of \citet{Gladman2009}, we expected to find about three thousand asteroids down to a limiting magnitude fainter than 25.0, and to confirm their reality by detecting them on both nights. On April 07-08, we intended to recover most of the newly discovered asteroids and calculate approximate orbits from the resulting 8-9 day arcs. 

We were blessed with mostly clear weather on all four nights, but the seeing during the recovery run was extremely poor, in the 3-4 arcsecond range. Background-limited sensitivity to faint objects has the same dependence on seeing as it has on telescope aperture: thus, a 4-meter telescope in 4 arcsecond seeing is no better than a 1-meter telescope in 1 arcsecond seeing. Our sensitivity on April 07 and 08 is at least one magnitude worse than during our discovery run, and approximate orbits will be calculable only for a minority of the new asteroids found in our discovery run. Fortunately, we can use the rotational reflex velocity (RRV) method \citep{curves, LinRRV} to calculate the distances and absolute magnitudes of all the newly discovered asteroids using only the data from March 30 and 31, when the seeing was 1-1.5 arcsec (Figure \ref{fig:seeing}). Our 50\% completeness limit on these nights was fainter than 25th magnitude, representing a regime of flux and absolute magnitude that has never before been systematically analyzed in the main belt. Hence, we focus our current analysis on these nights, deferring to a future paper the analysis of orbital statistics for the brighter subset of asteroids recovered on April 07-08.

\begin{figure*} 
\plottwo{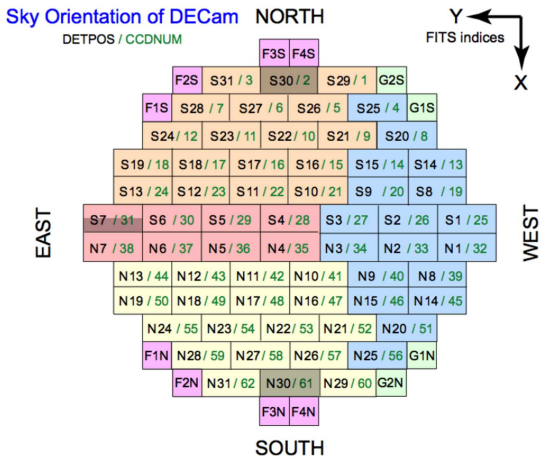}{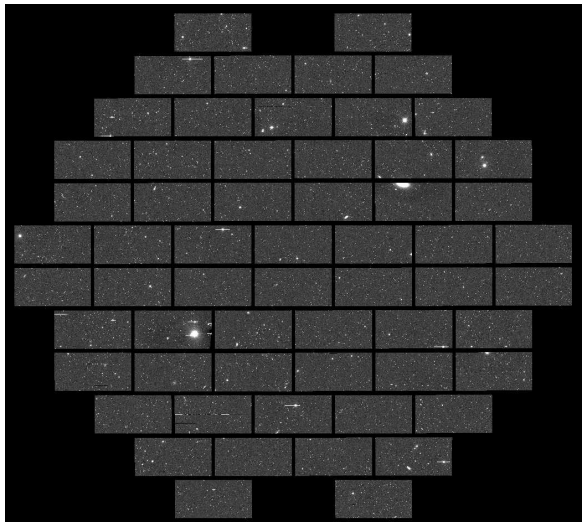}
\caption{{\em Left:} Diagram of the DECam focal plane, taken from the NOAO Data Handbook. CCDs labeled with F or G (colored magenta or green) are for focus or guiding, respectively, while those whose labels begin with N (north) or S (south) are science CCDs which get their own extensions in the multi-extension FITS images that are DECam's raw output. The colors of the science CCDs (orange, pink, blue, yellow) indicate which of four sets of readout electronics controls them. {\em Right:} A single 90-second exposure from our data set, with the images from all detectors repixellated onto a master astrometric grid. Note that detectors N30 and S30 from the diagram are dead, but all other science detectors are fully functional. \label{fig:DECamdet}}
\end{figure*}

\begin{figure*}
\plottwo{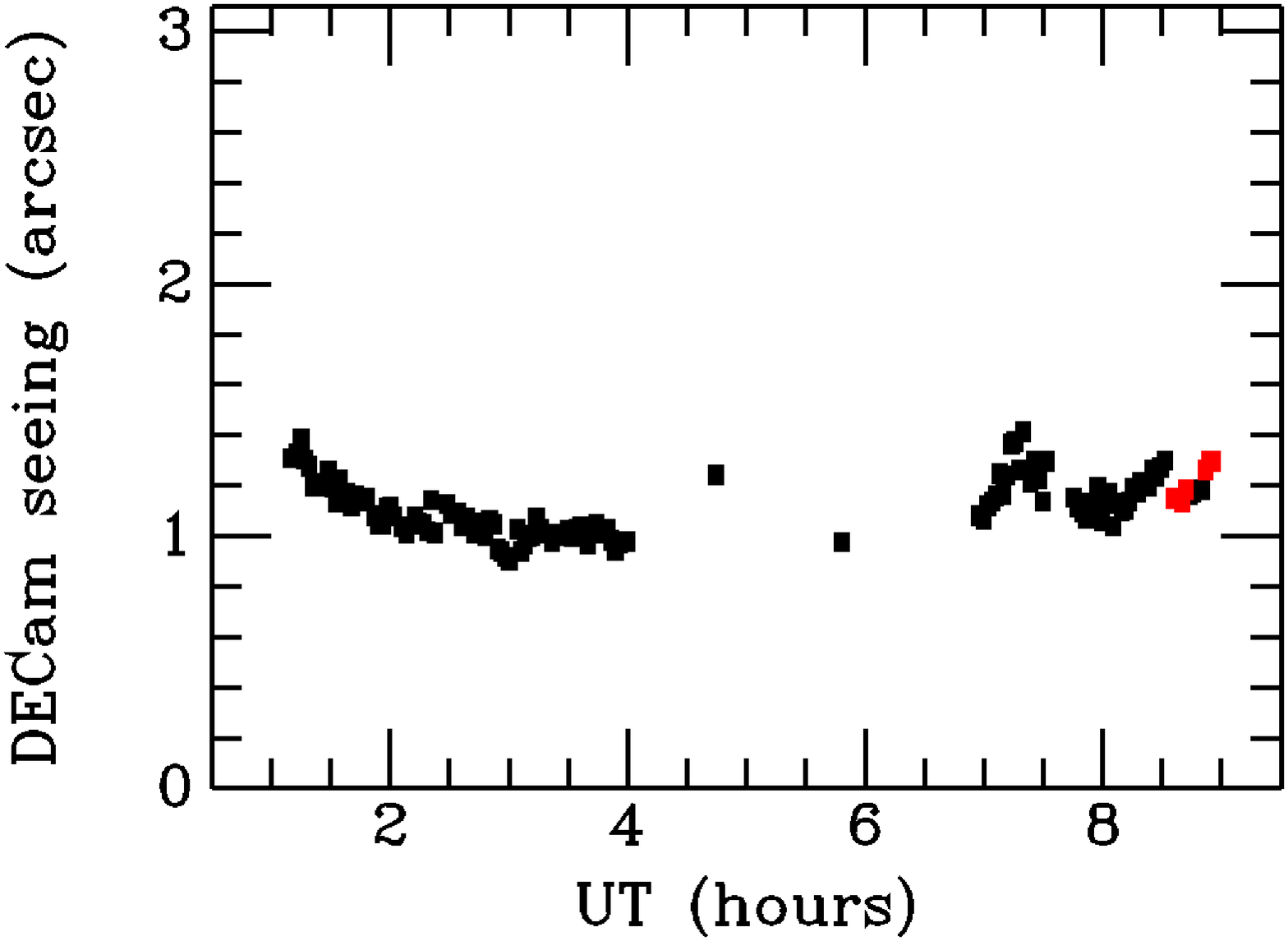}{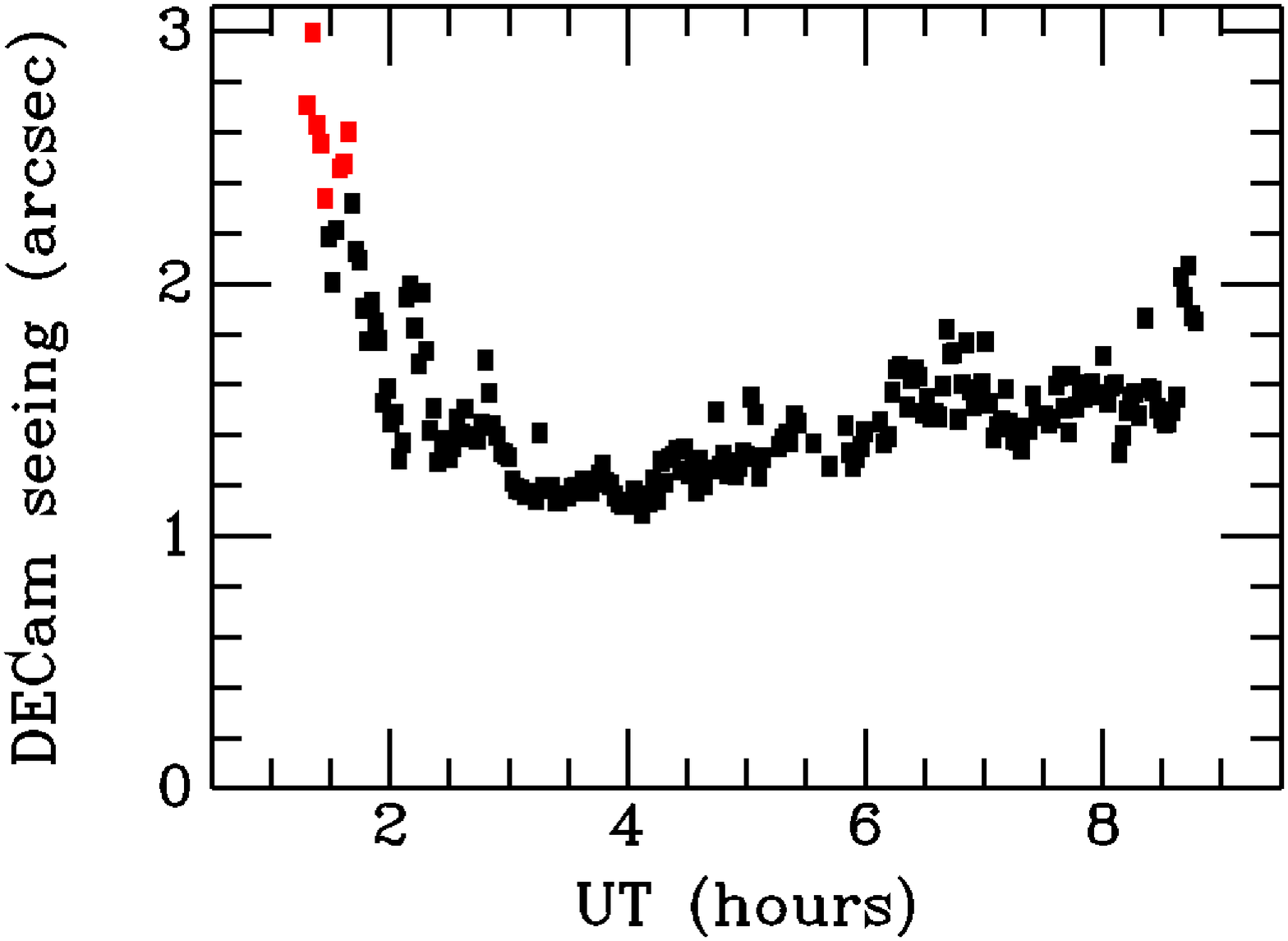}
\caption{{\em Left:} Seeing as a function of time for all VR images of our antisolar field obtained on March 30. The large gap between UT 4 and UT 7 is due to the failure of two of the control computers for DECam. The instrument was restored to full operation before dawn. {\em Right:} The same plot for our March 31 observations. Red points indicate poor quality images not used in our final digital tracking stacks. Bad images on March 31 were mostly due to poor seeing. Note that seeing is directly proportional to the PSF radius $r$ used to calculate the noise parameter $\xi$.  \label{fig:seeing}}
\end{figure*}

\begin{figure*}
\plottwo{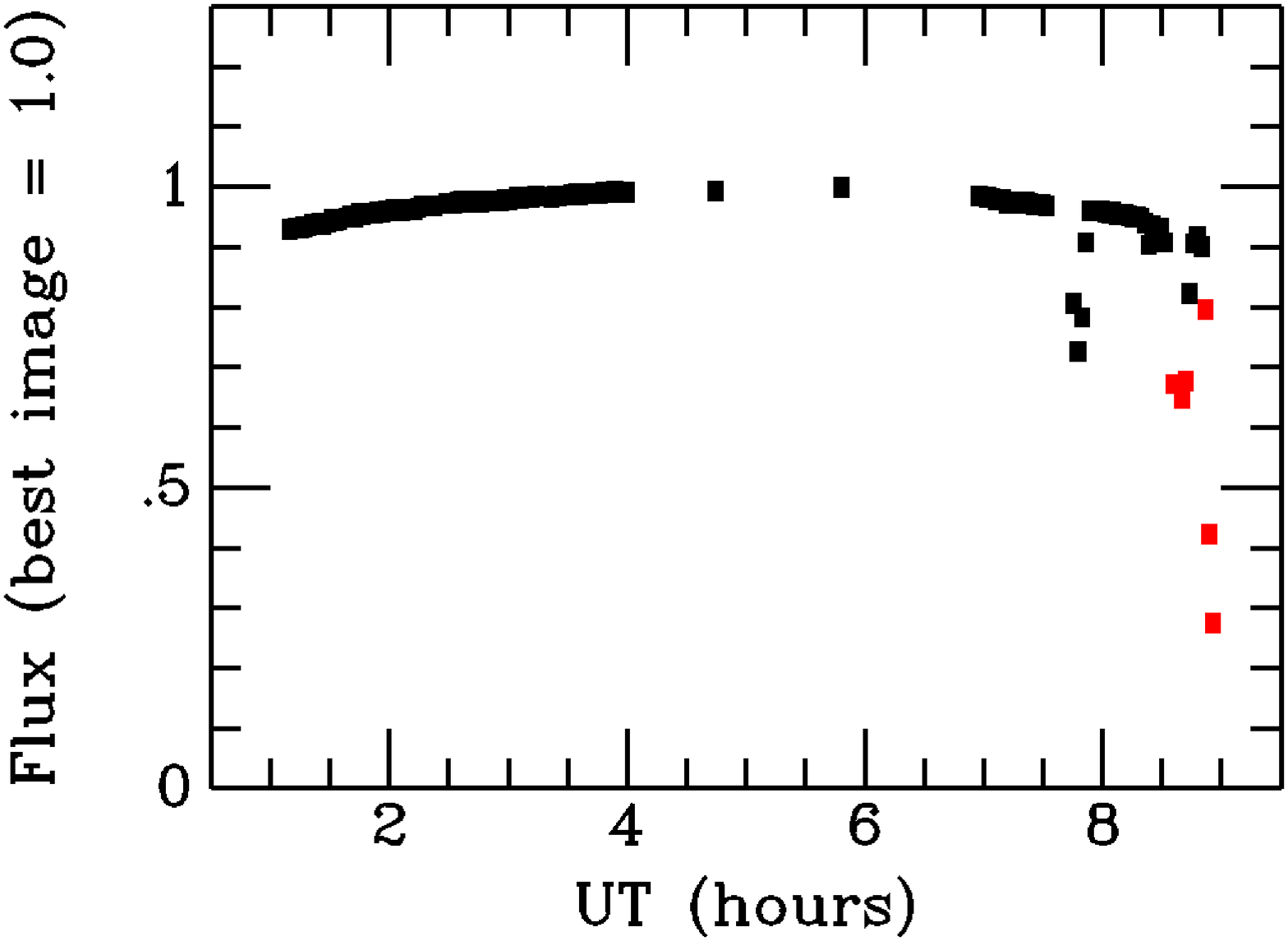}{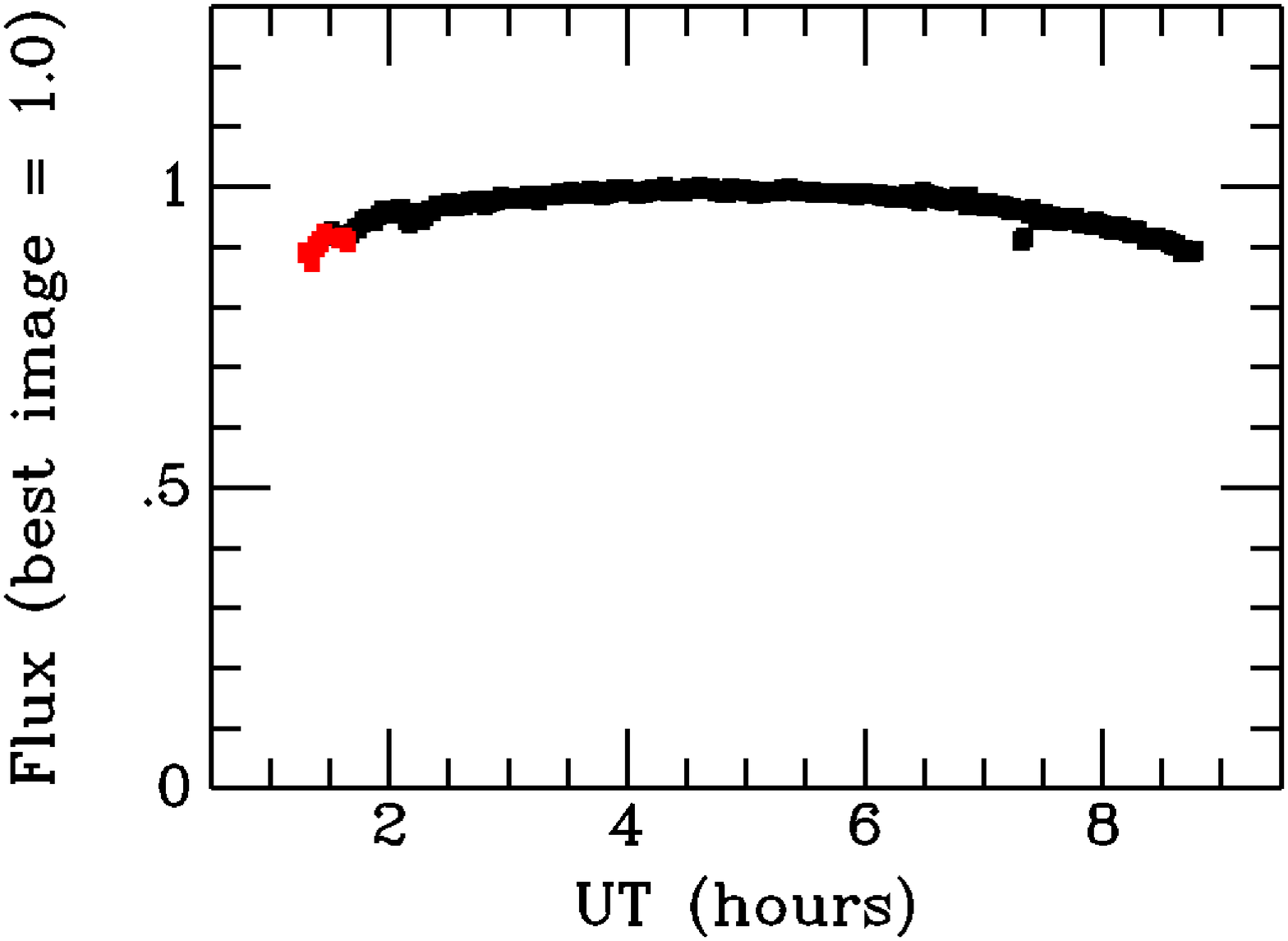}
\caption{{\em Left:} Atmospheric transmission $\epsilon$ relative to the best image for all VR images of our antisolar field obtained on March 30. {\em Right:} The same plot for our March 31 observations. Red points indicate poor quality images not used in our final digital tracking stacks. The overall curvature on both nights is the effect of airmass. The sky was clear all night on March 31, but on March 30 some of our latest images were affected by clouds and rejected for this reason.
\label{fig:flux}}
\end{figure*}

\begin{figure*}
\includegraphics[scale=5.6]{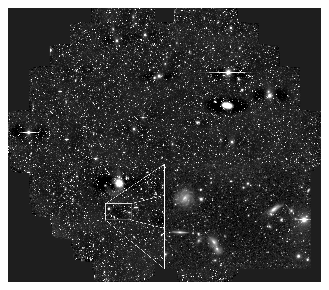}
\caption{A star-registered all-night integration from March 31, created by median-stacking the 202 images we ultimately used for asteroid detection for that night. Sensitivity to stars and galaxies extends well past $R=25$ mag, and digital tracking enabled us to achieve comparable sensitivity for unknown asteroids. Even the inset only hints at the tens of thousands of faint galaxies visible in the full-size image.
\label{fig:fullstack}}
\end{figure*}

\subsection{Image Selection for Digital Tracking Analysis} \label{sec:imcull}

We obtained 147 images of our target field on March 30 and 225 on March 31. To ensure uniform image stacks, we applied our digital tracking analysis only to images acquired in the wide VR filter, restricting us to 131 images on March 30 and 210 on March 31. 

To identify any additional images that should be rejected from the final stack, we define a noise parameter $\xi$ that is inversely proportional to the SNR (or, equivalently, directly proportional to the fractional uncertainty on the measured flux) of a faint source whose detection is entirely background-limited. The background noise goes as the square root of the number of sky photons that overlap with the point spread function (PSF) of the source: hence, $\xi$ is proportional to the square root of the sky brightness $B$ times the effective radius $r$ of the PSF, which in turn is proportional to the measured seeing on the image. We take the seeing measurement $s_i$ on a given image to be the full width at half maximum (FWHM) of stars on that image (Figure \ref{fig:seeing}). The measured flux from the source varies directly with the atmospheric transparency, which we have called $\epsilon$ and plotted in Figure \ref{fig:flux}. Hence, our noise parameter for image $i$ is given by:

\begin{equation}
\xi_i  = \frac{s_i \sqrt{B_i}}{\epsilon_i}
\label{eq:noise}
\end{equation}

For optimal detection of faint, non-moving objects, we would create a stack of images weighted by $1/\xi^2$. To simplify the analysis, and in particular the determination of precise angular velocities, we elect not to apply such a variable weighting to our digital tracking stacks. We are therefore interested in using $\xi$ not to weight images, but rather to identify any that are of such poor quality that including them in an unweighted stack would \textit{reduce} the ultimate sensitivity. By interpreting $\xi$ as the fractional uncertainty on the measured flux of a hypothetical point source, we can identify a threshold value of $\xi$ that separates useful images from those that would only contribute noise. We sort the images in order of increasing $\xi$, and then explore the predicted fractional uncertainty on a uniformly weighted stack of these images, truncated at image $m$ before the end of the list:

\begin{equation}
\xi_{stack}  = \frac{\sqrt{\sum_{i=1}^m \xi_i^2}}{m}
\label{eq:stacknoise}
\end{equation}

Under Equation \ref{eq:stacknoise}, the images at the beginning of the ordered list (that is, those with the lowest $\xi_i$) contribute the greatest reduction in the value of $\xi_{stack}$, but poorer quality images continue to make some contribution through the bulk of the list. However, we find a small number of images at the end of the list actually do cause $\xi_{stack}$ to increase when they are included. The threshold values for $\xi$ are 66.0 and 81.0 for March 30 and 31, respectively, where the higher threshold for March 31 is due to the poorer seeing on that night. We reject images with $\xi_i$ exceeding these thresholds from our digital tracking analysis. Of a total 131 target-field images taken in the VR filter on March 30, we reject 8 and retain 123. Atmospheric transparency is the dominant cause of image rejection on this night, due to light clouds blowing across our target field late in the data sequence (see Figure \ref{fig:flux}). For March 31, of a total 210 images taken in the VR filter, we reject 8 and retain 202. The dominant cause of rejection is bad seeing near the beginning of the data sequence (see Figure \ref{fig:seeing}). 

Our digital tracking analysis proceeds with 123 images for March 30 and 202 for March 31. As each image is a 90 second exposure, the cumulative integrations are 3.08 hr and 5.05 hr, respectively. The temporal span from the first image to the last is 7.67 hr on March 30 and 7.30 hr on March 31. The mean seeing over accepted images was 1.11 arcsec on March 30 and 1.46 arcsec on March 31. Figure \ref{fig:fullstack} illustrates the clean, deep image obtained by stacking the 202 images used for asteroid detection on March 31.

\section{Data Processing} \label{sec:dataproc}

\subsection{Basic Processing}

We begin our processing of DECam data by reading the compressed, multi-extension fits.fz file for each image using CFITSIO, and converting it into multiple, single-extension, uncompressed fits files, each corresponding to a single detector. Our processing then proceeds for several steps almost as if each individual detector were a separate CCD imager.  A standard sequence of calibration images, including biases and flats in all relevant bands, was obtained on each night of our observations.  For each night, we create a master bias frame for each detector using a median stack.  We create flat frames in the same manner, after bias-subtraction and normalization.  However, the normalization is not performed independently for each detector.  Instead, all the subframes corresponding to a given flat image are normalized by the same factor: the factor required to obtain a unit mean on detector N04, which we choose as our reference detector because of its central position (see Figure \ref{fig:DECamdet}).  By this procedure, we allow the flatfield itself to remove detector-to-detector variations in quantum efficiency.



We proceed to bias-subtract and flatfield all of our science frames independently. We then subtract the sky backgrounds using a polynomial model of the sky emission on each detector.  A two-dimensional quadratic model produces excellent results, after the stars are iteratively rejected from the fit.  As the DECam detectors show no noticeable fringing in our wavelengths of interest, more advanced methods of sky subtraction are not necessary.

\subsection{Astrometric Re-pixellation} \label{sec:repix}

We begin the task of re-pixellating our images on a consistent astrometric grid by by fitting RA and DEC as quadratic functions of x and y pixel position on each detector, using Gaia DR1 as our astrometric reference catalog. We use the resulting astrometric fit to place the data from each subframe of a given image into a single large, astrometrically registered array, which has dimensions of 32,000$\times$28,000 pixels and an invariant pixel scale of 0.2625 arcsec/pixel. We use the same simple map projection as \citet{digitracks}: given a reference RA and Dec $\alpha_0$ and $\delta_0$, and a reference pixel location $x_0$ and $y_0$ (both corresponding to the exact center of the repixellated image), the $x,y$ pixel coordinates of a given $\alpha,\delta$ position are given by:

\begin{equation} \label{eq:astrom}
\begin{array}{lcl}
x-x_0 &=& (\alpha_0 - \alpha) cos(\delta) / s_{pix} \\
y-y_0 &=& (\delta - \delta_0) / s_{pix} \\
\end{array}
\end{equation}

Where $s_{pix}$ is the invariant pixel scale. More sophisticated projections would also be compatible with digital tracking. We select this relatively crude option because in \citet{digitracks} we intensively analyzed its behavior with respect to digital tracking, demonstrating that it would produce good results as long as the target declination is less than 60 degrees from the celestial equator. As the current data span a declination range of only -5.1 to -3.0 degrees, the distortion is very small.

In constructing the repixellated image, we trim off and discard the outermost 10 pixels of each subframe to avoid edge anomalies present on the detectors. Due to variations in the pixel scale of raw DECam images, our re-pixelization creates systematic photometric variations from center to edge across the full DECam field. As the full extent of these photometric effects is less than 1\%, they are negligible for our purposes.

Repixelation using Gaia delivered systematic astrometric error so small it is difficult to measure. We have determined that the median systematic error is less than 0.007 arcsec. The maximum systematic error occurs near the image edges, and appears to be about 0.013 arcsec. These errors are negligible compared to random errors for our target asteroids, and will make no appreciable contribution to errors on distances computed using the RRV method. 

The effects of saturated bleed-streaks from bright stars must be handled during repixelation in a way that enables us to remove them without leaving artifacts. We find that pixels immediately adjacent to bleeds typically have anomalous values, so we mark all such pixels as saturated. Where a bleed-streak reaches the edge of DECam detector, it broadens out as the electrons `pool' at the detector edge, and induces a variable negative offset in the pixel values over a wide region surrounding the pool. Since this offset affects only a tiny fraction of our data, we elect simply to mask the affected pixels.

Our repixelation uses bilinear interpolation, but reverts to nearest-neighbor sampling whenever any of the four pixels being interpolated is masked or saturated.  This preserves hard edges in the resampled images, and enables us to remove all saturated bleeds at the image subtraction stage.

\subsection{Photometric Calibration} \label{sec:photcal}

Photometric calibration of our data is complicated by the use of the nonstandard, wide VR filter. Our objective is to report magnitudes that correspond as closely as possible to the magnitudes that the same objects would have in the Johnson $R$ filter. We choose this filter because it overlaps essentially completely with the VR filter; and because it was the filter used by \citet{Gladman2009}, to which, as the most rigorous survey of faint MBAs prior to this work, we are interested in comparing our results.

For photometric calibration purposes, we obtained DECam images of a field centered at 14:42:00, -00:05:00 which is a standard calibration field for DECam because it contains many thousands of well-measured stars with SDSS photometry. We observed this field in the $g$, $r$, $i$, and VR filters, but elected in the end to extract our calibration from the VR observations only.

We used photometric transformations given at the SDSS website\footnote{\url{http://classic.sdss.org/dr4/algorithms/sdssUBVRITransform.html}} to convert from cataloged magnitudes in the SDSS filters to $R$ magnitudes. We experimented with two sets of transformation equations. One from \citet{Jester2005}, derived for stars with $Rc-Ic < 1.15$, is given in Equation \ref{eq:photcal1}. The other, derived by Robert Lupton in 2005, is unpublished but derived from a catalog query documented on the SDSS website. It results in Equation \ref{eq:photcal2}. 

\begin{equation} \label{eq:photcal1}
\begin{array}{lcl}
V      &=&    g - 0.59*(g-r) - 0.01 \\
V-R    &=&    1.09*(r-i) + 0.22 \\
\end{array}
\end{equation}

\begin{equation} \label{eq:photcal2}
   R = r - 0.2936*(r - i) - 0.1439
\end{equation}

We find that Equations \ref{eq:photcal1} and \ref{eq:photcal2} both give satisfactory results in their ability to predict the measured fluxes of stars on our VR images of the SDSS field, and the resulting magnitude zeropoints differ by only 0.026 magnitudes. The consistency of Equation \ref{eq:photcal1} appears slightly better when comparing the mean results from different starfields, and we adopt it. Thus, we are able to map $R$ band magnitudes derived from SDSS $gri$ photometry to fluxes measured on our VR images, and hence to calculate the approximate $R$ magnitudes of other objects based on our VR images. 

We do this using aperture photometry with a large aperture of radius 20 pixels (5.25 arcsec), deriving magnitude zeropoints of 25.798 at airmass 1.21 on March 30, and 25.771 and 25.756 at airmass 1.16 and 1.35, respectively, for March 31. Since these zeropoints are derived for repixelated images processed through the customized methodology described above, they would not be expected to correspond exactly to those determined for other pipelines. The difference of 0.027 magnitudes in the low-airmass zeropoints for March 30 and 31 appears to represent a genuine difference in atmospheric transparency during the measurements of the SDSS calibration field on the respective nights, in the sense that the atmosphere was slightly less transparent for the March 31 images. This was not true of March 31 in general, as we discuss below.

For final calibration of asteroid magnitudes, we wish to use stars measured on the same images as the asteroids. Hence, we use the SDSS-derived calibrations to measure the magnitudes of stars in the science fields, based on averaged results from subsets of 7 science images taken at approximately the same airmass as the SDSS calibration images. Final magnitudes of the asteroids are then based on these stars, without the need for further inter-image comparison. We select the set of stars to be measured on the science images according to strict criteria that will also make them appropriate templates for generating fake asteroids (see  \S \ref{sec:fakepix}). We selected them from among stars in Gaia DR1 with magnitudes ranging from 16 to 20. After rejecting all saturated or near-saturated objects, we extract postage stamps centered on the remaining stars and manually examine all of them, rejecting those that have neighbors contributing significant flux within a 40x40 pixel area. We perform this selection independently for the March 30 and 31 science fields, producing catalogs totaling 2651 and 3507 stars, respectively.  

We use these catalogs of bright, isolated stars with known $R$ magnitudes to calculate magnitude zeropoints for each of our science images. We fit these photometric zeropoints as a function of airmass to find slopes of $0.0874 \pm0.0040$ and $0.0804 \pm 0.0014$ mag/airmass in the VR filter on March 30 and 31, respectively. The slope derived for March 31 is consistent with the difference in the magnitude zeropoints independently determined from the SDSS field at two different airmasses on that night. The similarity in the airmass slopes for the two nights shows that the brighter zeropoint (hence, lower atmospheric transparency) measured for the SDSS field on March 31 was not a global property of that night: if it were, the airmass slope should also be steeper by about 0.027 mag/airmass on March 31. On the contrary the slope for March 31 is shallower, suggesting slightly ($\sim 0.008$ mag) better average atmospheric transparency. We discuss the issue of relative photometric calibration for the two nights further in \S \ref{sec:fakepix}.

Since some observers \citep[e.g.][]{Terai2013} have explored faint asteroids using the $r$ filter, we also performed a calibration mapping VR fluxes directly to SDSS $r$ magnitudes. We found a difference of +0.26 mag relative to our approximate $R$ magnitudes. Our 50\% sensitivity limit of $R=25.3$ mag therefore corresponds to $r=25.56$ mag. However, since both \citet{Gladman2009} and \citet{Yoshida2003} used the $R$ band (and deriving either $R$ or $r$ mags from VR fluxes is an approximation), we focus our analysis on $R$ magnitudes. Interested readers can convert to $r$ band by adding 0.26 mag.

\subsection{Subtracting Stars and Other Stationary Objects} \label{sec:starsub}

We subtract stars, galaxies, and other stationary objects from our science images prior to the digital tracking analysis, using the method of \citet{Alard1998}. In this method, a template image of the same target field is subtracted from each science frame after being convolved with a kernel designed to make the PSF of the two images identical. In our case, the template image is a stack of other images taken far enough away in time (i.e. 10 minutes) that the slowest-moving asteroids don't self-subtract. The convolution kernal is constructed from a sum of basis functions that have the form of radial Gaussians multiplied by polynomials in $x$ and $y$. \citet{Alard1998} used three Gaussians with $\sigma = 1$, 3, and 9 pixels, respectively, and multiplied them by polynomials of order 6, 4, and 2. This produces a total of 49 basis functions, with each of which the template image must be independently convolved --- a procedure that is very computationally demanding. Fortunately, we find that such a large number of polynomials is not necessary for our DECam data. 

For our March 31 data, we adopt two Gaussians, with $\sigma = 1$ pixel and $\sigma = 2$ pixels respectively, each multiplied by a 2nd order polynomial for a total of only 12 basis functions. Subtracting the 210 images of our March 31 data took about 2 weeks using a high-end desktop workstation with 32 cores. For our March 30 data, which was taken in somewhat better seeing, we adopt three Gaussians with $\sigma = 0.6$, $1.2$, and $2.4$ pixels, yielding 18 basis functions. The image-subtraction took somewhat longer, but the higher quality of the data and smaller number of images justified it. Apart from the constraints of processing time, we would have used more basis functions for March 31 also. The improved subtraction would have little practical effect, however: we deliberately targeted a field far from the Milky Way, and the fractional area affected by subtraction residuals is miniscule in both cases. A larger set of basis functions might well be necessary for otherwise-identical data targeting the richer star fields at low galactic latitude.

In order to follow PSF and sky background variations across the images, we perform the subtraction independently in blocks of $4000\times4000$ pixels. For each block on each science frame, we create a customized template image. The customization is intended to enable better subtraction with our limited set of basis functions. Hence, we construct the template by stacking other images whose PSF is similar to that of the science frame: specifically, the $n$ best-matching images where $n$ is defined by the minimum number required to meet two conditions. These conditions impose minimum acceptable values for the minimum coverage (\texttt{mincov}) and the mean coverage (\texttt{meancov}) of the stack, where coverage is defined as the number of images that contribute data at a given pixel location. The thresholds \texttt{mincov} and \texttt{meancov} are designed to ensure that sufficient images go into the template stack to render its background noise much lower than that of the science frame.

We initially attempted the subtraction with \texttt{mincov} $=5$ and \texttt{meancov} $=12$, drawing images for the template stack from the pool of all science images separated by at least 0.15 hr in time from the one being processed.  This was very effective, but too slow, as the level of PSF similarity between different images had to be evaluated so many times. We decided instead to draw the subtraction images from a smaller set of 21 images, each made by stacking an independent subset of 10 science images.  Further, we restricted the set of images available for subtracting any given science image to seven, rather than the full 21.  In this context, we set \texttt{mincov} $=2$ and \texttt{meancov} $=2$: such small values do not incur more background noise than the previous settings, since now each subtraction image is the stack of ten science images. 

We address sky background variations independently after the star subtraction, by fitting and subtracting a second order polynomial model of the residual sky background in each $4000\times4000$ pixel block. All saturated pixels, including bleed-streaks and regions of `pooled' electrons from saturated stars, are masked on the subtracted frames.

The subtraction completely eliminates faint stars and galaxies from our images, but residuals remain around the cores of bright stars. It makes sense to mask these based on some criterion external to the subtracted image itself, to avoid interpreting bright asteroids as star residuals and hence masking them. Previously \citep{digitracks} we have masked all pixels brighter than a certain threshold on the master stacked image (e.g. Figure \ref{fig:fullstack}), which of course shows only stationary objects. Unfortunately, this produces excessive masking around large, smooth galaxies, which subtract much better than stellar PSFs of equal surface brightness. Thus, instead of a master stack of unsubtracted images, we stack all the {\em subtracted} images and find the RMS through the stack at each pixel. Large values in the RMS map accurately indicate stars too bright to subtract cleanly. We mask pixels on the subtracted science images if their value on the RMS map is greater than 15 ADU, scaled by an adjustment factor calculated independently for each $4000\times4000$ pixel block in each science frame. The adjustment factor is designed to reduce masking on images with unusually good subtraction and increase masking if the subtraction was poor. We obtain it by taking the ratio of the absolute value of pixels on the subtracted science image to the same pixels on the RMS map, averaged over a limited range in pixel brightness designed to reject both pure sky noise and excessively bright stars. The 15.0 ADU threshold is divided by this mean ratio. For a well-subtracted image, the residuals are low relative to the RMS map, so the mean ratio will be less than 1.0. Hence, the masking threshold is greater than 15 ADU and only small regions around bright stars are masked, preserving as much data as possible. By contrast, a poorly subtracted image gets a lower threshold, resulting in wider masked regions around bright stars.

The final output of our processing is a set of repixellated, subtracted, and masked images suitable for digital tracking analysis, to be described below.

\section{Digital Tracking Analysis} \label{sec:digitracks}

\subsection{Introduction to Digital Tracking} \label{sec:introtracks}
In \citet{digitracks} we have given a detailed description of digital tracking (also called synthetic tracking; \citet{Shao2013,Zhai2014}) and a review of past applications. We provide a much briefer introduction here. 

Digital tracking is the technique of shifting and stacking a large number of images to discover moving objects much fainter than could be seen in a single image.  Since the angular velocities of the objects to be discovered are initially unknown, it is necessary to probe a range of trial velocities, which we refer to as `trial vectors', and produce a separate `trial stack' for each trial vector. The trial vectors, each of which corresponds to a two-dimensional angular velocity in the plane of the sky, must be chosen to span the range of `angular velocity phase space' inhabited by the objects of interest. The density of trial vectors must be chosen so that for any object in the region of angular velocity phase space being explored, there will be at least one trial stack that accurately registers all its images and renders it as a sharp point source. 

As in \citet{digitracks}, here we search a rectangular region of angular velocity parameter space defined as specific ranges in eastward and northward angular velocity, chosen to include most MBAs in our field. This rectangular region is illustrated in Figure \ref{fig:angphase}.

\begin{figure*}
\includegraphics{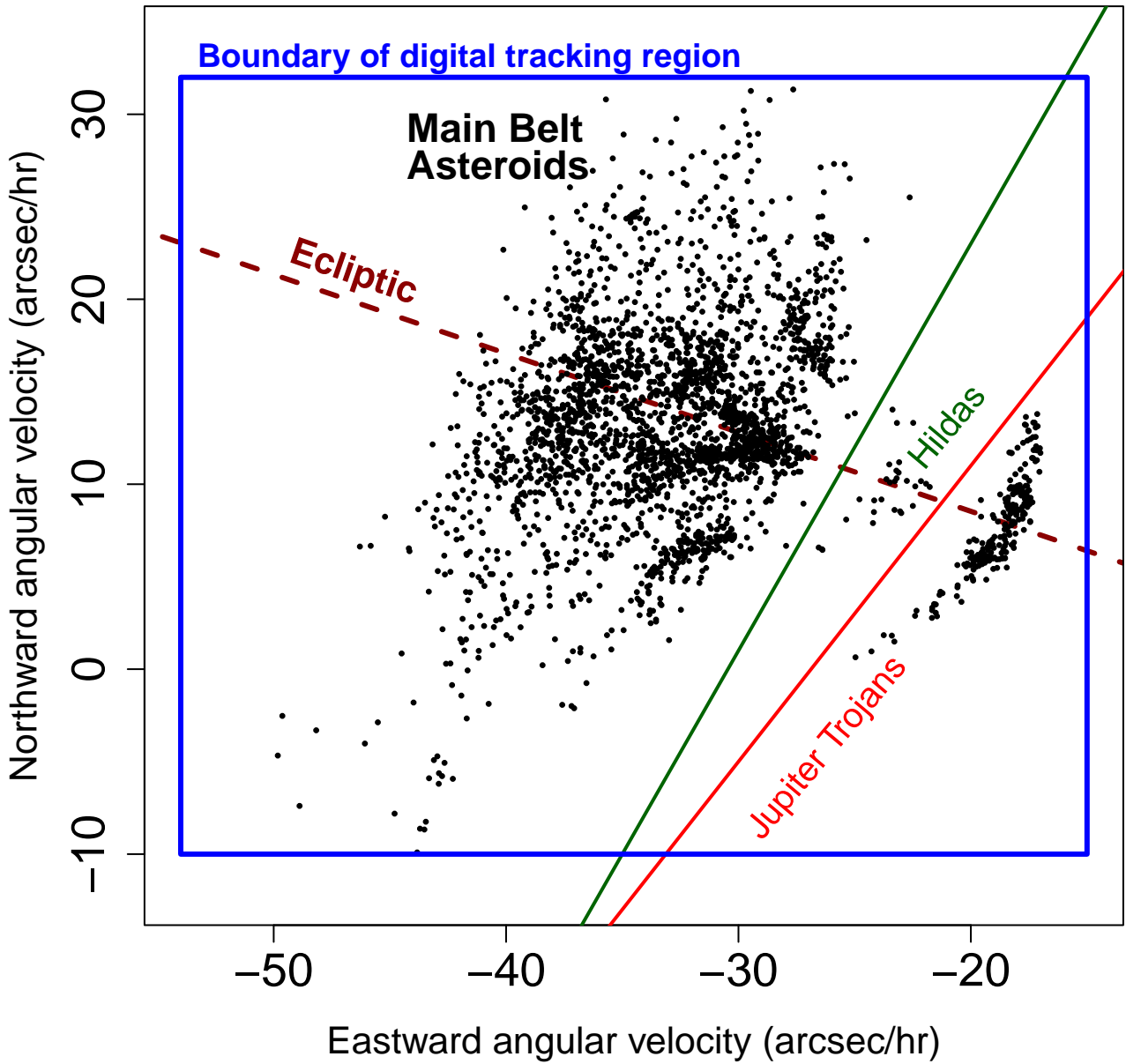}
\caption{The angular velocity phase space searched in our digital tracking analysis, with the phase space positions of asteroids detected in our March 31 data plotted. The dark red dashed line indicates motion parallel to the ecliptic. The green line indicates the boundary in angular velocity space that we use in \S \ref{sec:skydens} to separate the Hilda asteroids from the MBAs; while the red line is the boundary between Hildas and Jupiter Trojans. We excluded both Hilda asteroids and Jupiter Trojans from our statistical analysis of the main belt because of their uneven distributions in ecliptic longitude.
\label{fig:angphase}}
\end{figure*}

The optimal spacing of trial vectors in angular velocity phase space is given by Equation \ref{eq:blur} \citep{digitracks}.

\begin{equation}
\Delta_m  = \frac{\sqrt{2} \; b_{max}}{t_{int}}
\label{eq:blur}
\end{equation}

Where $b_{max}$ is the maximum permissible blur in arcseconds, and $t_{int}$ is the temporal span of the digital tracking integration. For March 30 and 31 we have $t_{int}$ = 7.67 hr and $t_{int}$ = 7.30 hr, respectively. We may reasonably set $b_{max}$ equal to the seeing, which for the two nights was 1.11 arcsec and 1.46 arcsec, respectively. This produces $\Delta_m$ = 0.205 arcsec/hr for March 30 and $\Delta_m$ = 0.283 arcsec/hr for March 31. We conservatively adopt a spacing of 0.20 arcsec/hr for March 30 and 0.25 arcsec/hr for March 31.

We search a rectangular region in angular velocity phase space that extends from -54 to -15 arcsec/hr toward the east, and -10 to +32 arcsec/hr toward the north. This was chosen to include the vast majority of known MBAs in the vicinity of our field at the time of our observations. As shown in Figure \ref{fig:angphase}, the actual distribution is elliptical rather than rectangular, and a considerably smaller area in angular velocity phase space could have spanned all the real objects. We could have predicted this region based on known objects and reduced the digital processing runtime by 30-60\% by targeting an optimized elliptical region or regions. However, we consider the regions of `white space' in the angular velocity plot to be a useful check on false positives in our analysis: a spurious asteroid would be equally likely to occur at any angular velocity, so the absence of detections in the corner regions of the plot suggests there are very few false positives. We address this question more rigorously in \S \ref{sec:fp}, but the blank areas of the Figure \ref{fig:angphase} are a good initial indication. We note also that the generous rectangle we searched in angular velocity phase space enabled us to detect a considerable number of Jupiter Trojan asteroids, which we had not anticipated.

\subsection{Computational Requirements}

We performed our digital tracking analysis primarily using idle time on two processing nodes of the computing cluster belonging to the Asteroid Terrestrial-impact Last Alert System \citep[ATLAS,][]{Tonry2018}. Each node has 24 cores and 128 GB of memory.

\subsubsection{Memory}
As mentioned above, each of the input images to our digital tracking analysis has dimensions of 32,000$\times$28,000 pixels: 896 megapixels in all. They are saved on disk as floating-point (32 bit) FITS files with a size of 3.4 GB. Loading all of them into memory would require 690 GB. Our computing nodes have only 128 GB of memory, and yet efficient stacking requires that all of the images be read only once. The fact that our data are highly oversampled, with 0.2625 arcsecond pixels in 1.1 arcsecond seeing, enables us to use 2$\times$2 binning in the digital tracking analysis and hence cut down the data volume by a factor of four while preserving Nyquist sampling. This binning is done internally when the files are read by our digital tracking code, without altering the images on disk. 

Even after binning, the data volume is still too large at about 170 GB. We obtain an additional factor of two reduction by storing the images in memory at half-precision (16 bits per pixel), a capability that was acquired only with the expenditure of considerable programming effort and finesse. The approach of loading the images at half precision was adopted only after careful validation confirmed that its output was functionally equivalent to the results of full-precision storage. We use nearest-neighbor interpolation to shift the images, obviating any need to use separate memory to store shifted images prior to stacking. For stacking, we use a 5$\sigma$ clipped median, and the pixel values are converted back to full floating-point precision prior to stacking. This does not use excessive memory since it can be done one pixel at a time. With half-precision storage, we can load all of the images from March 31 using only about 70\% of our computers' 128 GB of memory, and hence can perform the digital tracking analysis efficiently.

\subsubsection{Processing} \label{sec:proc}
The angular velocity phase space region and step sizes determined in \S \ref{sec:introtracks} require us to probe 41356 angular velocity vectors for our March 30 data and 26533 for March 31. In \citet{digitracks}, we parameterized the computational cost of a digital tracking analysis in terms of vector-pixels, where the number of vector pixels is the number of trial vectors times the number of pixels per image times the number of images. Using this terminology, given our input images of 32,000$\times$28,000 pixels which were internally binned 2$\times$2 by the digital tracking code, we probed $1.14\times10^{15}$ vector-pixels for our March 30 data and $1.20\times10^{15}$ for March 31. On the ordinary, 24-core nodes of the ATLAS computing cluster, we obtained typical processing rates of about $1.3\times10^{12}$ vector-pixels per hour per node. Thus, we could complete the digital tracking analysis in about 80 node-days. However, as described below we actually analyzed the entire data set three times: once to detect real asteroids, once to probe false positives, and once with fake asteroids inserted to probe our completeness rate. The overall analysis thus required the equivalent of 240 node-days, although some of it was performed on an experimental, 1024-core supercomputer, which we found to be equivalent (for purposes of digital tracking) to about 7.5 ordinary cluster nodes.

\subsection{Detection of Candidate Asteroids} \label{sec:digidet}
As in \citet{digitracks}, our digital tracking analysis does not save the trial stacks, which would require prohibitively large volumes of storage. Instead, our code automatically detects sources on each trial stack and writes a detection log. This detection log is the primary output of the digital tracking analysis, and must then be further analyzed to identify and precisely measure the real asteroids, as described below.

Our digital tracking code operates on images binned 2$\times$2 relative to the full-size repixellated images, and hence the pixel size of the output trial stacks is 0.525 arcsec. Asteroid candidates are detected on this trial stack using the same methodology as \citet{digitracks}, which we describe here for the reader's convenience and because the detailed parameters are slightly different. 

Our code smoothes the stack with a square boxcar of size 3$\times$3 pixels. This correponds to 1.575 arcseconds, which is somewhat larger than the FWHM of a typical stellar PSF but is appropriate for the detection of asteroids that may not be perfectly registered even on the best trial stack. We create maps of the sky background and noise of this smoothed image, at 1/3 resolution. We obtain the value for each pixel in these maps by calculating the mean and standard deviation of the smoothed image in a surrounding block of 55 pixels square. To preserve independence, we use only every ninth pixel (every third pixel in each dimension) within this block. To ensure that the sky brightness and noise measurements for a given location are not affected by the possible presence of an asteroid at that location, we exclude all pixels within 16 pixels of the center in either dimension, turning the block into a square annulus. In general, the three-pixel sampling would probe $19^2 = 361$ individual pixels within the block, but the 16-pixel central exclusion removes 121 pixels, leaving 240 within the square annulus (fewer if we are near a masked region or image edge).

To reduce the influence of other asteroids that may be present in the square annulus, both the sky brightness and noise estimation are performed using a trim mean, which iteratively rejects the 10\% most deviant pixels. The sky brightness $B$ is then taken to be the mean of the surviving pixels, and the noise $\kappa$ is their standard deviation. We note that the rejection of deviant points by the trim mean causes $\kappa$ to underestimate the true noise: in the case of Gaussian statistics with 10\% trim mean rejection, $\kappa = \sigma/1.267$. Note that estimating $\sigma$ by $1.267\kappa$ would be like taking a robust standard deviation, which is very similar conceptually to what we have done. We detect asteroids by finding pixels in the smoothed trial stack whose brightness exceeds the sky brightness $B$ by at least $10 \kappa$, which is equivalent to $7.89 \sigma$ in Gaussian terms. Bright asteroid images create many pixels brighter than our detection threshold, so we reduce redundant detections by deleting all candidates within a `redundancy radius' of a more significant detection. To guard against the loss of actual distinct asteroids, we conservatively set the redundancy radius to just 5 pixels on the trial stacks: that is, 2.625 arcsec. The redundant detections that survive this cull are dealt with in post-processing of the digital tracking detection log, as we describe below in \S \ref{sec:cluster}.

\section{Post-processing of Digital Tracking Detections} \label{sec:post}

The output of the digital tracking analysis described in \S \ref{sec:digitracks} is a detection log listing all the candidate asteroid detections. The log supplies five numbers for each detection: the eastward and northward angular velocities of the trial stack on which it appeared; its x and y coordinates on a reference image to the nearest pixel; and its significance in units of $\kappa$ (see \S \ref{sec:digidet}). The reference image (equivalently the reference time) is chosen to be near the temporal center of the observations: e.g., near the average of all the image acquisition times. Our reference times are 4.752800 (04:45:10) UT and 5.149811 (05:08:59) UT for March 30 and March 31, respectively. These times are based on the start-of-exposure timestamps in the DECam image headers, corrected to mid-exposure by the addition of 45 seconds. In terms of the images used in our final digital tracking analysis, the reference times correspond to image 76 of 123 for our March 30 data and image 108 of 202 for March 31. They do not fall exactly in the center of the ordered sequences because the images are not evenly spaced in time. Uneven spacing arises due to images taken in filters other than the VR filter; images that do not pass the quality thresholds described above in \S \ref{sec:imcull}; and in the case of March 30, the three hours lost due to the DECam computer failure.

The digital tracking detection log is the product of the most computationally intensive stage of our analysis. Although it is extremely valuable, it does not in itself provide measurements whose precision and accuracy realize the full potential of our data. The angular velocity values are quantized by the step sizes used in the digital tracking analysis, and the position measurements are given only to whole pixels on the trial stacks, which are binned $2\times2$ relative to the original input images. 

Effectively, the detection log serves as a `treasure map' indicating where in the immense, four-dimensional parameter space probed by the the digital tracking analysis the asteroids are to be found. The current section describes how we dig up the treasure: that is, how we extract precise measurements of every detected asteroid at full resolution from the input images.

\subsection{Spurious Detections at the Edge of the Image} \label{sec:edgedets}
Our detection logs have large numbers of spurious detections near the edge of the region of valid data. These do not correspond to the edges of individual DECam detectors (those vanish completely, thanks to our well-optimized dither), but rather to the boundary beyond which we acquired no data with any detector. This boundary is rendered approximately circular by the layout of the DECam detectors, and has nonzero width since the actual edge of the data does not occur at the same pixel coordinates for all trial stacks.

We do not know in detail the cause of the spurious detections at the edge of our data region. They appear despite several aspects of our code that are designed to prevent noise at the edges of a stack from being erroneously recorded as a significant detection. The large numbers of spurious edge detections first appeared when we began storing the digital tracking images in memory at half precision, but we have not been able to identify the aspect of half-precision storage that causes them. As described above, we have extensively validated the output of our half-precision code relative to previous full-precision results for actual asteroid candidates: the only significant difference is the appearance of this vast number of false positives at the images edges.

Since real asteroids could not be usefully detected at the extreme edge of the images anyway, we have elected simply to cull the detection logs using a map of the edge regions where false detections occur. We construct this map from the detection logs themselves, first plotting the pixel coordinates of all the detections, and then using a combination of blurring and boundary-detection to identify the full extent of the false-detection zone while encroaching as little as possible on the valid data interior to it. We find additional use for this map later in our analysis: it provides a measurement of the area on the sky over which we had good sensitivity.

Our detection logs from March 30 and 31 originally contained 7.5 million and 6.4 million candidates, respectively. After removing the spurious edge-detections, the numbers dropped to 3.3 million and 2.4 million, the vast majority of which are redundant detections of bright asteroids, as we describe below.

\subsection{Removing Duplicates} \label{sec:cluster}

Millions of duplicate detections of real asteroids survive our edge-culling and the simple redundancy cut described in \S \ref{sec:digidet} --- especially since bright asteroids can be detected (as long streaks) even on trial stacks far from their actual angular velocity. To remove duplicate detections without incorrectly deleting true, distinct asteroids, we use the fact that once the `best' (i.e., brightest and correctly angular-velocity-matched) detection of a given asteroid has been identified, the locations of possible duplicate detections in four-dimensional digital tracking parameter space can be accurately predicted. For example, given the 7-hour temporal span of our data, a bright asteroid will leave a streak 70 arcseconds long on a trial stack at a velocity 10 arcsec/hr away from its true motion. The position of this streak may be precisely calculated, and detections falling along it assigned as duplicates.

We take advantage of this by first sorting the detection log in descending order of detection significance. The first detection should then correspond to the brightest asteroid, and it will be correctly motion-matched because any velocity error would reduce its brightness. We identify all subsequent detections that are duplicates of the first detection, based on their proximity to its predicted trail, and remove them from the list. Then we proceed to the second surviving detection --- a motion-matched detection of another bright asteroid --- and identify its duplicates, and so on to the end of the list. We refer to this process as `asteroid clustering', since we are trying to form clusters of detections that all refer to the same asteroid.

The exact parameters we use in asteroid clustering constitute a step-wise function in two dimensions: distance from the primary detection in angular velocity phase space (which determines the length and relative brightness of its trailed image), and distance in pixels from the predicted trail. A detection is included in the cluster if its significance relative to the primary detection falls below a threshold given by this stepwise function. We use three steps in terms of pixel distance, which each lead to six steps of angular velocity distance, producing 18 distinct thresholds in total. This stepwise function was carefully optimized through an extensive manual procedure. For the March 30 analysis, our clustering analysis reduced the 3.3 million input candidates to 4,179 distinct clusters, the largest of which contained almost 40,000 individual detections. For March 31, our analysis reduced 2.4 million detections to 4,624 clusters, with the largest containing about 15,000 detections. In both cases, subsequent manual investigation found the majority of clusters corresponded to real asteroids.

\subsection{Images for Asteroid Analysis} \label{sec:checkim}

Given the four dimensional location of an asteroid candidate in the detection log, we can calculate its location on each of the input images, extract small `postage stamps' centered on this location from each image, and stack these stamps to produce an image of the asteroid for manual examination. This image is superior to the trial stack evaluated internally by the digital tracking code because it is made at the full resolution of the input images; because the images are registered at sub-pixel precision using bilinear rather than nearest-neighbor interpolation; and because the image data have not had to be loaded into memory at half precision. We use such stacks in several ways for manually vetting the asteroid candidates identified by the clustering process described in \S \ref{sec:cluster}.

First, for each asteroid candidate we create several postage-stamp stacks that cover a grid of angular velocities with the nominal angular velocity at the center. This procedure aids in determining the asteroid's true angular velocity with a finer resolution than the sampling used by the digital tracking code. The stamp-stacks at different angular velocities can be tiled together to make a single image easy to interpret by eye (Figure \ref{fig:checkim}). We refer to these as `check images' since their basic purpose is to enable a manual check that a candidate asteroid behaves plausibly in angular velocity space. We also use the image to obtain a quantitative measurement of the angular velocity by measuring the flux on each stamp-stack and performing a two-dimensional quadratic fit to the measured flux as a function of angular velocity. The flux is measured around a centroid determined by iterative Gaussian centroiding, which is also used to refine the exact position of the asteroid on the reference image. We measure the flux within a 5-pixel (1.3 arcsecond) radius of the centroid, after multiplying by a radial Gaussian with a FWHM of about 1 arcsecond: this ensures that the flux drops significantly as soon as there is any blurring of the image due to angular velocity mismatch\footnote{We use this type of flux calculation {\em only} in the current context, not for actual photometry, which we discuss later.}. The new angular velocity measurement is simply the point in angular velocity space where the quadratic fit to the flux reaches its maximum. 

\begin{figure*}
\includegraphics[scale=0.66]{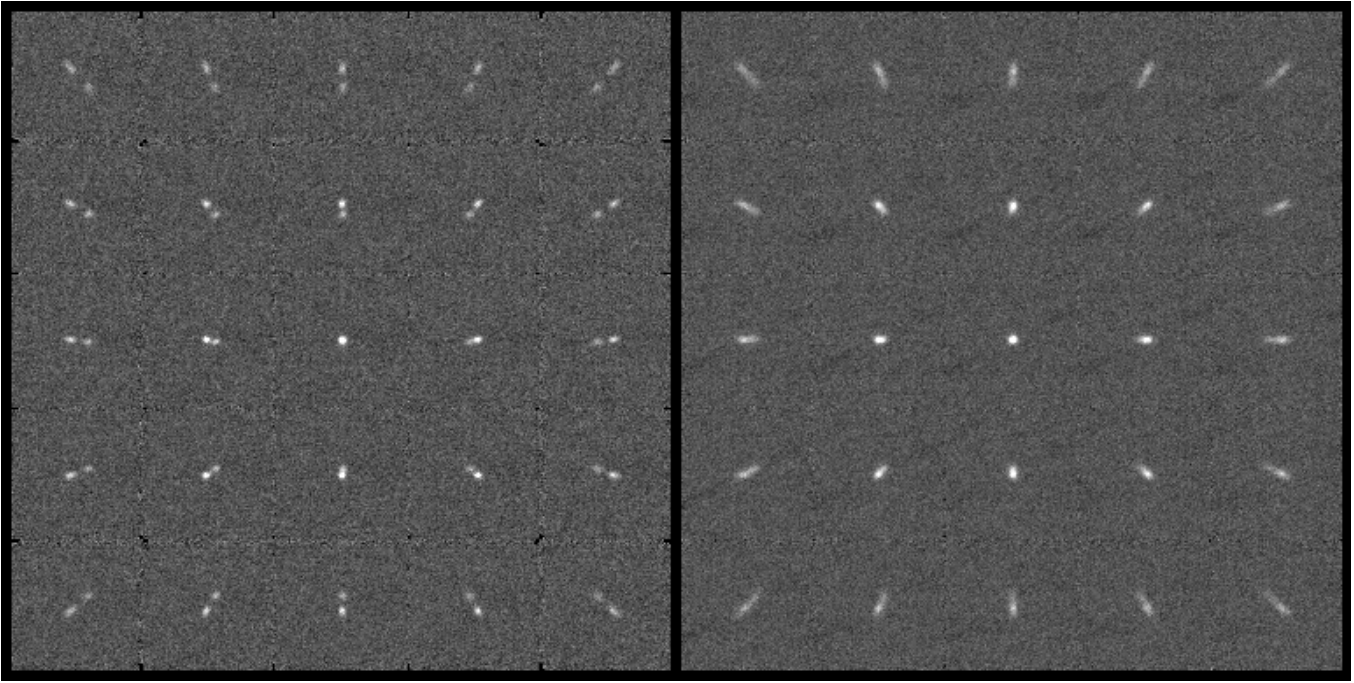}
\caption{Examples of the `check images' described in the text, which we use to check the reality of candidate asteroids and refine our angular velocity measurements. The tiles in each panel cover a grid in angular velocity space with a spacing of 0.5 arcsec/hr: optimal for illustrative purposes but coarser than the sampling we ultimately used for precise velocity determination. Each tile has a width of 151 pixels (40 arcsec). The data shown are from March 30 (left) and March 31 (right). These are relatively bright asteroids in our context: each is the 600th entry in the significance-ranked list output by the cluster analysis for its respective night. The trails in the March 30 images have time-gaps due to the DECam computer failure. The March 31 images are fuzzier but have a less noisy background relative to March 30: the seeing was better on March 30, but more good images were available to be stacked on March 31.
\label{fig:checkim}}
\end{figure*}

We apply a second type of verification by making multiple stamp-stacks at the same angular velocities, but with different subsets of the data, created by even splitting of the temporally ordered set of images. For manual examination, we arrange these in a pyramid shape: the stack of all the images is at the top of the pyramid; below come two stacks each made of half the images; below that three stacks each made with a third of the data, etc. Real asteroids should be consistently detected in multiple layers of the pyramid, until the stacks become too shallow and they fade into the noise. In some cases an asteroid may change significantly in brightness due to rotation (or move off the image or into a masked region near a bright star), and hence will be absent on several pyramid tiles. Our general criterion for confirming an asteroid as real is that it must appear fairly consistently on at least two tiles in {\em some} row of the pyramid image: equivalently, it must be detectable in two independent subsets of the data.

Figure \ref{fig:testim} shows examples of two pyramid images, probing much fainter asteroids than Figure \ref{fig:checkim}. The asteroid in the second panel of Figure \ref{fig:testim} gets passed by another, much brighter object at a distance of less than 6 arcsec on the sky and 2.9 arcsec/hr in angular velocity. Both asteroids were independently recovered by our asteroid clustering analysis. The fact that the fainter one was not consumed by the brighter object's cluster despite their proximity in all four dimensions is a testimony to the accurate tuning of our clustering thresholds, as described in \S \ref{sec:cluster}.

\begin{figure*}
\includegraphics[scale=0.66]{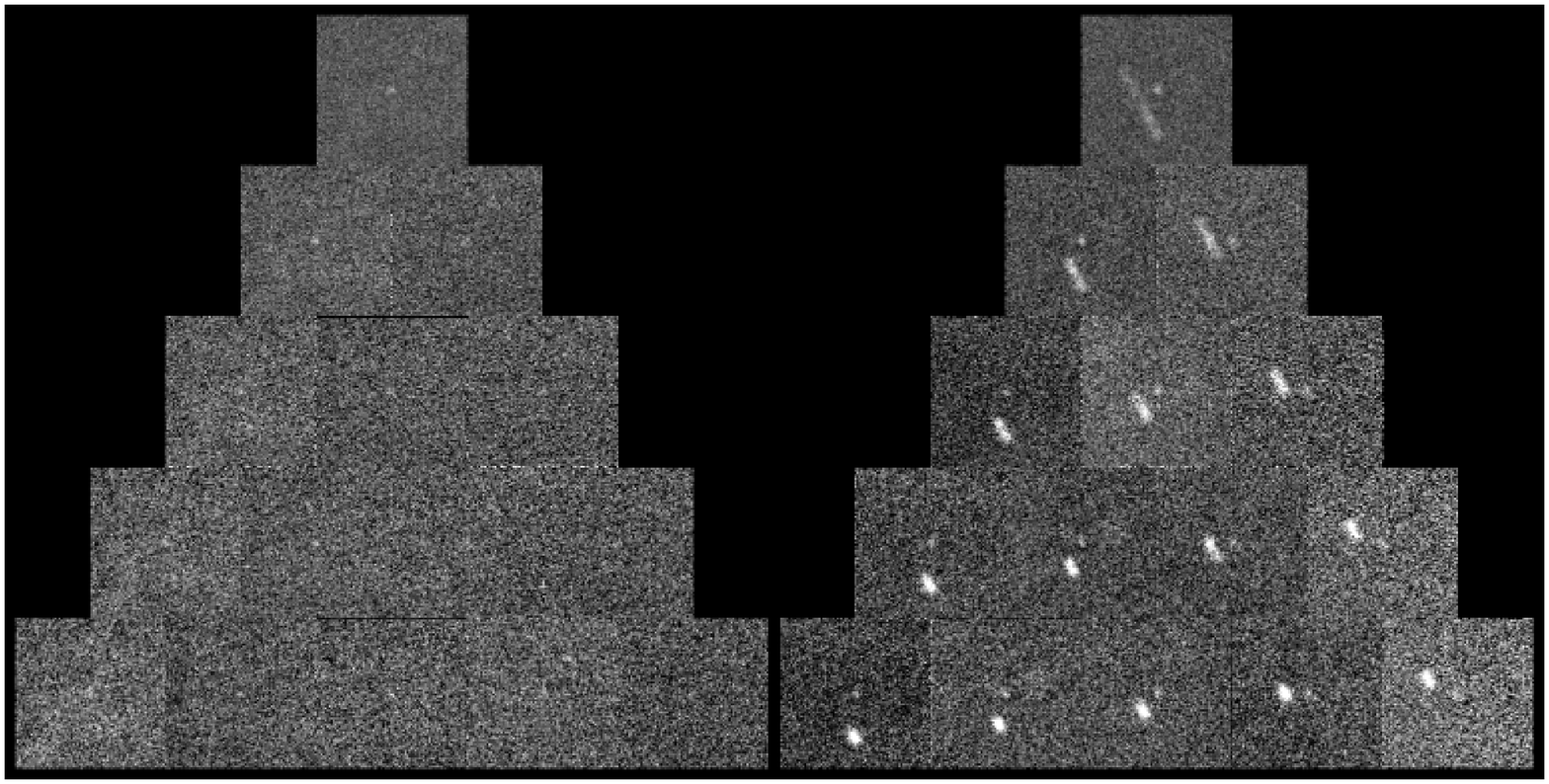}
\caption{Examples of the pyramid images described in the text, which we use to confirm the reality of asteroid candidates. Each panel represents a distinct asteroid from March 31, while successive rows show the division of the data into multiple independent, temporally ordered subsets. Tiles are 40 arcsec on a side. The left panel shows asteroid candidate 3,561 from the significance-ranked list output by our clustering analysis: this asteroid is definitively real (confirmed in the March 30 data) but is near the faint limit of our sensitivity. The right panel shows another real asteroid, ranked 2,189 in significance, which entertainingly gets passed at a distance of less than six arcseconds by a much brighter asteroid whose angular velocity differs by only 2.9 arcsec/hr. The brighter asteroid is independently detected as number 517 in the ranked list. 
\label{fig:testim}}
\end{figure*}

Besides the creation of the check images and pyramid images, we also created larger images showing many asteroids at once, typically $10\times10$ tilings of stamp-stacks at the nominal parameters. This allowed us to rapidly screen 100 asteroids at a time by eye, and aided in catching classification mistakes or glitches in our fits for refining the pixel and angular velocity coordinates of each object.

\subsection{Creation of Final Asteroid Lists} \label{sec:flist}
Using tiled arrays of stamp-stacks; check images; and pyramid images, we visually screened the candidate lists output by our clustering analysis. Of the original lists of 4,179 and 4,624 candidates for March 30 and 31, respectively, we produced culled sets totalling 2,798 and 3,091 objects. Most of the rejected candidates turned out to be duplicates of bright asteroids or very tenuous, unconfirmable detections bordering on pure sky noise. A few were noise artifacts associated with masked regions near bright stars.

Among the objects passed by our manual screening, we allow two distinct classes. Class `a' means the object looks completely real, with a symmetrical and star-like PSF. Class `b' (a small minority) means the image looks morphologically peculiar in some way, but yet we did not feel it could be definitively rejected.

The tenuous/non-existant detections rejected by our manual screening highlight the unsophisticated nature of the signficance calculation employed (by computational necessity) in our digital tracking code. Hence, we re-calculate the significance of all the candidates passed by manual screening using a more sophisticated method. We do this on stamp-stacks using aperture photometry. We first perform photometry centered on the asteroid, and then we measure the sky background noise {\em on the scale of the PSF} (and hence properly accounting for correlations between pixels), by performing identical forced photometry in a large number of non-overlapping sky apertures near the asteroid. These apertures are constrained to be centered within a 100 pixel (26.25 arcsec) box centered on the asteroid, and yet must not overlap a disk of 12 pixel (3.15 arcsec) radius around the object. The sky noise is evaluated as the STDEV of the forced photometry after iterative rejection of the 10\% most deviant points, and is converted to Gaussian $\sigma$ using the factor of 1.267 discussed in \S \ref{sec:digidet}. The significance of the detection is simply the measured flux for the asteroid divided by the derived $\sigma$ value of the sky noise.

The effective FWHM of our asteroid images can vary for several reasons. The fastest-moving asteroids will be slightly ($\lesssim 1$ arcsec) blurred on our individual exposures. Additionally, a slightly different set of individual images contributes to the final stacked image of each asteroid (due to asteroids falling in gaps between DECam detectors, traversing masked regions near bright stars, exiting the field, or simply being very faint on some images due to rotational flux variations). Since the seeing FWHM of the stellar PSF varies from image to image, the final FWHM of the stacked asteroid images will also vary. Hence, to identify the true significance of an asteroid detection, we probe a range of apertures and identify the one that produces the highest significance value. Our apertures range from 1.5 pixel to 6.0 pixel (0.4 to 1.6 arcsec) in radius with a spacing of 0.1 pixels. The number of non-overlapping sky apertures that can be measured depends on the aperture radius being probed (at larger radii not as many apertures can be fit in the allowed region), but is typically a few hundred. The median optimal radius is 3.1 pixels (0.81 arcsec) for March 30 and 3.7 pixels (0.97 arcsec) for March 31, consistent with the generally better seeing on March 30.

If the internal significance calculation in our digital tracking code were perfect, all of the detections it output would have signifance at least 7.89 $\sigma$. In fact, our more sophisticated analysis finds that the vast majority of detections passed by our manual screening do have at least this significance: only 2.2\% and 7.7\% fall below it for March 30 and 31, respectively. Whether these relatively low-significance detections correspond to real asteroids is a question we address further below.


\subsection{Magnitude Measurements} \label{sec:precmeas}

We extract optimized photometry for each asteroid from its stamp-stack using a method similar to our significance calculation: by testing a large number of apertures and selecting the optimal one. The background noise is also calculated in the same way. However, the optimization is different because we now seek to minimize the fractional uncertainty on the flux, and this includes an additional source of uncertainty beyond the sky background noise: the uncertainty on the aperture correction. The aperture corrections and their uncertainties were determined by measuring 100 bright asteroids at both the large (20 pixel radius) aperture used for the stellar photometric calibrations and the range of smaller apertures referred to above. Since the aperture correction has a larger fractional uncertainty for small apertures, the optimal aperture for photometry is typically larger than for the significance calculation. The median optimal photometric aperture has a radius of 3.7 pixels (0.97 arcsec) for March 30 and 4.3 pixels (1.13 arcsec) on March 31. These values can be compared with the optimal radii of 0.81 arcsec and 0.97 arcsec for the significance calculation on the respective nights.


\section{False Positive Analysis} \label{sec:fp}

In order to extract accurate information on asteroid populations from our data, it is essential to analyze both the rate of false positives and that of false negatives (survey incompleteness). We describe in this section the intensive analyses we have carried out to determine detection thresholds for a negligible false-positive rate. In the next section we will quantify our magnitude-dependent completeness.

Digital tracking enables a very realistic probe of the false positive rate. By randomly re-assigning (i.e. scrambling) the time-stamps of our images, we can repeat the digital tracking analysis using identical data and methodology and yet with the certainty that any detections will be false. This is because scrambling the image times makes it impossible for real asteroids to be registered in any of the trial stacks, and hence they vanish in the clipped median combine.

We have performed a full digital tracking analysis of our images with scrambled timestamps, down to the creation of check images and the sophisticated significance calculation described above. We find that the most significant candidate detection in these scrambled data sets is $7.43 \sigma$ for March 30 and $7.28\sigma$ for March 31. Under manual examination, these detections appear indistinguishable from real asteroids with the same significance.

To determine if the detection of spurious sources at these significance levels is surprising, we estimate the number of noise realizations of the PSF that our analysis has probed. As described in \S \ref{sec:proc}, our analysis probed 41,356 trial stacks for our March 30 data and 26,533 for March 31. Each trial stack produced an image of dimensions $16000\times14000$ binned pixels, with the region of valid data approximating an ellipse inscribed inside this rectangle and hence having an area of about $1.8 \times 10^8$ pixels. Our digital tracking analysis assumed that the effective area of a PSF is 9 binned pixels, in which case the valid data on each trial stack covers the size of 20 million PSFs. Multiplying this by the number of trial stacks, we find that we have effectively sampled about $7.3 \times 10^{12}$ and $4.7 \times 10^{12}$ realizations of the PSF-scale noise on our two nights, respectively.

The expected number of spurious detections from pure Gaussian noise equals unity for a threshold $x_{lim}$ beyond which the one-sided tail of a normalized Gaussian integrates to the inverse of these numbers of realizations. Thus, for $N_{real} = 7.3 \times 10^{12}$ and $4.7 \times 10^{12}$ on the two nights respectively, we seek $x_{lim}$ such that:

\begin{equation}
\int_{x_{lim}}^{\infty} \frac{e^{-x^2/2 \sigma^2}}{\sigma \sqrt{2 \pi}} = \frac{1}{N_{real}}
\label{eq:fpthresh}
\end{equation}

Solving Equation \ref{eq:fpthresh} yields $x_{lim} = 7.31 \sigma$ and $7.25 \sigma$ for March 30 and 31, respectively. These numbers are close to the values of $7.43 \sigma$ and $7.28\sigma$ for the most significant detections actually found in the time-scrambled digital tracking analysis, so we conclude that such detections are not surprising --- and that the methods we have employed for rejecting spurious sources leave behind something quite close to pure Gaussian noise. This quantitative conclusion aligns with the visual impression that our optimized dithering, star subtraction, masking, shifting, and clipped-median stacking do indeed produce images more free of artifacts than is usually possible with astronomical data.

We desire an extremely low false-positive rate in order to ensure that the scientific conclusions we ultimately draw about the population of small MBAs will not be invalidated by false positives. Hence, we adopt the maximum significance values seen in the scrambled-time analysis ($7.43 \sigma$ and $7.28\sigma$ for March 30 and 31, respectively) as our minimum thresholds for the independent detection of real objects. A few of the real detections manually classified as `a' (and many of those classified as `b') fall below this threshold. Class `a' candidates falling below the thresholds are reclassifed as `subsig', and are not considered as definitive, independent detections (even though their much greater abundance relative to sources of similar significance in the time-scrambled data indicates most of them are real). Only the `a' class objects with significance above the threshold are considered to be definitively, independently detected on a single night. The totals in various categories are given in Table \ref{tab:detclass}. With the thresholds we have adopted, we expect no more than one false positive per night --- a rate far too low to meaningfully affect scientific conclusions drawn from our analysis.

\begin{deluxetable}{llrrc}
\tablewidth{0pt}
\tablecaption{Asteroid Candidates Detected with Digital Tracking\label{tab:detclass}}
\tablehead{ & & & \colhead{Number} & \colhead{Match}\\
\colhead{Date} & \colhead{Category} & \colhead{Number} & \colhead{matched\tablenotemark{a}} & \colhead{rate}}
\startdata
March 30 & `a' & 2760 & 2720 & 98.6\% \\
March 30 & `subsig' & 16 & 11 & 68.8\% \\
March 30 & `b' & 22 & 0 & 0 \\
March 30 & total & 2798 & 2738 & 97.9\% \\[0.1in]
March 31 & `a' & 2973 & 2902 & 97.6\% \\
March 31 & `subsig' & 37 & 14 & 37.8\% \\
March 31 & `b' & 81 & 1 & 1.2\% \\
March 31 & total & 3091 & 2917 & 94.4\% \\
\enddata
\tablenotetext{a}{`Number matched' is the count of asteroids in each category that were detected and unambiguously matched in data from the other night. It includes objects that were automatically detected on both nights as well as those that were initially detected in only one night's data but were then successfully recovered in data from the other night at a sky position and angular velocity predicted based on the original detection (see \S \ref{sec:2night}).}
\end{deluxetable}

\section{Completeness analysis} \label{sec:fn}

Because the main science result of our survey is the statistical distribution of asteroids as a function of flux, we put a great deal of effort into a rigorous evaluation of our fractional detection rate as a function of apparent magnitude. To accomplish this, we repeated our entire digital tracking analysis on images to which we had added fake asteroids. We took pains to make sure these fake asteroids would be subject to all aspects of our analysis that could affect our sensitivity to real ones. We placed them in the images prior to the subtraction of stationary sources, so that if (contrary to our expectations) any real asteroids were dimmed by self-subtraction, so might be the fake ones. We placed them at locations based on realistic dynamical orbits so that if (again, contrary to expectations) nonlinearity in the sky motion of real asteroids reduced our sensitivity, the fake asteroids would experience the same effect. Finally, we calculated the pixel locations where our fake asteroids should be placed not by taking our astrometric mapping (Equation \ref{eq:astrom}) at face value, but by obtaining an entirely new, 5th order global astrometric fit to the repixellated images. Thus, in the unlikely event that errors in our astrometric repixelation using Equation \ref{eq:astrom} were large enough to cause blurring and loss of sensitivity in our stacked images of real asteroids, the fake asteroids should experience similar degradation.

\subsection{Orbital Population of Fake Asteroids}

As we developed our strategy for the fake asteroid test, we realized that by judicious choice of the input population of asteroids, we could link the statistics of our actual detected asteroids to the \textit{total} population of asteroids in the whole asteroid belt that had the same characteristics. Our first step in doing this was to simulate the entire population of small MBAs. We based this simulation on the known orbits of MBAs with orbital semimajor axis between 1.7 and 4.1 AU and absolute magnitude H$<14.0$. We chose this fairly bright absolute magnitude threshold to ensure a sample of known asteroids that would be complete for all main-belt orbits, avoiding a statistical bias against orbits in the outer belt where small objects are harder to see. We chose the semimajor axis range to include the entire main belt, but to reject Jupiter Trojans and NEOs (i.e., all objects with perihelia inside of 1.3 AU). Mars-crossing and Jupiter-crossing asteroids were retained, since their orbital distributions appeared to be contiguous with non-planet-crossing MBAs.


We obtained our input list of asteroids from JPL's Solar System Dynamics website\footnote{\url{http://ssd.jpl.nasa.gov/dat/ELEMENTS.NUMBR}}. The final list, after culling as described above, comprised 25409 orbits. We wished to use these orbits of well known MBAs as a basis for generating a much larger, but statistically identical, set of simulated MBA orbits. We found that although two of the Keplerian orbital elements, argument of perihelion and mean anomaly at the epoch, are uniformly distributed, the other four are not and have substantial correlations with one another. Rather than attempt to arrive at analytical approximations of the very complex distributions of these four parameters, we elected to base each simulated asteroid on one of the real asteroids. The two elements mentioned above would be chosen from a uniform distribution, but the other four elements would be adopted from the real asteroid's orbit and modified by a Gaussian fuzz with amplitude chosen to randomize the orbits as much as possible without erasing the structure and correlations of the original distributions.


We simulated asteroid orbits according to this protocol, setting the epoch for each orbit to JD 2456746.5: that is, 00:00 UT on 2014 March 30, near the start of our observations. For each asteroid orbit we calculated the corresponding RA and Dec at 00:00 UT on March 30. We terminated the simulation when a sufficient number of simulated asteroids had been found to be within 1.7 degrees of our field center at this time. This is significantly larger than the 1.1 degree radius of DECam's field of view, ensuring that all simulated MBAs that could appear on our DECam images on either March 30 or March 31 would be included.

We aimed to simulate 20000 asteroids within the 1.7-degree target region. Since our input orbital distribution is carefully chosen not to be biased against objects in distant orbits, such objects will be better-represented in the simulation than in our actual survey, which, being flux-limited, has reduced sensitivity to more distant objects in the outer main belt. This bias would produce a considerably faster average angular velocity for real asteroids detected in our images relative to our simulated asteroids. Since detection efficiency might depend on angular velocity, we wished to `re-bias' the simulated asteroids to more closely imitate the real population. Hence, for asteroids more than a threshold geocentric distance $d_{thresh}$ of 1.5 AU (chosen as a representative distance for the inner half of the main belt), we assigned a probability of being visible given by $P_{vis} = 10^{-\alpha \; dM}$, where $\alpha$ is the slope of the MBA apparent magnitude distribution found by \citet{Gladman2009}, and $dM$ is the difference between the asteroid's actual apparent magnitude and the magnitude it would have at the threshold distance (note that, as defined, $dM$ is positive for all the relevant asteroids). For asteroids near opposition, which are the only ones relevant here, $dM$ is given to sufficient accuracy by:

\begin{equation}
dM = 2.5 \; \log_{10} \left( \left( \frac{\Delta^2}{d_{thresh}^2} \right) \cdot \left( \frac{r^2}{(d_{thresh}+1)^2} \right)  \right)
\label{eq:dthresh}
\end{equation}

Where $\Delta$ and $r$ are the geocentric and heliocentric distances of the asteroid, respectively, in AU. For asteroids with $\Delta > d_{thresh}$, we randomly assign a status of visible or invisible according to probability $P_{vis}$. We allowed the simulation to proceed until it generated 20009 visible asteroids in the target area, at which point it had generated 35275 total asteroids in the target region and $5.43 \times 10^7$ total asteroids anywhere in the Solar System.

For the purpose of inserting fake asteroids into our data, only the 20009 visible asteroids should be considered. The statistical value of the remaining simulated asteroids, including those flagged as invisible, will be considered in our companion paper (Heinze et al., in prep.) on the absolute magnitude distribution and total population of the main belt. 

\subsection{Placing Fake Asteroids in the Images} \label{sec:fakepix}

To insert the simulated asteroids into the actual images, we use the catalog of bright, isolated, and unsaturated field stars described in \S \ref{sec:photcal}. We remind the reader that these consist exclusively of stars that have been confirmed by manual examination not to have neighbors contributing significant flux within a 40$\times$40 pixel (10.5$\times$10.5 arcsec) area, and that independent screened catalogs with 2651 and 3507 stars were derived for March 30 and March 31 respectively. It is these stars that we use as templates for our fake asteroids.

The magnitude distribution we selected for our simulated asteroids, given in Table \ref{tab:fakemags}, is not intended to imitate the expected distribution of real objects but rather to provide optimized statistical power to model our loss of sensitivity at faint magnitudes. Thus, the majority of the simulated asteroids are concentrated in the magnitude range from 24.5 to 26.0, over which our completeness goes from near 100\% to near zero. A relatively small number of brighter asteroids are included, to diagnose any unexpected detection failures for objects that should be glaringly obvious.

\begin{deluxetable}{cc}
\tablewidth{0pt}
\tablecaption{Magnitude Distribution of Simulated Asteroids \label{tab:fakemags}}
\tablehead{\colhead{Magnitude Range} & \colhead{Fraction of Asteroids}}
\startdata
20.0--21.0 & 0.02 \\
21.0--22.0 & 0.02 \\
22.0--23.0 & 0.02 \\
23.0--24.0 & 0.09 \\
24.0--24.5 & 0.10 \\
24.5--25.0 & 0.20 \\
25.0--25.5 & 0.30 \\
25.5--26.0 & 0.25 \\
\enddata
\end{deluxetable}

We simulated the blurring due to motion of each simulated asteroid on each image by calculating the asteroid's position at 10 instants spanning the 90-second exposure time of the image, interpolating from the 5-minute sampling of the simulated ephemeris. For each of these ten positions, we identified the nearest object from our list of isolated stars that could be translated to the asteroid's position (within a tolerance of 0.1 pixels) by shifting an integer number of pixels. Thus, we were able to build up the asteroid images without blurring due to to interpolation. Since, on average, less than 100 stars would have to be examined to find one that could be moved to the asteroid's location within the specified tolerance, we expect that the stars used to build up each asteroid's image will generally come from near the asteroid in the DECam field. We have not attempted to set any maximum distance, since the PSF is remarkably consistent across the DECam field. 

Once a suitable star is identified to construct part of a simulated asteroid's image, a region 40x40 pixels centered on the star is copied, scaled by the magnitude difference, further scaled to account for the fact that it will only contribute one tenth of the asteroid's final flux, shifted, and added back in to the original image. The star positions are measured individually on each image. As mentioned above, however, the asteroid positions are converted from the celestial coordinates of the ephemeris file to pixel positions using a 5th order fit based on Gaia astrometry, which is entirely independent of the astrometric fit originally used to repixellate the image. Thus, any flaws in the original astrometric solution, which could blur the stacked images of real asteroids, can affect the simulated asteroids in a similar way. The simulated asteroids, like real objects, follow coherent orbits between the two nights. Hence, the same fake asteroid can be found on both nights and its distance can even be calculated using the RRV method and compared with the actual distance, which in the case of the fake asteroids is exactly known by construction.

Comparison of fake asteroid detections across the two nights made us aware of a photometric inconsistency in the star catalogs used as templates for the fake asteroids (and photometric calibration for real asteroids as well). To explore this, we identified 1491 stars in common between the March 30 and March 31 catalogs. We found that on average, we had measured the same star to be 0.053 magnitudes fainter on March 30 relative to March 31. Three effects contribute to this surprisingly large offset. First, it is due in part to the difference in SDSS-derived zeropoints described in \S \ref{sec:photcal}. This accounts for 0.0271 magnitudes. Then, we had made an error in the airmass match for March 31 (corrected in the input of the fake asteroids, but not at the level of the star catalogs) which amounted to 0.0152 mag. The remaining 0.0107 mag is presumably due to a slightly poorer mean atmospheric transparency (during observations of the science field) for March 30 relative to March 31. This is consistent both with the presence of some clouds on March 30 (See \S \ref{fig:flux}), and with the slightly steeper airmass slope we find for the March 30 observation in \S \ref{sec:photcal}. 

To correct this 0.053 mag error, we elected not to make a judgment on which night's SDSS calibration was more accurate. Hence, we corrected the March 31 reference star magnitudes by +0.0152 mag to remove the known airmass error, and split the remaining 0.0378 mag evenly between the two nights. These adjustments (by construction) brought the mean magnitudes of reference stars measured on both nights into perfect agreement. We applied corresponding corrections to the measured magnitudes of both real and fake asteroids measured on the respective nights, and to the nominal input magnitudes of the fake asteroids.

\subsection{Matching Detected Fake Asteroids to Input} \label{sec:fakematch}

We analyzed the fake asteroids using the same steps as for the real asteroids, including manual screening by means of stamp-stacks, tiled images, check images, and pyramid images. As for the real asteroids, we created lists of fake asteroids with classifications of `a', `subsig', and `b'. Up to this point, the fake asteroid test was blind. Only when the classified lists were complete did we compare them with the input asteroids. These lists are provided in Table \ref{tab:fakedetclass}, which is exactly analogous to Table \ref{tab:detclass} for the real objects.

\begin{deluxetable}{llr}
\tablewidth{0pt}
\tablecaption{Detections in our Fake Asteroid Analysis\label{tab:fakedetclass}}
\tablehead{\colhead{Date} & \colhead{Category} & \colhead{Number Found}}
\startdata
March 30 & `a' & 4424 \\
March 30 & `subsig' & 23 \\
March 30 & `b' & 71 \\
March 30 & total & 4518 \\
March 31 & `a' & 4379 \\
March 31 & `subsig' & 29 \\
March 31 & `b' & 59 \\
March 31 & total & 4467 \\
\enddata
\end{deluxetable}

Comparing to the input catalog, we find that for March 30, 13 of the 23 detections classified as `subsig' and 3 of the 71 classified as `b' actually corresponded to input asteroids. For March 31, 11 of the 29 `subsig' detections and 2 of the 59 `b' detections corresponded to input asteroids. Thus we see that the `subsig' class is much more likely than the `b' class to correspond to genuine objects, but both are full of false positives. Embarassingly, we found that 5 of the 4424 class `a' detections for March 30 did {\em not} correspond to input asteroids. Examining them revealed four of them to be obvious mistakes in the form of motion-mismatched detections of other fake asteroids. We note that we would not expect such mistakes to occur for the real asteroids, which were subjected to more intensive manual screening and double-checking. The final unmatched `a' detection in March 30 looks plausibly genuine and is possibly a false positive arising from the noise distribution. However, it is almost 1$\sigma$ more significant than the most significant detection from the time-scrambled data set. This is especially odd since the fake asteroids were added to the time-scrambled images, and hence the same false positives should exist in both analyses. Direct comparison of stamp-stacks of time-scrambled images without and with fake asteroids show the false positive source is far more significant in the latter case: hence, something about the addition of the fake asteroids seems to have boosted its significance to values not representative of the actual false positive rate. There were no analogous mistakes or false positives in the March 31 fake asteroid analysis: every one of the class `a' detections was matched to an input fake asteroid.

\subsection{Survey Completeness from Fake Asteroids} \label{sec:completeness01}

We calculate our flux-dependent detection completeness for each night individually, using only fake asteroids classified as `a' in Table \ref{tab:fakedetclass}, which totalled 4424 and 4379 for March 30 and 31, respectively. For comparison, the numbers of fake asteroids input on the respective nights were 6595 and 6688, where these totals include only objects input within sky area over which could detect asteroids, as defined by the boundary map we described in \S \ref{sec:edgedets}. The ratio of detected to input fake asteroids in each magnitude bin gives our detection completeness, which is illustrated for March 30 in Figure \ref{fig:completeness}. The March 31 results are very similar.

\begin{figure*}
\plottwo{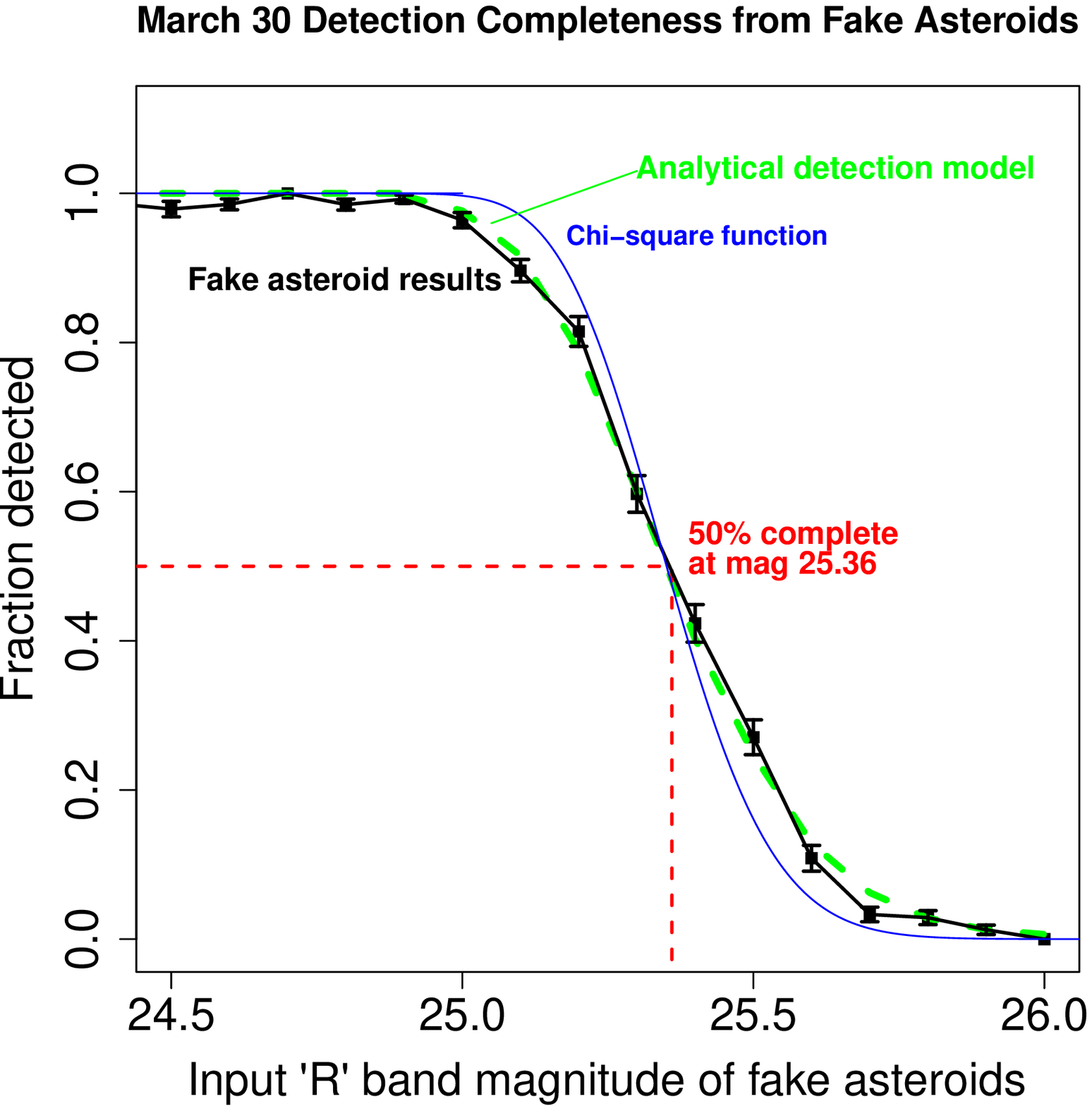}{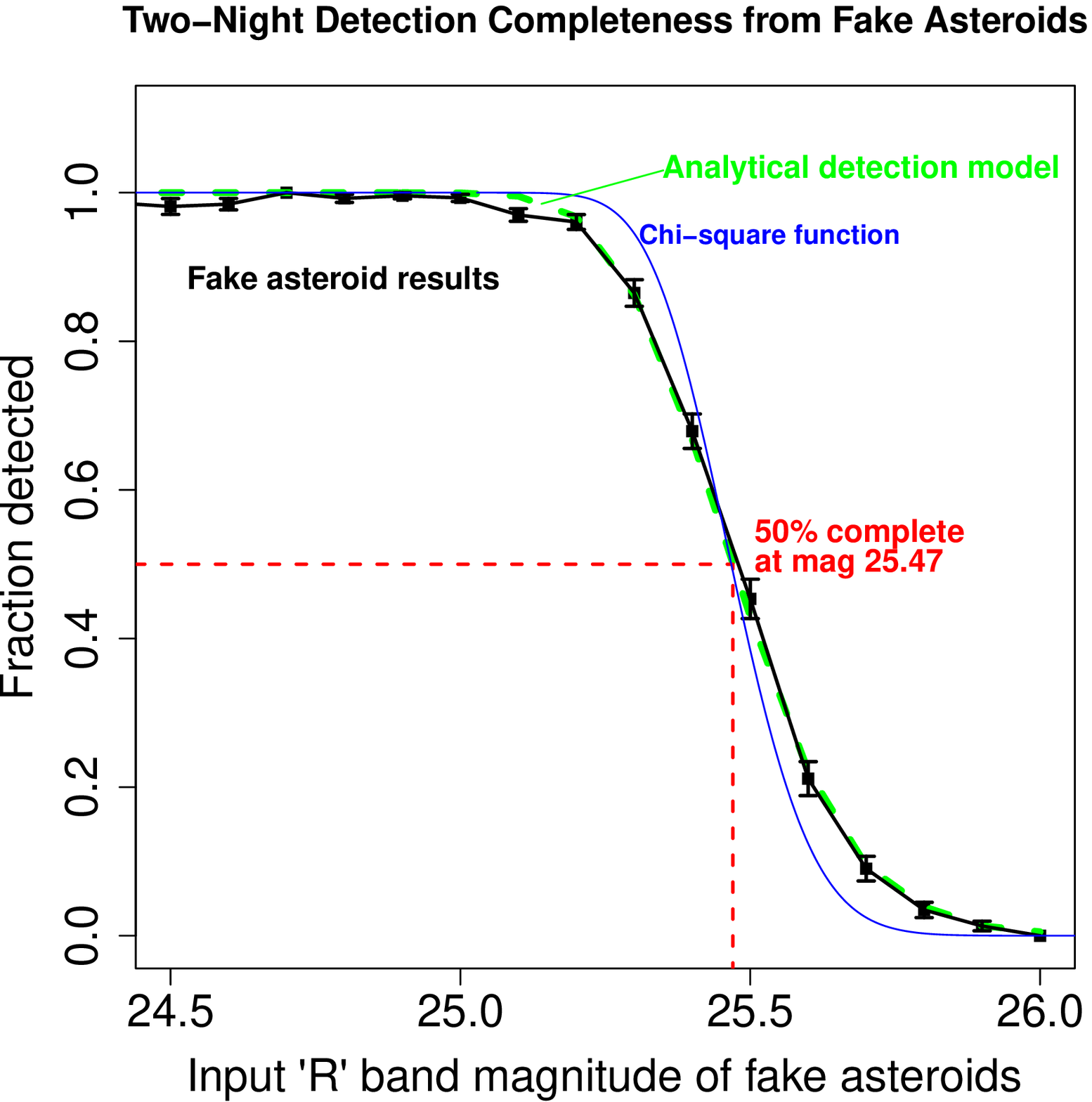}
\caption{Detection completeness for faint asteroids as a function of magnitude. The black line and points with error bars show the histogram-based completeness from the fake asteroid simulation. Results for March 30 and for the two-night analysis (\S \ref{sec:2night})are shown as examples; March 31 was analyzed in the same way. The completeness is above 98\% out to magnitude 24.9. Our limiting magnitude is fainter than 25.3 on both nights individually, and fainter than 25.4 for the two-night analysis thanks to the greater sensitivity enabled by confirming asteroids on both nights. The long-dashed green curves and the thin blue curves are from the more sophisticated analytical model described in \S \ref{sec:analyticalmod}. Results from March 31, not shown here, are very similar to March 30, differing only in that a softer shoulder at magnitude 25.0--25.1 causes the analytical model (green curve) to fit less well, and the limiting magnitude is fainter by 0.03 mag.
\label{fig:completeness}}
\end{figure*}

We find that our completeness is near 100\% out to a magnitude of about 24.9, and drops to essentially zero by magnitude 26. Using the standard definition --- that is, the magnitude at which 50\% completeness is reached --- our limiting magnitude is fainter than 25.3 on each night. A small number of bright fake asteroids are missed, less than 0.5\% down to magnitude 23.0 and less than 1.4\% to magnitude 24.9. It is not clear that these losses represent a meaningful incompleteness measurement that would also apply to the real asteroids. Some of the non-detections are due to an asteroid being crossed by the trail of a brighter object, which would occur somewhat more frequently among the fake asteroids because of their greater numbers. Hence, we adopt a nominal completeness of 100\% brightward of magnitude 24.9, while recognizing this may result in a $\sim 1$\% under-counting of the real asteroids. 

For asteroids of magnitude 24.9 and fainter, we apply a completeness correction in bins of 0.1 magnitude width, corresponding to the points plotted in Figure \ref{fig:completeness}. The correction consists of simply dividing the number of real asteroids in each bin by the corresponding detection fraction from the fake asteroid simulation. 

\section{Asteroids Detected on Both Nights} \label{sec:2night}

\subsection{Matching Asteroids from Night-to-Night} \label{sec:match}

By its nature, digital tracking analysis produces accurate angular velocity measurements for every asteroid. We can therefore extrapolate the motion of an asteroid detected in one night's data to predict where it should be in the data from the other night. However, a simpleminded extrapolation will produce predictions offset to the west of the asteroids' actual locations if looking forward in time (i.e., from March 30 detections to predictions for March 31), and offset to the east if we look back in time. The reason for this systematic error is the rotational reflex velocity \citep[RRV; see][]{curves,LinRRV}: that is, the part of the asteroid's angular velocity that is due neither to the orbital velocity of the asteroid nor that of the Earth, but rather to the observer's rotation about the geocenter. This rotational velocity averages to zero over a full day, but throughout the nightime hours it maintains a positive component in the same direction as the Earth's orbit. Asteroids at opposition are all moving westward on the sky because they are in their retrograde loop as the Earth overtakes them due to its faster orbital motion around the Sun, and the RRV term simply increases this westward angular velocity. If this too-fast westward motion is used to extrapolate forward or backward by one day, it produces the errors described above.

To extrapolate asteroid positions more accurately from one night to the next, we wish to estimate and remove the RRV contribution to our measured angular velocity. The RRV contribution is equal to the rotational velocity of the observer (projected on the plane of the sky and  averaged over the time span of the observations) divided by the distance to each asteroid. Since our observations were designed so that the target field transited the meridian near the center of our observing sequence, a simplified calculation yields an excellent approximation for the projected, averaged velocity. If $t_{obs}$ is the time interval over which the digital tracking observations were obtained (7.30 and 7.84 hr, respectively, for our two nights), the angle through which the Earth rotates between the start and midpoint (or equivalently midpoint and end) of the observations is $\theta_{obs} = 2 \pi (\frac{1}{2} t_{obs}) / (1 \hspace{0.05 in} \mathrm{day})$, and the velocity in question is given by:

\begin{equation} \label{eq:rotvel}
v_{rot} = v_{eq} \cos(\theta_{lat}) \frac{\sin(\theta_{obs})}{\theta_{obs}}
\end{equation}

\noindent where $v_{eq}$ is the equatorial rotation velocity of the Earth and $\theta_{lat}$ is the latitude of the observatory. For our March 31 observations, Equation \ref{eq:rotvel} gives $v_{rot} = 1386$ km/hr.

It remains to estimate $\Delta$, the asteroid's distance from Earth, which can be done only very approximately from a single night's data. Fortunately, a rough approximation will suffice to remove the majority of the RRV offset. If all asteroids moved in circular orbits at zero inclination, their distances could be calculated exactly from their angular velocities, and we base our distance estimates on this simplified scenario\footnote{Note that the distance estimate described here is {\em not} equivalent to the RRV method of \citet{curves} and \citet{LinRRV}. RRV distances are far more accurate and have no intrinsic systematic bias, but they require asteroids already to be linked over two nights. The night-to-night linkage we describe here is a prerequisite for the accurate RRV distances we will present in our companion paper (Heinze et al., in prep).}. For known asteroids in our data, this approximation systematically underestimates $\Delta$ by 10\%, with an RMS scatter of 14\%. Though better approximations could be devised, this one proved completely sufficient for our purposes, enabling us to predict second-night asteroid positions with an RMS error of less than 4 arcsec and a mean systematic offset of 0.5 arcsec (see below).}

Given $v_{rot}$ from Equation \ref{eq:rotvel} and $\Delta$ from our circular-orbit approximation, we take the RRV contribution to the angular velocity to be $v_{rot}/\Delta$, entirely in the RA direction. This is applied as a correction to remove the RRV signal from the measured angular velocity, and the corrected angular velocity is then used to extrapolate forward or backward by one day. For a typical main belt asteroid with $\Delta = 2.0$ AU, the RRV contribution on March 31 is 0.96 arcsec/hr, and hence the positional correction over 24 hours is 23 arcsec.

We linked asteroids from one night to the next based on their angular velocities as well as predicted positions, using a match radius of 40 arcsec for position and of 1.0 arcsec/hr for angular velocity. These radii are small enough that there is usually only one potential match for each asteroid (the few cases with multiple match candidates were checked by hand and could always be confidently resolved), yet the radii are much larger than typical matching errors in the respective dimensions. Of 2525 asteroids immediately matched between the two nights, the mean absolute deviations in RA and Dec between predicted and actual positions are 3.4 and 0.7 arcsec, respectively, while the mean systematic offsets are only 0.5 and 0.2 arcsec. The average change in angular velocity from March 30 to March 31 for matched asteroids is -0.10 arcsec/hr eastward and 0.05 arcsec/hr northward, and the respective standard deviations are 0.09 arcsec/hr and 0.07 arcsec/hr.


For each asteroid that was not matched (i.e., that was automatically detected on only one night), we performed a further manual investigation at the predicted position and angular velocity for the night on which it was not detected. We did this by creating `check images' (see \S \ref{sec:checkim}) centered on the predicted location and velocity of each un-matched asteroid. For these images, we used a half-width of 150 pixels (40 arcsec), and we probed 9 different angular velocities in a 3$\times$3 grid centered on the predicted velocity, with a grid spacing of 0.2 arcsec/hr. 

Manual examination of these images produced a clearly-detected match in most cases, although these manually recovered second-night detections often had a significance below the automated detection threshold of our digital tracking search. We decided whether these detections were confidently real using a threshold of $5.6\sigma$, which we obtained by solving Equation \ref {eq:fpthresh} with $N_{real} = 9\times10^7$. The value of $N_{real}$ comes from the fact that each of the `check images' used to match asteroids has an area about $9\times10^4$ times larger than the $\sim 3\times3$ pixel effective area of the PSF, and we multiply this value by 1000 to find a threshold that should produce less than 1 false positive over the entire survey. This is conservative because the `check images' are generously sized and the asteroids are usually found near the center; and because the total number of single-night asteroids that we attempted to match based on predicted positions was not 1000 but only about 600. About two dozen manually matched detections had significance below 5.6$\sigma$ and were ultimately rejected (even though most were probably real) to ensure a negligible rate of false positives. For the remaining matches, all with $>5.6\sigma$ significance, successful detection at a predicted location in an independent data set confirms the reality of the asteroids beyond reasonable doubt --- though most of them were already known to be real based on single-night detections above the $\sim 7\sigma$ thresholds quoted above.

\subsection{Discussion of Detected Two-Night Asteroids} \label{sec:count2}

Table \ref {tab:detclass} gives the numbers of objects in various categories that were detected, either automatically or manually, on a second night. In category `a' (morphologically normal-looking detections above the significance limit) less than 2.5\% of all detections failed to be matched. Out of a total of 103 candidates in category `b' (detections that looked morphologically peculiar under manual screening) only one object was matched on a second night: the manual screener's impression of something being wrong with a given detection was usually accurate. In the `subsig' category (morphologically normal but below the single-night significance limit), slightly under 50\% of detections were confirmed on a second night. These candidates are thus confirmed as real asteroids for purposes of our two-night analysis, though they do not count as independent single-night detections.

Among the small number of confidently-detected single-night asteroids that could {\em not} be manually matched, the cause in many cases was obvious: the asteroid had moved out of our target field. There were a few cases where the image quality seemed good but the `check image' showed apparently empty sky or only a very low significance ($<5\sigma$) detection. We believe that in most of these cases, the asteroid was real and present in the sky area covered by the image, but had simply become too faint for detection due to rotational brightness variations. Observations of Near-Earth asteroids --- in the same size range as the main belt objects we have detected --- show that they often vary by more than 0.7 magnitudes in brightness as they rotate \citep[e.g.][]{Hergenrother2011}. Such variations could easily cause an object that was detected at good significance (e.g. 8$\sigma$) on one night to fade below our detection limit on the other. 

We find that 2525 asteroids were automatically, independently detected on both nights; 206 asteroids were automatically detected on March 30 and manually recovered in the March 31 data; and 392 asteroids were automatically detected on March 31 and manually recovered in the March 30 data. Thus, a total of 3123 asteroids were each detected on both nights. Table \ref{tab:matchdet} gives the breakdown of these numbers in terms of manual classifications. In addition to the 3123 two-night asteroids, 40 and 71 detections on March 30 and 31, respectively, could {\em not} be recovered on a second night but were confidently real. Hence, we have detected a total of 3234 distinct asteroids. 

\begin{deluxetable*}{p{180pt}rcccccc}
\tablewidth{0pt}
\tablecaption{Confirmed Asteroids Detected on Both Nights\label{tab:matchdet}}
\tablehead{ &  & \multicolumn{6}{c}{Breakdown of manual classifications\tablenotemark{d}} \\
 &  & \multicolumn{3}{c}{March 30} & \multicolumn{3}{c}{March 31} \\
\colhead{Description} & \colhead{Total} & \colhead{`a'} & \colhead{`subsig'} & \colhead{`b'} \hspace{0.2in} & \colhead{`a'} & \colhead{`subsig'}  & \colhead{`b'} }
\startdata
Automatically detected on both nights & 2525 \hspace{0.3in} & 2518 & 7 & 0 \hspace{0.2in} & 2520 & 5 & 0 \\[0.10in]
Automatically detected only on March 30\tablenotemark{e} & 206 \hspace{0.3in} & 202 & 4 & 0 \hspace{0.2in} & NA & NA & NA \\[0.10in]
Automatically detected only on March 31\tablenotemark{f} & 392 \hspace{0.3in} & NA & NA & NA \hspace{0.2in} & 382 & 9 & 1 \\[0.10in]
All two-night detections\tablenotemark{g} & 3123 \hspace{0.3in} & 2720 & 11 & 0 \hspace{0.2in} & 2902 & 14 & 1 \\
\enddata
\tablenotetext{d}{The `a', `subsig', and `b' classifications are explained in \S \ref{sec:flist}. Briefly, `a' and `subsig' both mean the detection looked genuine under manual screening, but `subsig' means its formal significance (see \S \ref{sec:flist}) fell below the thresholds of $7.43 \sigma$ and $7.28\sigma$ derived in \S \ref{sec:fp} for March 30 and 31, respectively. A classification of `b' means the detection looked peculiar under manual screening, regardless of its significance.}
\tablenotetext{e}{All of these detections were manually recovered in the March 31 data.}
\tablenotetext{f}{All of these detections were manually recovered in the March 30 data.}
\tablenotetext{g}{This is the total count of asteroids that were each detected on both nights, regardless of whether the detections were automatic or manual.}
\end{deluxetable*}

\subsection{Survey Completeness for Two-Night Asteroids} \label{sec:2dens}

Since our fake asteroids were placed using self-consistent orbits from night to night, fake asteroids can be linked from one night to the next in the same way as real asteroids. Hence, we can determine the fraction of fake asteroids that were detected on both nights, and create a new completeness curve that corresponds to two-night detections. We obtain the effective magnitude of asteroids detected on both nights, whether real or fake, by simply averaging the measured magnitudes for each individual night. We compare the fake asteroids detected on both nights to the input fake asteroids that {\em could in principle} have been detected on both nights, rather than to the total number of input fake asteroids, some of which moved into or out of our field of view from one night to the next. The resulting completeness curve is shown in the right panel of Figure \ref{fig:completeness}. The fact that lower-significance asteroids could be confirmed by detection on a second night contributes to higher completeness at faint magnitudes relative to our single-night results.

\begin{figure}
\includegraphics[scale=0.5]{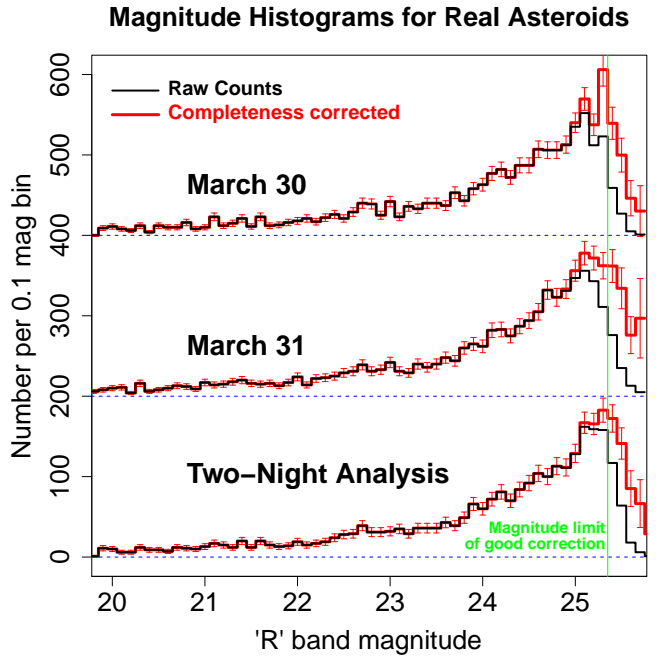}
\caption{Histograms of apparent $R$ magnitude for real main-belt asteroids showing both raw counts and counts corrected for survey incompleteness based on the completeness curves from the fake asteroid test (see Figure \ref{fig:completeness}). The correction fails for magnitudes fainter than 25.3, for reasons described in \S \ref{sec:statbias}. Note that the greater sensitivity enabled by confirming asteroid detections on both nights (\S \ref{sec:2night}) produces significantly higher completeness at faint magnitudes for the two-night analysis.
\label{fig:realhist}}
\end{figure}

\section{Abundance and Magnitude Distribution of Faint Asteroids} \label{sec:counts}

We have described in \S \ref{sec:fp} and \S \ref{sec:2night} how we arrived at lists of real asteroids independently detected on each night, totalling 2760, 2973, and 3123 objects for March 30, March 31, and the combined two-night data set, respectively. Hence, we now have three different sets of detected asteroids, each with an associated completeness curve from our fake asteroid test. Of these three sets, those for March 30 and March 31 have been measured entirely independently (though necessarily they include many of the same objects); while the two-night set, though not independent, is more sensitive and is subject to different selection effects. Therefore, results that are found consistently for all three samples can reasonably be regarded with additional confidence.

In the analysis below, we apply each probe of the on-sky density and apparent magnitude distribution of MBAs to each of our three data sets in turn\footnote{The only exception is our power law fits to the cumulative magnitude distribution, which we analyze only for the single-night data sets, finding almost identical results for each night.}. In the interests of space, we illustrate certain types of analysis using example figures showing only one or two of our sets of asteroids, but we provide numerical results for all three data sets in tabular form. 

\subsection{On-Sky Number Density} \label{sec:skydens}

We begin by simply calculating the number of asteroids per square degree, down to various magnitude limits.

\subsubsection{Effective Sky Area Probed}

The effective sky areas probed by our observations, calculated using the boundary maps described in \S \ref{sec:edgedets}, are 2.932 square degrees for March 30 and 2.989 for March 31. To calculate the effective area for our analysis of asteroids detected on both nights, we use the areas of sky within the boundaries covered by our single-night observations, combined with the known motions of real asteroids detected within these fields on both nights. We find the arc on the sky through which each asteroid moved between March 30 and March 31. We then shift the March 30 field in real celestial coordinates according to this arc, and calculate its overlap with the March 31 field. The effective sky area is the field overlap averaged over the motion arcs of all the asteroids, and is equal to 2.848 square degrees. Thanks to our well-optimized choice of target fields, which offset the field centers from one another by an amount that accurately matched the average daily motion of our target asteroids, this is only 3\% smaller than the area covered on March 30. The 3\% reduction represents the statistical fraction of detectable main belt asteroids that, being near the edge of the field in one night, would not be in the field at all on the other.

\subsubsection{Culling the Lists of Real Asteroids}

We exclude the Hilda asteroids and the Jupiter Trojans from our final lists, for several reasons. First, we wish to probe the main-belt asteroids specifically, and the Hildas and Trojans constitute distinct populations with larger average distance and different size-frequency distributions \citep{Terai2018}. Second, we wish to obtain results directly comparable to other surveys \citep[e.g.][]{Gladman2009} that targeted MBAs. Finally, we want our measurements of the on-sky density of asteroids at opposition to be substantially independent of the time of year (i.e. ecliptic longitude) and the position of Jupiter. This is expected if we confine our analysis to MBAs, which are distributed fairly uniformly in ecliptic longitude, but not if we include the Jupiter Trojans (which are strongly clustered around Jupiter's L4 and L5 Lagrange points) or the Hilda asteroids (which are significantly concentrated at ecliptic longitudes 180 and $\pm60$ degrees from Jupiter).

We identify Hilda asteroids and Jupiter Trojans using boundaries in angular velocity phase space indicated by green and red lines in Figure \ref{fig:angphase}. On March 30 and 31 respectively, we reject 231 and 242 Jupiter Trojans, and 28 and 29 Hilda asteroids, leaving 2501 and 2702 independently detected main belt asteroids.

Prior to removing the Jupiter Trojans and Hilda asteroids from our list of 3123 two-night asteroids, we perform preliminary culling to ensure the final list can be used for statistical analysis herein and in our companion paper (Heinze et al., in prep.) on the absolute magnitude distribution. We remove 170 two-night asteroids that were manually recovered only near the very edge of our images on one of the nights: they were outside the nominal boundary of valid data described in \S \ref{sec:edgedets}. Automated detections are impossible outside the boundaries, but the objects in question were found manually at positions predicted based on automatic detections from the other night. We remove them because we cannot accurately quantify the sensitivity or the effective sky area probed by such detections. Seventeen additional two-night asteroids were removed due to astrometric uncertainties so large that their RRV distances would have greater than 10\% errors\footnote{This cull is needed because the current analysis will be foundational for our companion paper on absolute magnitudes. Many of the culled objects may also have been Jupiter Trojans, whose RRV signals are hard to measure due to their distance. The comparison with fake asteroids is not affected because none of the latter had such large distance uncertainties.}. Finally, we remove 241 Jupiter Trojans, and 27 Hildas, leaving 2668 two-night asteroids detected under conditions of well-quantified sensitivity --- a list directly comparable to the similar one we constructed for the two-night fake asteroids.

We plot the distributions of these objects' apparent $R$ band magnitudes in Figure \ref{fig:realhist}.

\subsubsection{Calculated On-Sky Number Densities}

Interesting magnitude thresholds for our sky density calculation include the value of 23.0 for which \citet{Gladman2009} found 210 asteroids per square degree; the value of 24.4 for which \citet{Yoshida2003} found $\sim290$ while \citet{Gladman2009} extrapolated their power law to estimate $\sim500$; 25.0 because it is the integer magnitude nearest the sensitivity limit of our survey; 25.3 because it is our limiting magnitude for 50\% completeness, and 25.6 because it is the extreme limit of our sensitivity, beyond which our completeness is less than 10\%.

The completeness curves derived from our fake asteroid simulation indicate that our detection rate is nearly 100\% out to $R$ mag 24.9. At mag 25.0 it is still 96.4\% and 94.2\% for March 30 and 31 respectively, and 99.3\% for the two-night analysis. Hence, we will undercount the true numbers of asteroids only slightly if we simply sum our detected asteroids out to the respective thresholds. Table \ref{tab:skydens} presents raw counts as well as completeness-corrected values, and the former are likely to be as accurate as the latter out to magnitude 25.0. Fainter than 25.0, the incompleteness becomes significant and only the corrected numbers should be regarded as meaningful measurements. It is important to note that all of these on-sky number densities apply only to asteroids at opposition on the ecliptic (i.e., near the antisolar point). If we move away from the antisolar point but stay on the ecliptic, the sky density would be expected to drop quickly due to the rapid falloff in asteroid brightness with increasing phase angle. Moving off the ecliptic, the density should drop even faster because there are fewer asteriods in highly inclined orbits \citep[e.g.][]{Terai2007,Terai2013}.

\begin{deluxetable*}{lccccc}
\tablewidth{0pt}
\tablecaption{On-Sky Number Density of Asteroids\label{tab:skydens}}
\tablehead{ & & \multicolumn{2}{c}{\textbf{Raw Counts}} & \multicolumn{2}{c}{\textbf{Completeness Corrected}} \\ 
 & \colhead{Magnitude} & \colhead{Total} & \colhead{Asteroids} & \colhead{Total} & \colhead{Asteroids} \\
\colhead{Date} & \colhead{threshold} & \colhead{detected} & \colhead{per deg$^2$} & \colhead{detected} & \colhead{per deg$^2$}}
\startdata
March 30 & 23.0 & 603 & $206 \pm 8$ & 603 & $206 \pm 8$ \\
March 31 & 23.0 & 623 & $208 \pm 8$ & 623 & $208 \pm 8$ \\
Two-Night & 23.0 & 587 & $206 \pm 9$ & 587 & $206 \pm 9$ \\
Average\tablenotemark{a} & 23.0 & \nodata & $207 \pm 8$ & \nodata & $207 \pm 8$ \\
March 30 & 24.4 & 1326 & $452 \pm 12$ & 1326 & $452 \pm 12$ \\
March 31 & 24.4 & 1384 & $463 \pm 12$ & 1384 & $463 \pm 12$ \\
Two-Night & 24.4 & 1296 & $455 \pm 13$ & 1296 & $455 \pm 13$ \\
Average\tablenotemark{a} & 24.4 & \nodata & $457 \pm 13$ & \nodata & $457 \pm 13$ \\
March 30 & 25.0 & 1960 & $668 \pm 15$ & 1963 & $670 \pm 15$ \\
March 31 & 25.0 & 2078 & $695 \pm 15$ & 2084 & $697 \pm 15$ \\
Two-Night & 25.0 & 1916 & $673 \pm 15$ & 1917 & $673 \pm 15$ \\
Average\tablenotemark{a} & 25.0 & \nodata & $679 \pm 15$ & \nodata & $680 \pm 15$ \\
March 30 & 25.3 & 2350 & $802 \pm 17$ & 2431 & $829 \pm 17$ \\
March 31 & 25.3 & 2513 & $841 \pm 17$ & 2599 & $870 \pm 18$ \\
Two-Night & 25.3 & 2377 & $835 \pm 17$ & 2398 & $842 \pm 17$ \\
Average\tablenotemark{a} & 25.3 & \nodata & $826 \pm 17$ & \nodata & $847 \pm 17$ \\
March 30 & 25.6 & 2498 & $852 \pm 17$ & 2810 & $958 \pm 22$ \\
March 31 & 25.6 & 2691 & $900 \pm 17$ & 3022 & $1011 \pm 22$\\
Two-Night & 25.6 & 2657 & $933 \pm 18$ & 2883 & $1012 \pm 21$ \\
\enddata
\tablenotetext{a}{Unweighted numerical average over the three data sets. Uncertainty does not go down under the average since the individual uncertainties are mostly due to Poisson noise of counting a heavily-overlapping set of asteroids. Averages of total counts are not meaningful since the effective sky areas are not the same. At mag 25.6 the completeness correction fails and all values are underestimates, so the average would not be interesting.}
\end{deluxetable*}

Table \ref{tab:skydens} shows that our results are mutually consistent from night to night, and are in excellent agreement with those of \citet{Gladman2009} at $R$ = 23.0. At magnitude 24.4, our March 31 measurement of $463 \pm 12$ asteroids per square degree agrees much more closely with the extrapolated estimate of $\sim 500$ from \citet{Gladman2009} than with the value of 290 found by \citet{Yoshida2003}, indicating that the latter may have overestimated their completeness at faint magnitudes (as \citet{Gladman2009} had already suggested).

We expect the corrected counts in Table \ref{tab:skydens} to be accurate at magnitude 25.3, but not at magnitude 25.6, where Figure \ref{fig:realhist} indicates our completeness correction has become unreliable. While always mutually consistent within $2\sigma$ (except for $R=25.6$ where inaccuracy is expected), the asteroid counts are consistently higher for March 31 relative to March 30. This is plausibly a real effect due to the phase function of asteroids, which is steep enough near opposition that the fact that the March 30 field is centered $1.2^{\circ}$ away from the antisolar point would be expected to produce a small but significant reduction in the brightness of the asteroids. Since the March 31 observations were centered almost exactly on the antisolar point, they should give the most accurate measurement of the on-sky density of faint asteroids at opposition. On this date we find $870 \pm 18$ asteroids per square degree brighter than the $R=25.3$ mag limit of our completeness correction.

\subsection{Power Laws to Fit Asteroid Distributions} \label{sec:pwrlaws}

Published results on the magnitude and size distributions of faint asteroids have typically been presented in terms of a cumulative power law distribution over either magnitude or size. In terms of diameter $D$, the power law is given by Equation \ref{eq:sizepowercum} \citep[e.g.][]{Wiegert2007}:

\begin{equation}
N(>D) \propto D^{-b_c}
\label{eq:sizepowercum}
\end{equation}

In the approximation that the distribution of asteroid albedos is independent of diameter $D$, it follows from Equation \ref{eq:sizepowercum} that the cumulative distribution of asteroid absolute magnitudes is given by Equation \ref{eq:magpowercum} \citep[e.g.][]{Gladman2009}:

\begin{equation}
N(<M) \propto 10^{\alpha_c M}
\label{eq:magpowercum}
\end{equation}

\noindent where $M$ is the absolute magnitude, and the power law slopes $b_c$ and $\alpha_c$ are related by $b_c = 5 \alpha_c$. 

Equations \ref{eq:sizepowercum} and \ref{eq:magpowercum} describe the cumulative distribution. We can also explore the corresponding differential power laws:

\begin{equation}
\frac{dN}{dD} \propto D^{-b_d}
\label{eq:sizepowerdiff}
\end{equation}

\begin{equation}
\frac{dN}{dM} \propto 10^{\alpha_d M}
\label{eq:magpowerdiff}
\end{equation}

If the slope $b_d$ is constant over a large range in size, the cumulative and differential size slopes are related by $b_c = b_d-1$. In this case, the differential and cumulative magnitude slopes are the same, $\alpha_c = \alpha_d$, due to the exponential form of Equations \ref{eq:magpowercum} and \ref{eq:magpowerdiff}. If the same slopes described all main belt asteroids, these magnitude equations would refer equally to apparent magnitudes (e.g. $R$) and to absolute magnitude $H$, and the slope of the diameter power law ($b_d$ or $b_c$) could be calculated from the apparent magnitude distribution without ambiguity or error.

In fact, abundant evidence exists \citep[e.g.][]{Wiegert2007,Yoshida2007,Gladman2009} that the power law slopes are {\em not} constant. Hence, the cumulative slopes $b_c$ and $\alpha_c$ measured over a given range of size or magnitude constitute a type of weighted average over changing differential slopes $b_d$ and $\alpha_d$ for all the asteroids that contribute to the distribution --- i.e., not only those in the range being fit, but also objects larger/brighter that nevertheless add in to the cumulative total. This blending is a disadvantage of using the cumulative distributions. An advantage is the much larger number of asteroids per bin, and consequently smaller statistical errors. With over 2500 asteroids, we have a large enough sample to produce meaningful fits to the differential distributions, which we do in \S \ref{sec:diffpower01}. However, for consistency with previous results, we first consider fits to the cumulative distribution in \S \ref{sec:cumpower01}.

In the current work, we fit only the apparent $R$ magnitude distribution, reserving our fits to the distributions of absolute magnitude $H$ to a companion paper (Heinze et al., in prep.), in which we also describe how we obtained $\sim 1.5$\% accurate distances to the asteroids we detected on both nights. Since we are fitting the apparent magnitude distribution herein, the slopes we derive are a weighted average over the absolute magnitude distributions at various size ranges and distances in the main belt. For example, the range $R=24-25$ corresponds to $H=22.5-23.5$ (a size range of $70-100$ m for 15\% albedo) near the inner edge of the main belt at 1.0 AU from the Earth, but a range of $H=19.5-20.5$ (sizes of $270-430$ m) 2.4 AU from Earth in the outer main belt. This average nature of our results must be kept in mind when interpreting their implications for the size distributions.

\subsection{Cumulative Power Laws} \label{sec:cumpower01}

\begin{figure*}
\includegraphics{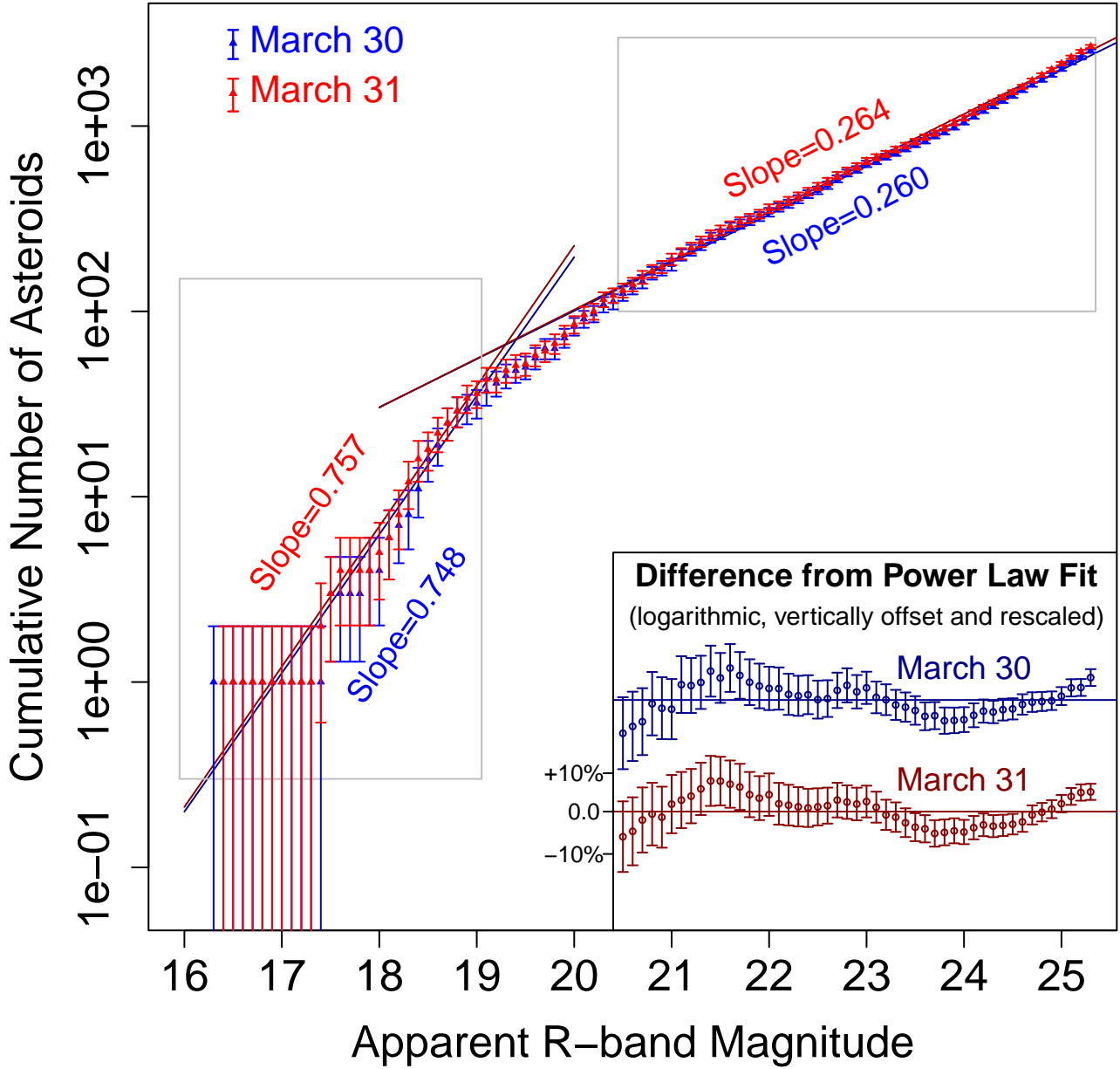}
\caption{Cumulative distribution of asteroids based on the corrected histograms shown in Figure \ref{fig:realhist}. Our sensitivity extends about one magnitude fainter than any previous survey. As discussed in \S \ref{sec:cumpower01} the slopes we derive over various magnitude ranges are mostly consistent with previously published results, though we disagree with \citet{Yoshida2003} in that we do not find a decrease in the power law slope for asteroids fainter than 23rd magnitude. On the contrary, the difference plot shown in the lower right suggests the slope may even steepen for the faintest asteroids we probe --- a possibility we explore further by fitting differential rather than cumulative power laws in \S \ref{sec:diffpower01}.
\label{fig:cumdist}}
\end{figure*}

Following \citet{Gladman2009}, we fit the cumulative distribution of $R$ magnitudes for our real asteroids using Equation \ref{eq:magpowercum}. This distribution is shown in Figure \ref{fig:cumdist}. We use weighted least-squares to fit a line to the quantity $\log_{10}(n)$, where $n$ is the number of detected asteroids. We calculate the uncertainty on the logarithm using Equation \ref{eq:logsig}:

\begin{equation}
\sigma_{\log_{10}(n)} = \frac{\sigma_n}{n \ln(10)}
\label{eq:logsig}
\end{equation}

\noindent Where we use $\sigma_n = \sqrt n$ for magnitudes brighter than $R=24.9$, but include the uncertainty on the completeness correction for fainter magnitudes. 


Consistent with \citet{Gladman2009}, we find a dramatic transition in the slope $\alpha_c$ at a relatively bright magnitude between $R=19$ and $R=20$. Brighter than $R=19$, we find very steep values of $\alpha_c = 0.75 \pm 0.02$ and $0.76 \pm 0.04$ for March 30 and 31, respectively. In this regime, \citet{Gladman2009} found $\alpha_c = 0.61 \pm 0.16$ and $0.56 \pm 0.15$ on two successive nights. Since both studies are plagued by small number statistics for these relatively rare, bright asteroids, the fact that we find steeper slopes than \citet{Gladman2009} may not be significant. The change in the slope of the apparent magnitude distribution that both we and \citet{Gladman2009} identify near $R=19$ mag probably corresponds to the dramatic change in the size distribution that \citet{Yoshida2019} detect at a diameter of about 6km in an analysis of completely independent data that includes space-based infrared measurements.

More interesting for our current analysis is the fainter regime. From $R=20.5$ to 22.5, \citet{Gladman2009} found $\alpha_c = 0.266 \pm 0.014$ and $0.265 \pm 0.014$ on two nights. Our values in the same regime are in full agreement: we find $0.267 \pm 0.002$ and $0.268 \pm 0.002$ for March 30 and 31. Translating to $b_c$ (subject to the caveats about averaging noted in \S \ref{sec:pwrlaws}), we find $b_c = 1.34 \pm 0.01$. For comparison, using observations in a similar magnitude range \citet{Wiegert2007} found $b_c = 1.35 \pm 0.02$ in the $g'$ band but a steeper slope of $b_c = 1.91 \pm 0.08$ in $r'$. We would have expected their $r'$ result to be more comparable to our own observations in $R$; note, however, that they calculated approximate distances and sizes for all their asteroids, and hence attempted to fit the actual size distribution rather than merely the apparent magnitude distribution as we have done. They also observed farther from opposition than we or \citet{Gladman2009}, so if there is a systematic difference in phase functions between large and small asteroids, this could explain the difference in measured power law slopes.

Extending our fit of the cumulative distribution from $R=20.5$ to our 50\% completeness limit at $R=25.3$, we find  $\alpha_c = 0.260 \pm 0.003$ and $0.264 \pm 0.002$ for March 30 and 31. Since our weighted fits emphasize the faint bins containing the largest numbers of asteroids, the fact that these slopes agree with \citet{Gladman2009} down to two magnitudes fainter indicates that there is no dropoff in the abundance of small asteroids at magnitudes fainter than the $R \sim 23.5$ limit probed by \citet{Gladman2009}, in mild disagreement with \citet{Yoshida2003}, who probed asteroids down to $R \sim 24.4$ and found evidence for such a dropoff. Converting from $\alpha_c$ to the size slope (again, subject to many caveats), we find $\beta_c = 1.31 \pm 0.01$, which disagrees with the value of $\sim 1.2$ reported by \citet{Yoshida2003}, but is consistent with the value of $\beta_c = 1.29 \pm 0.02$ found by \citet{Yoshida2007} based on observations that were sensitive down to $R \sim 24.2$.

The inset of Figure \ref{fig:cumdist} shows the logarithmic difference between the measured cumulative distribution and the best power law fit from $R=20.5$ to 25.3. Curvature in the difference plot shows the distribution is not a pure power law. The curvature at bright magnitudes may be due to the cumulative distribution's slow recovery from the large slope change near $R=19$, but the fact that the difference plot trends upward at the faintest magnitudes suggests a second transition, this time to a steeper slope for the faintest asteroids. This could be the first detection of increasing abundance of small MBAs below the strength/gravity transition, as predicted by \citet{Bottke2005b} and \citet{deElia2007}. We use the higher resolution enabled by the differential rather than cumulative distributions to explore this possibility further in \S \ref{sec:diffpower01}

\subsection{Differential Power Laws} \label{sec:diffpower01}

The cumulative distribution of our asteroid magnitudes (Figure \ref{fig:cumdist}) shows systematic deviations from the best-fit power law. To explore this further, we fit differential, rather than cumulative, power laws to the magnitude histograms. The differential distributions are noisier than the cumulative ones due to small number statistics, but they have the advantage that successive bins are independent. Where sufficiently large numbers of objects have been measured, the differential distribution offers much better resolution for determining transition points in the power law slope: by contrast, in the cumulative distribution the effect of such a transition is spread over a wide range (e.g. the slope transition at magnitude 19 affects the cumulative distribution out to at least magnitude 21).

\begin{figure*}
\plottwo{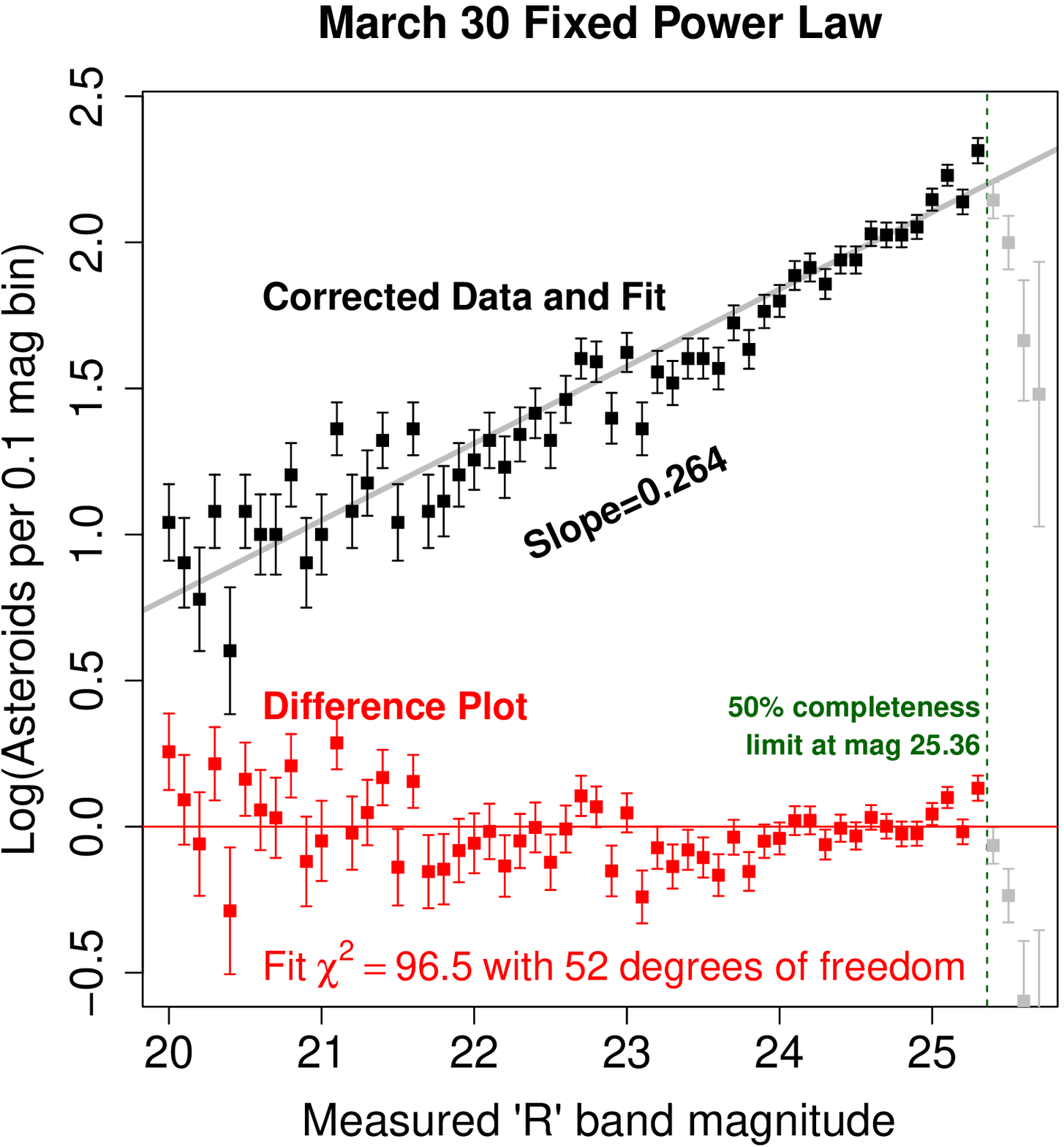}{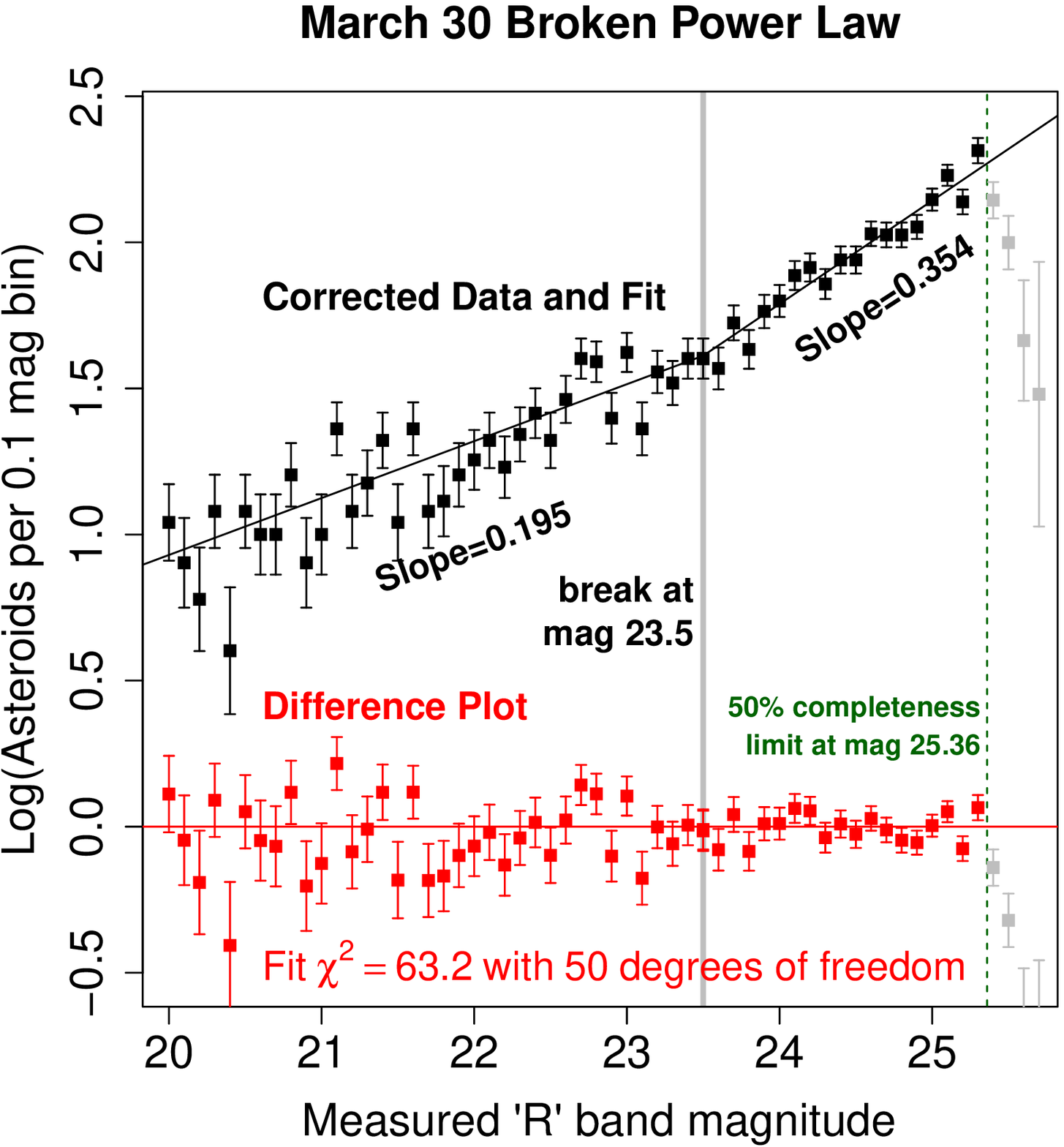}
\caption{Differential power law fits to the magnitude histogram of asteroids detected on March 30, showing that the data from $R=20.0$ to 25.3 can be fit well with a broken power law but not with a constant power law. Gray points at the faintest magnitudes were not included in the fit, since for them our completeness correction was unreliable. {\em Left:} Constant power law. Curvature in the difference plot is evident to the eye and manifests itself through the large $\chi^2$ value. The probability of getting $\chi^2 \ge 96.5$ with 52 degrees of freedom is $2 \times 10^{-4}$. {\em Right:} Broken power law with a slope transition at $R=23.5$. The difference plot is much flatter, and a $\chi^2$ value this large occurs with 10\% probability. \label{fig:diffcomp30a}}
\end{figure*}

\begin{figure*}
\plottwo{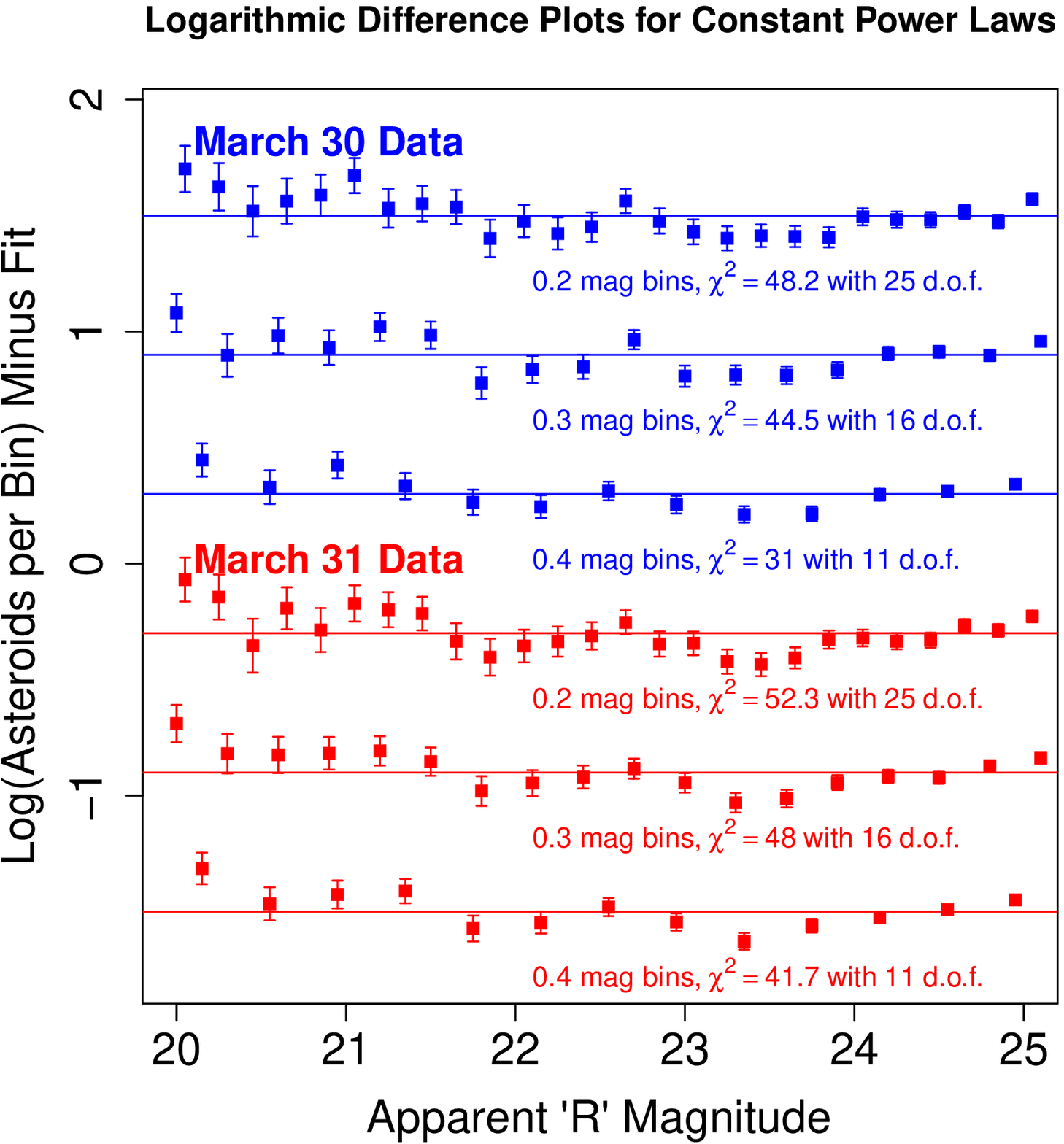}{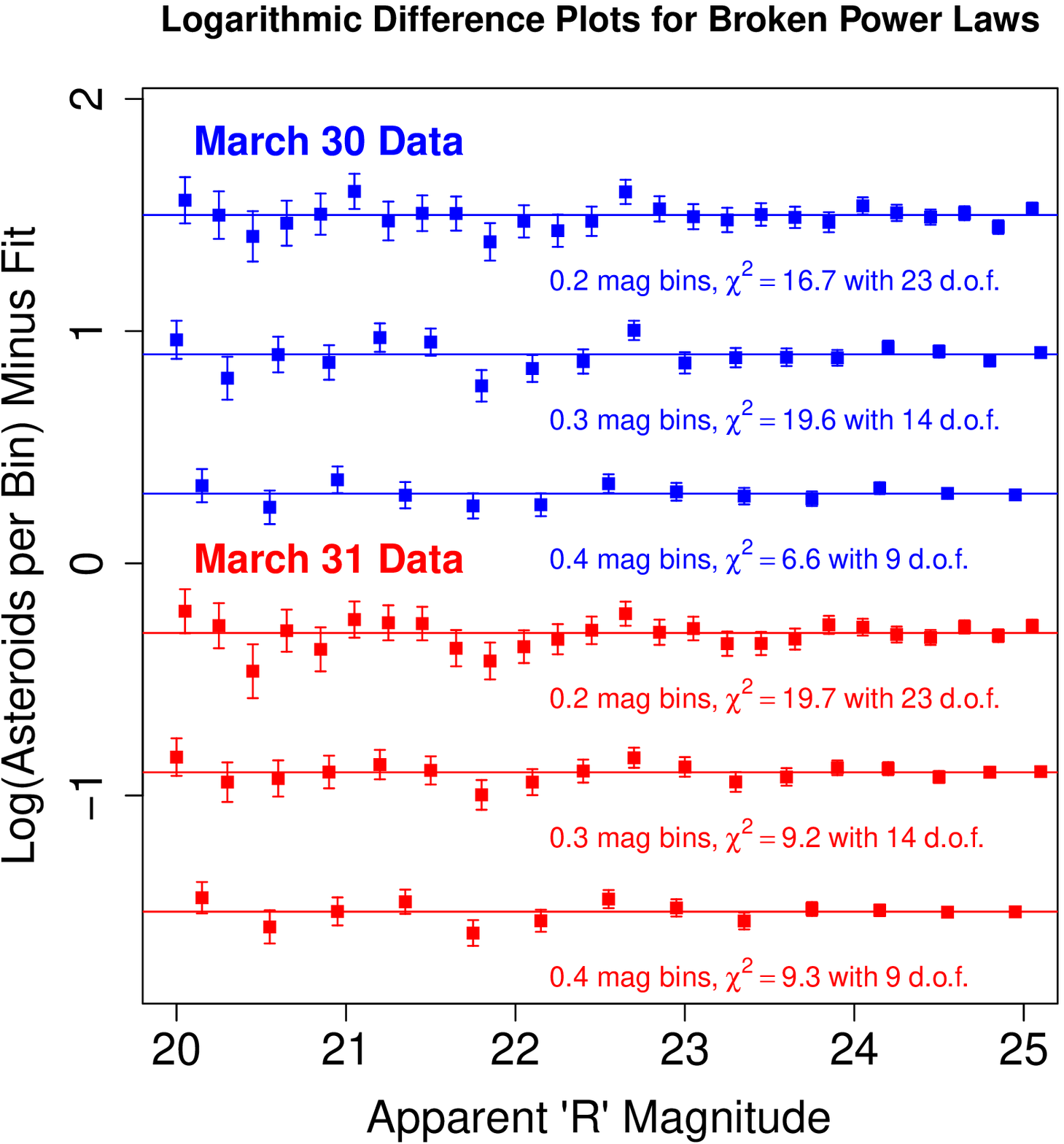}
\caption{Logarithmic difference plots for constant and broken power laws with asteroids binned in different ways, showing that the constant power laws consistently fail to yield a satisfactory fit while the broken power laws consistently succeed. {\em Left:} Constant power laws. The probabilities of getting the $\chi^2$ values as high as those listed from a true $\chi^2$ distribution with the correct number of degrees of freedom range from $5 \times 10^{-5}$ to $4 \times 10^{-3}$: they are all excluded with greater than 99.5\% confidence. {\em Right:} Broken power laws with the break at $R=23.5$. The probabilities of these $\chi^2$ values range from 14\% to 82\%: the fits are excellent and none would be rejected even at a 90\% confidence level. \label{fig:logdiff}}
\end{figure*}

We can fit the differential distribution of our asteroid magnitudes (i.e. the histograms of Figure \ref{fig:realhist}) from $R=20.0$ to 25.3 mag with single power laws having slopes of $\alpha_d=0.264$, 0.270, and 0.274 for the March 30, March 31, and two-night asteroids respectively. The fit for March 30 is shown as an example in the left-hand panel of Figure \ref{fig:diffcomp30a}. It and the corresponding fits for the March 31 and two-night analyses match the data approximately, but logarithmic difference plots for these fits (Figure \ref{fig:logdiff}) show an apparent lack of asteroids with magnitudes in the range 23--24 and/or an excess at magnitudes fainter than 24.5 --- consistent with the curvature seen in the difference plot for the cumulative distribution shown in Figure \ref{fig:cumdist}. These plots suggest a break in the power law slope at about magnitude 23.5, broadly consistent with predictions \citep[see, e.g.][]{Bottke2005b,deElia2007} of an increasing number of small asteroids in the regime of nonzero tensile strength.

Accordingly, we fit broken power laws to the observed magnitude histogram from $R=20$ to 25.3, exploring slope breakpoints in a wide range from $R=21.5$ to $R=24.0$ at 0.1 mag intervals. We do this using a weighted least-squares fit of a broken linear function to the logarithm of the completeness-corrected histogram, constraining the fit to be continuous at the breakpoint and calculating the uncertainties on the log of the histogram with Equation \ref{eq:logsig}. The right-hand panel of Figure \ref{fig:diffcomp30a} shows the resulting broken power-law fit for March 30, with the break at $R=23.5$, which yielded the lowest $\chi^2$ values. Results for March 31 and the two-night analysis, given in Table \ref{tab:powerfits}, are very similar. 

\begin{deluxetable*}{ccccccccccc}
\tablewidth{0pt}
\tablecaption{Differential Power Law Fits to the Magnitude Histogram\label{tab:powerfits}}
\tablehead{ & & \multicolumn{4}{c}{\textbf{Constant Power Law}} & \multicolumn{5}{c}{\textbf{Broken Power Law}} \\ 
\colhead{Date} & \colhead{Bin Size} & \colhead{Slope\tablenotemark{a}} & \colhead{$\chi^2$} & \colhead{d.o.f} & \colhead{Prob\tablenotemark{d}} & \colhead{Slope1\tablenotemark{b}}  & \colhead{Slope2\tablenotemark{c}} & \colhead{$\chi^2$} & \colhead{d.o.f} & \colhead{Prob\tablenotemark{d}}}
\startdata
March 30 & 0.1 mag & 0.2637 & 96.5 & 52 & $2 \times 10^{-4}$ & 0.1948 & 0.3539 & 63.2 & 50 & 10\% \\
March 30 & 0.2 mag & 0.2684 & 48.2 & 25 & $4 \times 10^{-3}$ & 0.2015 & 0.3567 & 16.7 & 23 & 82\% \\
March 30 & 0.3 mag & 0.2599 & 44.5 & 16 & $2 \times 10^{-4}$ & 0.2021 & 0.3442 & 19.6 & 14 & 14\% \\
March 30 & 0.4 mag & 0.2624 & 31.0 & 11 & $1 \times 10^{-3}$ & 0.2031 & 0.3546 & 6.6 & 9 & 68\% \\
March 31 & 0.1 mag & 0.2702 & 76.4 & 52 & 1.5\% & 0.2010 & 0.3596 & 41.4 & 50 & 80\% \\
March 31 & 0.2 mag & 0.2720 & 52.3 & 25 & $1 \times 10^{-3}$ & 0.2051 & 0.3586 & 19.7 & 23 & 66\% \\ 
March 31 & 0.3 mag & 0.2726 & 48.0 & 16 & $5 \times 10^{-5}$ & 0.2018 & 0.3737 & 9.2 & 14 & 82\% \\
March 31 & 0.4 mag & 0.2696 & 41.7 & 11 & $2 \times 10^{-5}$ & 0.2028 & 0.3725 & 9.3 & 9 & 41\% \\
Two-night\tablenotemark{e} & 0.1 mag & 0.2737 & 68.5 & 52 & 6.3\% & 0.2026 & 0.3605 & 34.6 & 50 & 95\% \\
Two-night\tablenotemark{e} & 0.2 mag & 0.2750 & 48.2 & 25 & $4 \times 10^{-3}$ & 0.2053 & 0.3605 & 15.7 & 23 & 87\% \\ 
Two-night\tablenotemark{e} & 0.3 mag & 0.2676 & 46.4 & 16 & $9 \times 10^{-5}$ & 0.2017 & 0.3606 & 14.8 & 14 & 39\% \\
Two-night\tablenotemark{e} & 0.4 mag & 0.2657 & 30.5 & 11 & $1 \times 10^{-3}$ & 0.2084 & 0.3530 & 8.6 & 9 & 48\% \\
\enddata
\tablenotetext{a}{$\alpha_d$ from Equation \ref{eq:magpowerdiff}. Applies from $R=20.0$ to 25.3}
\tablenotetext{b}{$\alpha_d$ from Equation \ref{eq:magpowerdiff} for $R=20.0$ to 23.5}
\tablenotetext{c}{$\alpha_d$ from Equation \ref{eq:magpowerdiff} for $R=23.5$ to 25.3}
\tablenotetext{d}{Probability that a true $\chi^2$ distribution with the same number of degrees of freedom would produce a value greater than or equal to the $\chi^2$ of the fit. All but one of the constant power laws are excluded with at least 95\% confidence (usually much more), and all of the broken power laws are accepted by the same criterion.}
\tablenotetext{e}{Two-night refers to our analysis that considered only asteroids that were each detected on both nights}
\end{deluxetable*}

We explored the robustness of the broken power law by re-binning the completeness-corrected histograms in larger bins of width 0.2, 0.3, and 0.4 mag and fitting both constant and broken power laws to each re-binned histogram. In all twelve cases (four bin sizes for three different data sets), the best fit was obtained for a power law break point at either $R=23.4$ or $R=23.5$ --- a remarkable level of consistency since the fit probed breakpoints from $R=21.5$ to 24.0. Since the majority of the fits had the break at $R=23.5$, we have adopted this value in all cases for the fits presented in Figure \ref{fig:logdiff} and Table \ref{tab:powerfits}. The best-fit slopes are also very consistent for different bin sizes. All but one of the constant power law fits are rejected with $>95$\% confidence (much greater in most cases), while all of the broken power law fits are accepted by the same criterion. Hence, this analysis strongly favors the broken power law. It also appears to validate the predictions of \citet{Bottke2005b}, \citet{deElia2007}, and others of an upturn in the size-frequency distribution of MBAs in the regime of nonzero tensile strength. We will present a fuller comparison of observations with theory, including an estimate of the actual size at which the transition occurs, in our companion paper on the absolute magnitude distribution (Heinze et al., in prep.).

We obtain our final measurements of the slopes and their uncertainties by taking averages and standard deviations of the values in Table \ref{tab:powerfits} for each data set. These averaged values are given in the `Log of Corrected Histogram' rows in Table \ref{tab:brokefit}. In every case the fitted slopes are consistent across the data sets. Averaging across all data sets, we find $\alpha_d = 0.268 \pm 0.005$ for the constant power law and $\alpha_d = 0.203 \pm 0.003$ and $0.359 \pm 0.008$ respectively for the bright and faint regimes of the broken power law.

The best-fit slopes in the brighter magnitude regime of the broken power law are remarkably shallow. Formally, $\alpha_d = 0.203$ implies a differential size slope $b_d = 2.013$ and (if the power law persisted) a cumulative size slope $b_c = 1.013$. No previous result has found a slope this shallow. However, this is easily explained by the fact that the shallow slope does not persist at magnitudes brighter than $R=20$ or fainter than $R=23.5$. Hence, it cannot be seen in the cumulative but only in the differential distribution. In fact, by integrating our best-fit broken power law for the differential distribution, we have determined that it is entirely consistent with the slope $\alpha_c \sim 0.27$ found for the cumulative distribution by both us and \citet{Gladman2009} in the same magnitude regime.

The implications of our very shallow slope measurement remain surprising. Since the apparent magnitude distribution is effectively a weighted average of absolute magnitude (and hence size) distributions over the main belt, our measurement of $\alpha_d \sim 0.203$ can only be explained if the absolute magnitude distribution has a slope at least this shallow somewhere in the main belt. This in turn requires that the differential size slope really is as shallow as $b_d \sim 2.013$, unless the true size distribution is masked by size-dependent variations in albedo.

\section{Analysis and Fits for the Faintest Asteroids} \label{sec:statbias}

Since our fake asteroid simulation indicated a detection rate above 10\% out to $R$ mag 25.6 (see Figure \ref{fig:completeness}), it seems intially surprising that the completeness correction would fail dramatically in the magnitude range 25.4--25.6, as Figures \ref{fig:realhist} and \ref{fig:diffcomp30a} demonstrate it does. \citet{Gladman2009} reported a similar effect near their completeness limit of $R=23.5$, and suggested the cause was the different handling of fake vs. real asteroids. In their analysis, real asteroids were detected by automated software but additionally screened by a human and rejected if the detections seemed to be dubious (i.e., had very low significance); while fake asteroids were not subjected to the same human screening. Hence, they suggested that very faint real asteroids would probably be discarded while fake asteroids at the same magnitude might be flagged as detected. This explanation does not appear viable for our data, since we subjected both fake and real asteroids to essentially the same manual screening, as well as identical significance thresholds for detection (see \S \ref{sec:fakematch}).

\begin{figure*}
\plottwo{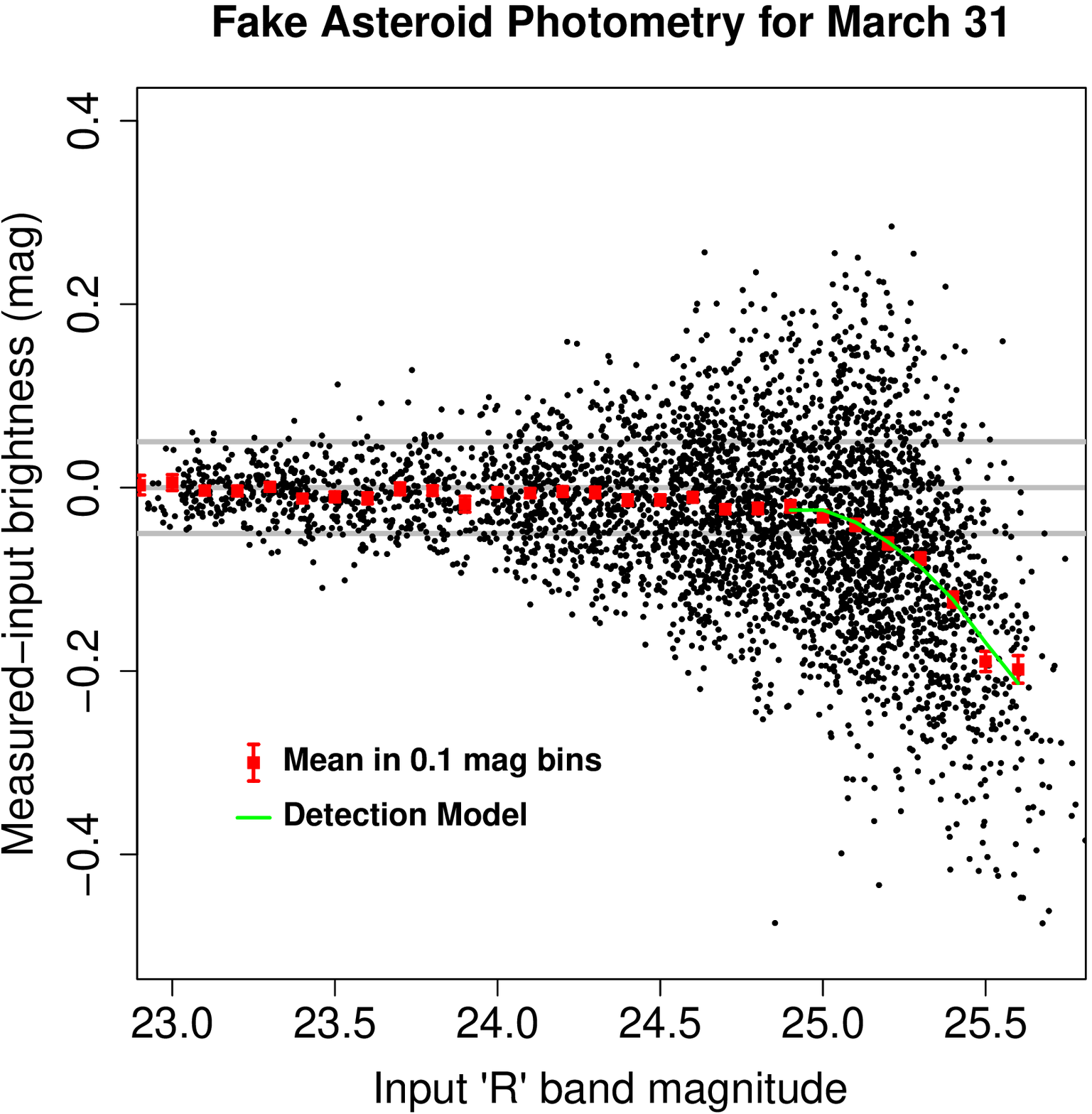}{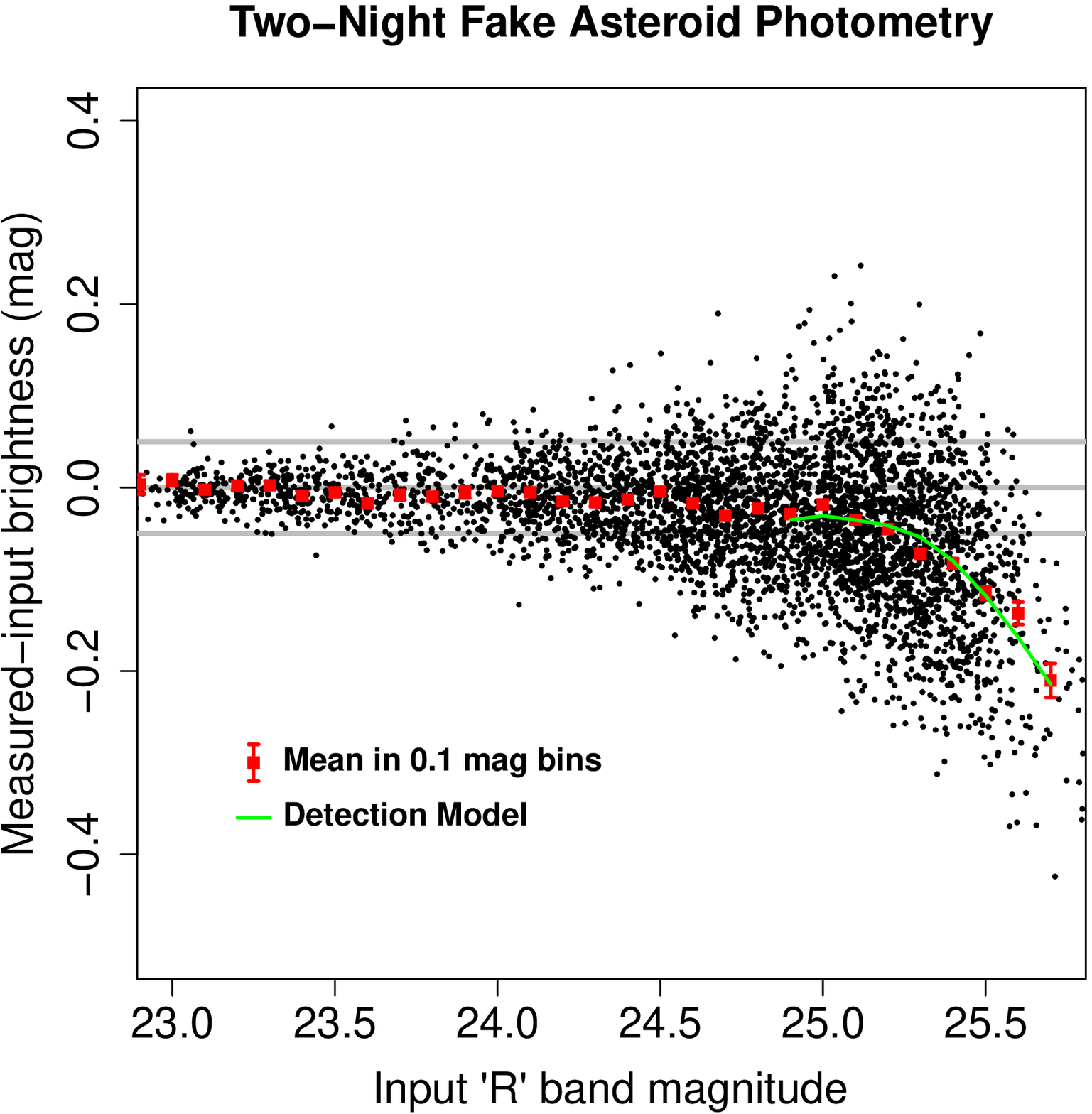}
\caption{Flux overestimation bias for measured magnitudes in our fake asteroid simulation. This is not a flaw of the simulation: instead, it reveals a measurement bias that affects the real asteroids too. A very faint asteroid will be detected only if it falls on a positive fluctuation of the sky noise, and hence its measured magnitude will be too bright. Gray lines indicate magnitude differences of 0.0 and $\pm 0.05$ mag. The green curves show the models of this effect that we describe in \S \ref{sec:analyticalmod}; the plot for March 30 (not shown) is very similar to March 31. Note that for the two-night fit the averaging of measured magnitudes has reduced the measurement scatter, and we have been able to extend the fit 0.1 mag fainter (to mag 25.7) relative to the single-night analyses.
\label{fig:iomag}}
\end{figure*}

The true explanation is the statistical bias that affects our fake asteroids at the faintest magnitudes, as illustrated by Figure \ref{fig:iomag}. This bias is not an error of our simulation; rather, it indicates the simulation correctly captured a known phenomenon called flux overestimation bias\footnote{\scriptsize{\url{https://old.ipac.caltech.edu/2mass/releases/allsky/doc/sec5\_3a.html}} provides a further interesting discussion of flux overestimation bias.} that affects real astronomical objects when brightnesses are measured near the detection threshold of a flux-limited survey. The bias arises from the fact that all measured fluxes are affected by random noise, and that objects fainter than the 50\% completeness limit are likely to be detected only if their fluxes are boosted by a large positive realization of the random noise. Therefore, the faintest objects will be detected only under circumstances that will also cause their measured fluxes to be offset brightward of the true values. 

It follows that our fake asteroid simulation correctly indicates our detection completeness as a function of true magnitude, but does not account for the fact that the faintest real objects, if detected at all, will be measured to be significantly brighter than they really are. Hence, for the faintest asteroids, the completeness correction is not an `apples to apples' comparison. Unavoidably, it compares input, fake-asteroid magnitudes that are free of flux overestimation bias with measured magnitudes of real asteroids that do experience the bias.

This is why the corrected counts of real asteroids fall far below the power law fit at magnitudes slightly fainter than 25.3. The completeness test accurately indicates that more than 10\% of real asteroids are detected even at magnitude 25.6, but does not account for the fact that these objects are measured as being brighter than they really are and hence do not populate the 25.6 magnitude bin in the histogram of real asteroids. It is interesting to speculate that the similar effect seen by \citet{Gladman2009}, which they attribute to the imperfect realism of their fake asteroid test, may in fact represent this same unavoidable bias.

\subsection{Modeling and Correcting Flux Overestimation Bias} \label{sec:analyticalmod}

We wish to correct the bias in our measured magnitudes in order to probe the statistics of the faintest asteroids. We do this by adopting a simple but physically motivated three-parameter analytical model of the detection process, and using statistical simulations to solve for the model parameters that best describe the results of the fake asteroid simulation. 

The dominant source of noise in our flux measurements of faint objects is the Poisson shot noise of the sky background. Even for a very dark sky with a brightness of 22 magnitudes per square arcsecond, the sky contributes 50 times more photons within a photometric aperture of radius 1 arcsecond than does a 25th magnitude star or asteroid. Thanks to this huge number of sky photons, the Poisson noise of the sky background can be well approximated by a Gaussian random variable.

In our model, we therefore allow the true flux of an asteroid to be modified by random sky noise with a Gaussian distribution: the standard deviation $\sigma_{sky}$ of this distribution is the first of the model's three parameters. Presumably, $\sigma_{sky}$ will be related to the square root of the typical number of sky photons detected over the angular size of the PSF. The asteroid is detected if its modified brightness exceeds $\sigma_{sky}$ by a factor $N_{thresh}$, which is the second parameter of our simulation and is expected to be in the range 7--8, since it should correspond approximately to the detection thresholds of $7.43 \sigma$ and $7.28\sigma$ derived in \S \ref{sec:fp}. However, we do not expect the likelihood of detection to drop discontinuously from 100\% to 0\% when the modified flux drops below $N_{thresh} \sigma_{sky}$. It should instead exhibit a gradual decrease in detection probablility --- albeit less gradual than the histogram-based completeness curves shown in Figure \ref{fig:completeness}, since they exhibit additional broadening due to the sky noise. 

We invested considerable thought in choosing a one-parameter mathematical form for this gradual sensitivity decrease, and we finally adopted a chi-square distribution. We choose the chi-square distribution because our detection process (\S \ref{sec:digidet} and \ref{sec:flist}) involves estimation of the sky noise by taking the standard deviation of flux measurements in a sample of image regions supposed to represent blank sky. It is a well known statistical result that if $S$ is the sample standard deviation for $n_{dof}$ samples drawn from a Gaussian distribution with true standard deviation $\sigma$, the quantity $(n_{dof}-1)S^2/\sigma^2$ is distributed according to a chi-square distribution with $n_{dof}-1$ degrees of freedom. Hence, $n_{dof}$ becomes the third and final parameter in our analytical model of the detection process.

Given an input `true' flux for a faint asteroid, the detection process is modeled as follows. First, the true flux is modified by the addition of a Gaussian random variable with mean zero and standard deviation $\sigma_{sky}$. Then, the detection threshold $N_{thresh} \sigma_{sky}$ is multiplied by a scale factor equal to $\sqrt{X_{chi}/(n_{dof}-1)}$, where $X_{chi}$ is a random variable drawn from a chi-square distribution with $n_{dof}-1$ degrees of freedom. Note that the expected value of the scale factor is unity. Finally, the asteroid is classified as detected if the modified flux is greater than or equal to the scaled detection threshold, and undetected otherwise.

To solve for the three parameters ($\sigma_{sky}$, $N_{thresh}$, and $n_{dof}$) of our analytical detection model, we first attempt a statistical fit to the results from the fake asteroid simulation. We perform a statistical simulation aimed at matching both the histogram-based completeness curves plotted in Figure \ref{fig:completeness}, and the mean offsets between measured and input magnitudes which we have plotted in Figure \ref{fig:iomag}. Judging the completeness to be near 100\% at magnitudes brighter than 24.9, and so low at magnitudes fainter than 25.6 that the fake asteroid results are unreliable, we attempt to fit the results only in the magnitude range 24.9--25.6. This range includes eight magnitude bins, in each of which we fit two data points (completeness and mean magnitude), so the three-parameter fit is heavily overconstrained. For the two-night data with its greater sensitivity, we fit 9 bins from magnitude 24.9--25.7.

We perform the fitting using a statistical simulation in which the actual input magnitudes of the $\sim 6600$ asteroids used in the fake asteroid simulation are taken as the `true' brightness. Each asteroid is cloned 160 times and each clone is subjected to an independent realization of probabalistic detection based on our analytical detection model. Finally, the histogram-based completeness and the mean measured magnitudes of detected asteroids are calculated and compared with the actual results from the fake asteroid simulation. We find evidence of small systematic errors in the fake asteroid photometry that do not arise from the flux overestimation bias, and to account for these, we allow a constant offset in the mean magnitudes produced by the statistical model. These biases amount to 0.0073 mag, 0.0309, and 0.0376 mag for March 30, 31, and the two-night data respectively. We do not expect the real asteroids to be subject to the same photometric offsets, and we speculate that the errors result from the fake asteroid images being very slightly blurrier than the real ones due to imperfections in the implantation process. Such errors would be expected to produce photometric variations that differ from night to night (because the average seeing was different) and affect faint asteroids more (because the optimized photometric apertures are smaller). They could arise from at least two different causes. As described in \S \ref{sec:fakepix}, our implantation of fake asteroids explicitly allowed registration errors up to 0.1 pixels. Alternatively, it is possible that despite our careful screening process some of the template `stars' were in fact very compact galaxies, and this fraction could have been different from night-to-night since the lists of template stars were independently constructed. In any case, the maximim photometric offset (0.0376 mag) is still less than half the smallest histogram bin (0.1 mag) used in our analysis.

In Table \ref{tab:ansim} we present the best-fit parameters for our analytical model applied to the fake asteroids in each data set, with $\sigma_{sky}$ converted to an equivalent magnitude. The analytical model fits to the detection rates and mean magnitudes are plotted as green curves in Figures \ref{fig:iomag} and \ref{fig:completeness}, respectively. The blue curves in Figure \ref{fig:completeness} also show our chi-square model of the detection rate as function of modified magnitude. The broader dropoff of the actual detection rate relative to this chi-square model is due to the sky noise (parametrized in our model by $\sigma_{sky}$), which is the actual cause of the flux overestimation bias. The best-fit values obtained for $\sigma_{sky}$ and $N_{thresh}$ are near the range we would expect, and lead to limiting magnitudes very close to those we derived from the histogram-based completeness curve. The best fit values of $n_{dof}$ are much smaller than the number of samples used to probe the sky noise ($\sim 200$; see \S \ref{sec:digidet} and \ref{sec:flist}). Since the relative width of the chi-square distribution gets narrower with increasing $n_{dof}$, the small best-fit values of $n_{dof}$ indicate that additional effects besides the variation in the sample standard deviation contribute to the gradual dropoff of detection efficiency, which is hardly surprising. It is probably best to think of $n_{dof}$ simply as parameterizing the gradual sensitivity decrease, and not as the number of degrees of freedom of in a literal chi-square distribution that is realized at any stage of our actual detection process. This does not detract from the value of our analytical model as a useful (though necessarily simplified) representation of our detection process.

\begin{deluxetable*}{llccc}
\tablewidth{0pt}
\tablecaption{Slope $\alpha_d$ for Best-Fit Power Laws\label{tab:brokefit}}
\tablehead{ &  & \colhead{Constant Power Law} & \multicolumn{2}{c}{Broken Power Law} \\
\colhead{Data Set} & \colhead{Method} & \colhead{$R>20$ mag} & \colhead{$R=20$ to 23.5 mag} & \colhead{$R>23.5$ mag}
}
\startdata
March 30 & Log of Corrected Histogram\tablenotemark{a} & $0.263 \pm 0.004$ & $0.204 \pm 0.004$ & $0.352 \pm 0.006$ \\
March 30 & Analytical Model\tablenotemark{b} & $0.275 \pm 0.013$ & $0.225 \pm 0.023$ & $0.325 \pm 0.025$ \\
March 30 & Average of Both Methods & $0.269 \pm 0.010$ & $0.215 \pm 0.018$ & $0.339 \pm 0.020$ \\
March 31 & Log of Corrected Histogram\tablenotemark{a} & $0.271 \pm 0.002$ & $0.203 \pm 0.002$ & $0.366 \pm 0.008$ \\
March 31 & Analytical Model\tablenotemark{b} & $0.280 \pm 0.018$ & $0.240 \pm 0.028$ & $0.310 \pm 0.028$ \\
March 31 & Average of Both Methods & $0.276 \pm 0.013$ & $0.222 \pm 0.024$ & $0.338 \pm 0.028$ \\
Two-night & Log of Corrected Histogram\tablenotemark{a} & $0.271 \pm 0.005$ & $0.205 \pm 0.003$ & $0.359 \pm 0.004$ \\
Two-night & Analytical Model\tablenotemark{b} & $0.283 \pm 0.020$ & $0.230 \pm 0.035$ & $0.330 \pm 0.033$ \\
Two-night & Average of Both Methods & $0.277 \pm 0.020$ & $0.218 \pm 0.026$ & $0.345 \pm 0.025$ \\
All & Overall Averages & $0.277 \pm 0.015$ & $0.218 \pm 0.026$ & $0.340 \pm 0.025$ \\
\enddata
\tablenotetext{a}{Uncertainties obtained for this method are probably unrealistically small.}
\tablenotetext{b}{This model, which corrects for flux overestimation bias, is described in \S \ref{sec:statbias}.}
\end{deluxetable*}

\begin{deluxetable*}{cccccc}
\tablewidth{0pt}
\tablecaption{Analytical Detection Models\label{tab:ansim}}
\tablehead{\colhead{Date} & \colhead{$\sigma_{sky}$} & \colhead{$N_{thresh}$} & \colhead{$n_{dof}$} & \colhead{phot offset} & \colhead{50\% completeness limit}}
\startdata
March 30 & 27.620 mag & 8.040 & 29 & 0.0073 mag & 25.357 mag \\
March 31 & 27.470 mag & 6.825 & 25 & 0.0309 mag & 25.385 mag \\
Both Nights & 27.880 mag & 9.200 & 49 & 0.0376 mag & 25.471 mag \\
\enddata
\end{deluxetable*}

\subsection{Apparent Magnitude Distribution from the Analytical Detection Model} \label{sec:pwrfit02}

We use the analytical detection models derived from our fake asteroid simulations to model the detection of real asteroids assuming that the true distribution is a power law (or broken power law) of the form given in Equation \ref{eq:magpowerdiff}. We attempt to match the histogram of real asteroids in 57 bins of 0.1 magnitude width covering the interval from magnitude 20.0 to 25.6. Since positive sky noise fluctuations can move faint asteroids into brighter bins on the magnitude histogram, we simulate an input power law distribution that extends all the way to magnitude 26.0. We perform fits using both a constant power law and a broken power law with a break point at $R=23.5$. For the constant power law, we solve for the single slope value $\alpha_d$ by simply probing a finely sampled range of values and selecting the one that produced the smallest $\chi^2$ value for the fit. For the broken power law, we solve for the values of $\alpha_d$ in the two magnitude regimes by a 2D grid search. As in the analytical detection model applied to the fake asteroids, hundreds of times more random asteroid clones are generated than are actually detected in the real data, in order to ensure the statistical noise from the simulation makes no significant contribution to the final uncertainty. 

By construction, the simulation explicitly models both incompleteness and photometric bias to predict the raw magnitude histogram that should be observed given an input power law. Hence, while the much simpler analysis of \S \ref{sec:diffpower01} fitted the logarithm of the completeness-corrected histogram without accounting for photometric bias, we now model the raw histogram itself. The handling of uncertainties is also necessarily different. In \S \ref{sec:diffpower01}, we used the uncertainty on the logarithm (Equation \ref{eq:logsig}), including a contribution from the completeness correction where appropriate. Now, however, we seek to match raw asteroid count $n$ in each bin, with uncertainty given by $\sigma_n = \sqrt n$.

\begin{figure*}
\plottwo{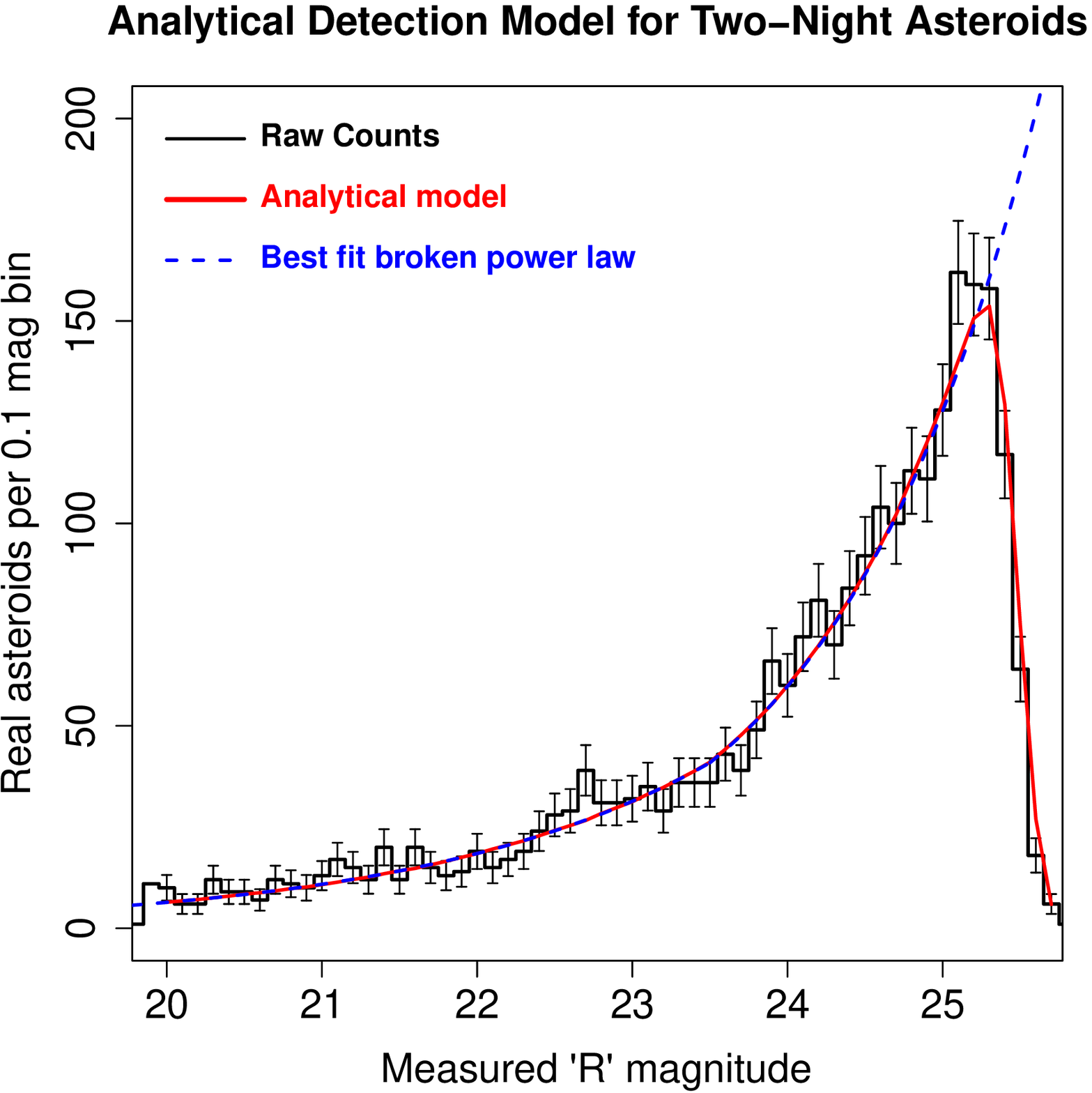}{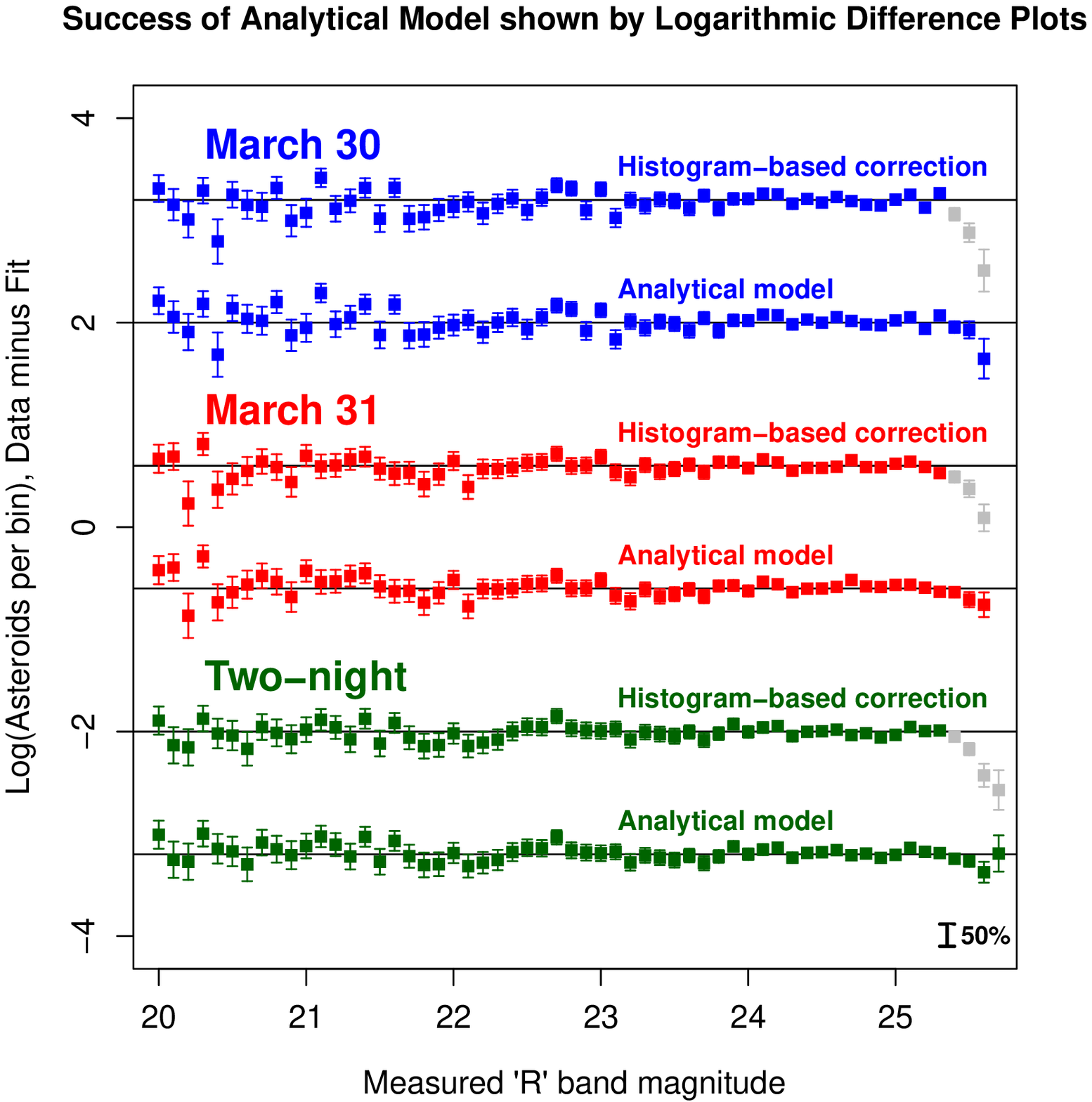}
\caption{Success of our analytical detection model (\S \ref{sec:analyticalmod}) at fitting the magnitude histograms of real asteroids. Unlike the simple histogram-based completeness correction (\S \ref{sec:diffpower01}), the model accounts for flux overestimation bias and successfully fits the dropoff in detections at faint magnitudes. We conclude the dropoff is caused by flux overestimation bias, rather than by a failure of the fake asteroid simulation or a real dearth of very small asteroids. {\em Left:} The same magnitude histogram of two-night asteroids as shown in Figure \ref{fig:realhist}, this time fit with the analytical detection model. The single-night results (not shown) look very similar.  {\em Right:} Logarithmic difference plots of the best-fit broken power laws using the old histogram-based completeness correction (e.g. Figure \ref{fig:diffcomp30a}) compared to those using the new analyical detection model.  With the new model, the faintest points are no longer statistical outliers: instead, the model enables us to probe the abundance of these extremely faint asteroids. 
\label{fig:ansim01}}
\end{figure*}

Figure \ref{fig:ansim01} shows the best broken power law fits obtained using our analytical model, demonstrating that our model has achieved its main objective: the faintest data points are no longer statistical outliers, indicating that we have successfully modeled the phenomena that lead to very low numbers of asteroids being measured in these bins. The model therefore represents a significant improvement over the simple, histogram-based completeness correction presented in \S \ref{sec:diffpower01}, and demonstrates that the failure of that correction at magnitudes fainter than 25.3 is indeed caused by flux overestimation bias and is not due to inaccuracy in the fake asteroid simulation, nor to a drop in the abundance of real asteroids.

The best fit constant power law slopes found with our new model are $\alpha_d = 0.275 \pm 0.013$, $0.280 \pm 0.018$, and $0.283 \pm 0.020$ for March 30, 31, and the two-night data respectively. The new values are consistent with but slightly steeper than the average value of $\alpha_d=0.268 \pm 0.005$ obtained for constant power law fits to the corrected histograms in \S \ref{sec:diffpower01}. The $\chi^2$ values of the new constant power law fits are 84.9, 64.2, and 58.4 for March 30, 31, and the two-night data respectively, with 55 degrees of freedom for the single-night data and 56 for the two-night fits, which reach 0.1 mag fainter. The probabilities of getting $\chi^2$ values at least this high with these numbers of degrees of freedom are $6\times10^{-3}$, 19\%, and 39\%, respectively. Hence, in contrast to the simpler, logarithmic fits of \S \ref{sec:diffpower01}, where the constant power laws consistently failed to produce an acceptable fit, both the March 31 and two-night fits are acceptable, though the March 30 fit is formally rejected with 99.4\% confidence.

\begin{figure}
\includegraphics[scale=0.5]{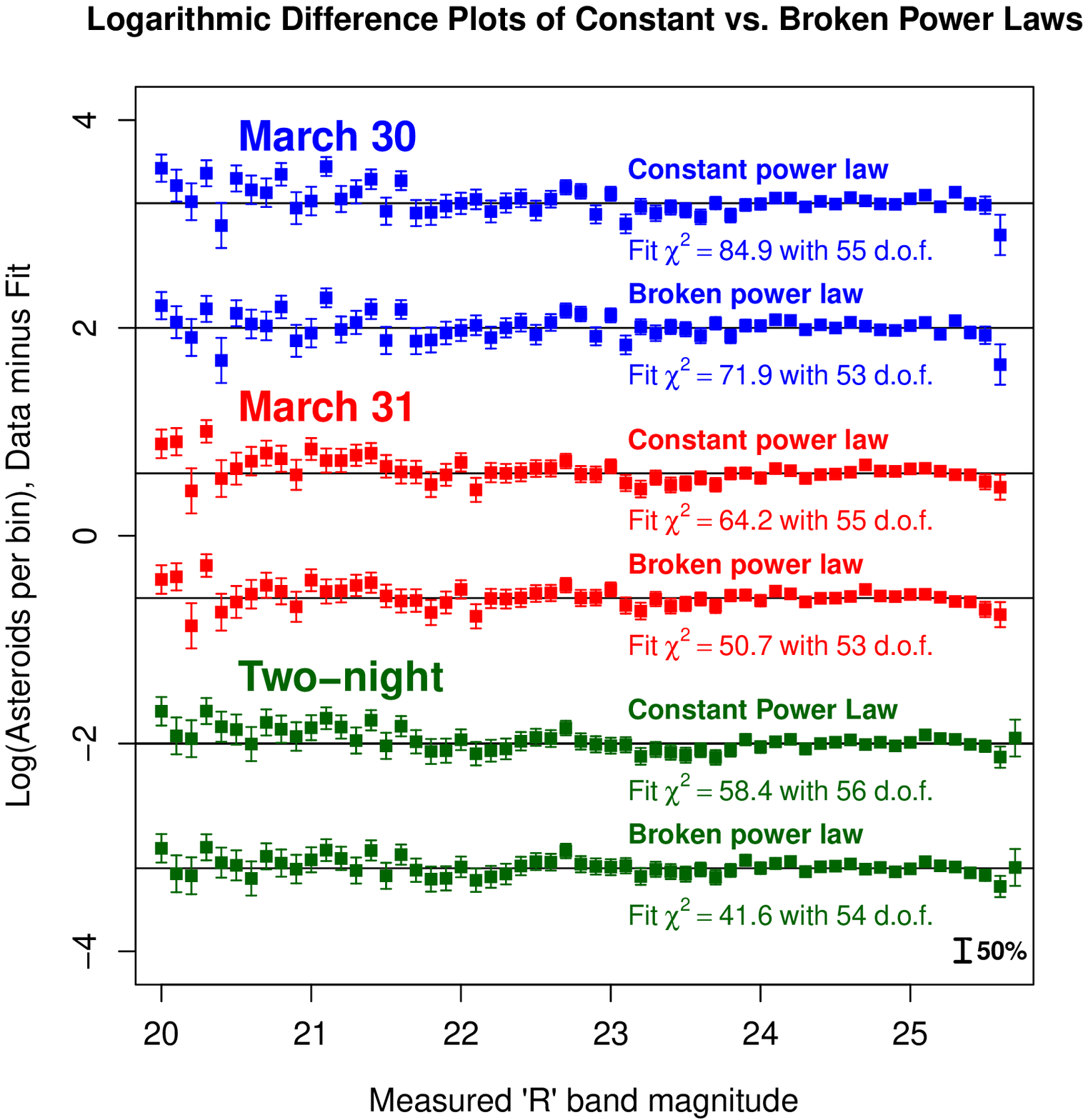}
\caption{Logarithmic difference plots using our analytical detection model (\S \ref{sec:analyticalmod}) to fit both constant vs. broken power laws and compare the fit quality, as we did with histogram-based completeness correction in (\S \ref{sec:diffpower01}). The broken power law fits are consistently better, although the differences in the fits are visually subtle, seen mainly in the fact that the data lie above the constant power law fits at bright magnitudes and below them near the break-point at mag 23.5. The probabilities of getting $\chi^2$ values as large as those observed are $6 \times 10^{-3}$, 19\%, and 39\% for constant power law fits to the respective data sets, and 4.3\%, 56\%, and 89\% for the broken power laws. Formally, one out of three fits is unacceptable at the 95\% confidence level in each case, so the broken power law does not exhibit the same level of dominance here as in our earlier analysis (e.g. Figure \ref{fig:logdiff}). Nevertheless, the broken power law is preferred in every case.
\label{fig:difflog02}}
\end{figure}

For broken power laws, we find in the bright regime ($R=20.0$ to 23.5 mag) that $\alpha_d = 0.225\pm 0.023$, $0.240 \pm 0.028$, and $0.230 \pm 0.035$ for March 30, 31, and the two-night data respectively, while in the faint regime ($R=23.5$ to 25.6 or 25.7 mag) we have $\alpha_d = 0.325 \pm 0.025$, $0.310 \pm 0.028$, and $0.330 \pm 0.033$ for the respective data sets. Interestingly, the slopes in both regimes have become less extreme than for the simpler fits in \S \ref{sec:diffpower01}. For the best broken power law fits using the new model, we have $\chi^2 = 71.9$, 50.7, and 41.6 for the respective data sets, where the single-night fits have 53 degrees of freedom and the two-night fit has 54. The probabilities of getting these values are 4.3\%, 56\%, and 89\% respectively. Hence, while the other two fits are excellent, the one for March 30 remains somewhat marginal even for a broken power law (though greatly improved relative to the constant power law).

Although the evidence for the broken power law does not seem as strong under the new type of fit, it clearly remains a better description of the data than a constant power law. The evidence for a break in the power law appears still more compelling when we consider the upturn in the cumulative distribution from Figure \ref{fig:cumdist}, which is hard to explain any other way and which happens consistently on both nights at a magnitude considerably brighter than the sensitivity limit. Finally, the fact that the new fits produce less extreme slopes in the bright regime makes the broken power law seem more believable. As mentioned in \S \ref{sec:diffpower01}, the very shallow slope of $\alpha_d \sim 0.203$ found by our earlier analysis seems somewhat improbable a priori.

In contrast to the steeper bright-end slopes, the faint-end slopes are considerably shallower under the new analysis. Concerned that this might indicate a bug in the analytical model (or a real dearth of extremely faint asteroids), we probed both possibilities. Constraining the fit to $R\le25.0$ mag still produced a shallower slope, and additionally tweaking the model to predict constant 100\% completeness everywhere also failed to change the result significantly. Hence, the slope differences are not due to anomalous behavior of the faintest magnitude bins nor to flaws in our analytical model. We believe, instead, that the difference arises from the fact that we previously fit the logarithm of the histogram, while the new model fits the actual observed counts. The uncertainties behave somewhat differently: in the former case they are given by Equation \ref{eq:logsig}; while in the latter we simply take the uncertainty to be $\sqrt n$. Both methods are approximations because they implictly assume symmetrical Gaussian uncertainties, when the number $n$ of asteroids per bin would actually be expected to have a Poisson distribution. The approximations are reasonable and difficult to avoid, but their differing imperfections likely explain the different slope and $\chi^2$ values of the fits in this section as compared to those in \S \ref{sec:diffpower01}.

Since it is not easy to tell which method is more `right', the slopes we quote in the abstract are the averages of the two values. The various slopes and our final adopted averages are given in Table \ref{tab:brokefit}.

Using our analytical detection model, we have successfully matched the observed histograms out to $R=25.6$. This gives us confidence that the best-fit power laws are meaningful out to this magnitude, and enables us to revisit the on-sky density calculations of \S \ref{sec:skydens}. We focus on March 31 as being most representative of the true sky density at opposition, since the field was accurately centered on the antisolar point. For our starting point, we take the corrected count of $697 \pm 15$ asteroids per square degree for $R < 25.0$ mag (Table \ref{tab:skydens}), since the completness there is still near 100\%. Integrating from mag 25.0 to 25.6 using the best-fit broken power law produced by our analytical model, we find $334 \pm 18$ asteroids per square degree in this magnitude range, and hence the total sky density of asteroids brighter than $R=25.6$ mag is $1031 \pm 23$ per square degree.

\section{Conclusion} \label{sec:conc}

We have used the technique of digital tracking to leverage the remarkable wide-field capability of DECam on the 4m Blanco telescope and perform the deepest ever survey of main belt asteroids. In a single DECam field (about 3 square degrees), we have detected 3234 distinct asteroids, of which 3123 are confirmed on two consecutive nights.

We have analyzed our detection rate as a function of magnitude using a carefully constructed fake asteroid simulation, which allows us to correct for incompleteness in our asteroid counts out to $R$ magnitude 25.3. Fainter than this, our completeness correction falters due to flux overestimation bias. This bias arises because the faintest objects are detected only if their flux is augmented by a positive realization of random noise (mostly Poisson noise from the sky backround in our case). Hence, when detected at all, the faintest asteroids are measured as systematically brighter than they really are. Building on our fake asteroid simulation, we construct an analytical model of our detection process that fits and corrects the flux overestimation bias, enabling our statistical analysis of asteroids down to $R$ magnitude 25.6.

We find a sky density of $697 \pm 15$ asteroids per square degree brighter than $R = 25.0$ mag, and $1031 \pm 23$ brighter than $R = 25.6$ mag. These numbers apply only at opposition from the Sun: away from opposition, asteroids become rapidly fainter due to their phase functions, while away from the ecliptic the sky density drops even faster due to the relative scarcity of asteroids in highly inclined orbits \citep{Terai2007,Terai2013}.

Consistent with \citet{Gladman2009}, we find the slope of the apparent magnitude distribution of asteroids is much steeper at magnitudes brighter than $R=19$ than it is at magnitudes fainter than $R=20$. From $R=20$ to the faint limit of our survey at $R \sim 25.6$, the differential magnitude distribution can be fit with a single power law (Equation \ref{eq:magpowerdiff}) with $\alpha_d \sim 0.28 \pm 0.02$. However, a better fit can be obtained with a broken power law that becomes steeper for magnitudes fainter than a break point at $R=23.5$. This implies that extremely faint asteroids are {\em more abundant} than extrapolating power laws fit to brighter objects would lead us to expect --- contrary to some reports claiming a reduction in asteroid abundance at magnitudes fainter than $23.5$. The average best fit slopes we find for this broken power law are $\alpha_d = 0.218 \pm 0.026$ for the bright regime ($R=20$ to 23.5 mag), and $\alpha_d = 0.340 \pm 0.025$ for the faint regime ($R=23.5$ to 25.6 mag).

If the broken power law is correct, there must be a range of size and distance somewhere in the main asteroid belt where the size distribution is very shallow (though not necessarily as shallow as our initial fits in \S \ref{sec:diffpower01} seemed to imply); and at smaller sizes the distribution must steepen significantly. A steepening of the distribution of main belt asteroids at very small sizes has been predicted (see, e.g., figures in both \citet{OBrien2005} and \citet{deElia2007} that show theoretical and observed asteroid size distributions). This steepening of the power laws is predicted because small asteroids are expected to have nonzero tensile strength and hence to be more difficult to disrupt \citep[per unit mass; see][]{Bottke2005a}. This transition in asteroid properties may produce a collisional wave, a possibility we explore further in our companion paper (Heinze et al., in prep.) on the absolute magnitude distribution of main belt asteroids. Herein, we have laid essential groundwork for the companion paper by identifying and characterizing our sample of two-night asteroids, for all of which we can calculate accurate distances (and hence absolute magnitudes) using the RRV method of of \citet{curves} and \citet{LinRRV}.

\section{Acknowledgments} 
We acknowledge essential support and assistance from John Tonry and Larry Denneau of the Asteroid Terrestrial-Impact Last Alert System (ATLAS). This work was substantially enabled by ATLAS computing resources, to which, as an ATLAS postdoctoral researcher, A. H. had access during periods of reduced data-processing demand from ATLAS itself. Support for the ATLAS survey is provided by NASA grants NN12AR55G and 80NSSC18K0284 under the guidance of Lindley Johnson and Kelly Fast.

Based on observations at Cerro Tololo Inter-American Observatory, National Optical Astronomy Observatory (NOAO Prop. ID: 2014A-0496; PI: A. Heinze), which is operated by the Association of Universities for Research in Astronomy (AURA) under a cooperative agreement with the National Science Foundation. 

This project used data obtained with the Dark Energy Camera (DECam), which was constructed by the Dark Energy Survey (DES) collaboration.
Funding for the DES Projects has been provided by 
the U.S. Department of Energy, 
the U.S. National Science Foundation, 
the Ministry of Science and Education of Spain, 
the Science and Technology Facilities Council of the United Kingdom, 
the Higher Education Funding Council for England, 
the National Center for Supercomputing Applications at the University of Illinois at Urbana-Champaign, 
the Kavli Institute of Cosmological Physics at the University of Chicago, 
the Center for Cosmology and Astro-Particle Physics at the Ohio State University, 
the Mitchell Institute for Fundamental Physics and Astronomy at Texas A\&M University, 
Financiadora de Estudos e Projetos, Funda{\c c}{\~a}o Carlos Chagas Filho de Amparo {\`a} Pesquisa do Estado do Rio de Janeiro, 
Conselho Nacional de Desenvolvimento Cient{\'i}fico e Tecnol{\'o}gico and the Minist{\'e}rio da Ci{\^e}ncia, Tecnologia e Inovac{\~a}o, 
the Deutsche Forschungsgemeinschaft, 
and the Collaborating Institutions in the Dark Energy Survey. 

The Collaborating Institutions are 
Argonne National Laboratory, 
the University of California at Santa Cruz, 
the University of Cambridge, 
Centro de Investigaciones En{\'e}rgeticas, Medioambientales y Tecnol{\'o}gicas-Madrid, 
the University of Chicago, 
University College London, 
the DES-Brazil Consortium, 
the University of Edinburgh, 
the Eidgen{\"o}ssische Technische Hoch\-schule (ETH) Z{\"u}rich, 
Fermi National Accelerator Laboratory, 
the University of Illinois at Urbana-Champaign, 
the Institut de Ci{\`e}ncies de l'Espai (IEEC/CSIC), 
the Institut de F{\'i}sica d'Altes Energies, 
Lawrence Berkeley National Laboratory, 
the Ludwig-Maximilians Universit{\"a}t M{\"u}nchen and the associated Excellence Cluster Universe, 
the University of Michigan, 
{the} National Optical Astronomy Observatory, 
the University of Nottingham, 
the Ohio State University, 
the University of Pennsylvania, 
the University of Portsmouth, 
SLAC National Accelerator Laboratory, 
Stanford University, 
the University of Sussex, 
and Texas A\&M University.

This publication makes use of the SIMBAD online database,
operated at CDS, Strasbourg, France, and the VizieR online database (see \citet{vizier}).

This publication makes use of data products from the Two Micron All Sky Survey, 
which is a joint project of the University of Massachusetts and the Infrared Processing
and Analysis Center/California Institute of Technology, funded by the National Aeronautics
and Space Administration and the National Science Foundation.

We have also made extensive use of information and code from \citet{nrc}. 

We have used digitized images from the Palomar Sky Survey 
(available from \url{http://stdatu.stsci.edu/cgi-bin/dss\_form}),
 which were produced at the Space 
Telescope Science Institute under U.S. Government grant NAG W-2166. 
The images of these surveys are based on photographic data obtained 
using the Oschin Schmidt Telescope on Palomar Mountain and the UK Schmidt Telescope.

Facilities: \facility{4m Blanco}

\end{document}